\documentclass[manuscript]{acmart} 
\usepackage{graphicx}
\usepackage{subcaption}
\usepackage{soul}
\usepackage[dvipsnames]{xcolor}
\usepackage{booktabs}
\usepackage{makecell}
\usepackage{array}
\usepackage{multirow}
\usepackage{siunitx}

\AtBeginDocument{%
  }

\acmSubmissionID{8264}

\begin{document}

\title{RF-Behavior: A Multimodal Radio-Frequency Dataset for Activity and Affective Behavior Analysis}

\author{Si Zuo}
\authornote{Both authors contributed equally to this research.}
\affiliation{%
  \institution{Aalto University}
  \city{Espoo}
  \country{Finland}
}
\email{si.zuo@aalto.fi}

\author{Yuqing Song}
\authornotemark[1]
\affiliation{%
  \institution{Aalto University}
  \city{Espoo}
  \country{Finland}
}
\email{yuqing.song@aalto.fi}

\author{Jin Han}
\affiliation{%
  \institution{Aalto University}
  \city{Espoo}
  \country{Finland}
}
\email{jin.1.han@aalto.fi}

\author{Sahar Golipoor}
\affiliation{%
  \institution{Aalto University}
  \city{Espoo}
  \country{Finland}
}
\email{sahar.golipoor@aalto.fi}

\author{Ying Liu}
\authornote{Ying Liu is the corresponding author}
\affiliation{%
  \institution{Aalto University}
  \city{Espoo}
  \country{Finland}
}
\email{ying.2.liu@aalto.fi}

\author{Xujun Ma}
\affiliation{%
  \institution{Southeast University}
  \city{Nanjing}
  \country{China}
}\email{xujunma@seu.edu.cn}

\author{Petter Holme}
\affiliation{%
  \institution{Aalto University}
  \city{Espoo}
  \country{Finland}
}
\email{petter.holme@aalto.fi}

\author{Stephan Sigg}
\affiliation{%
  \institution{Aalto University}
  \city{Espoo}
  \country{Finland}
}
\email{stephan.sigg@aalto.fi}

\renewcommand{\shortauthors}{Zuo et al.}

\begin{abstract}
We introduce \textit{RF-Behavior}, a multimodal radio frequency dataset for comprehensive human activity and affective behavior analysis. We collected data from 61~participants performing 21~gestures, 10~activities, and 6~affective behavior expressions from three environmental settings: lab, living room and industrial factory. Data were captured using synchronized sensors, including 13~radars (8 ground-mounted and 5~ceiling-mounted), 6 to 8~RFID tags (attached to arms and torso) and a LoRa transmitter/receiver. 
Inertial measurement units (IMUs) and 24~infrared cameras are used to provide precise motion ground truth. 
RF-Behavior comprises a unified multimodal dataset spanning a wide spectrum of human behavior, from brief gestures to activities and affective behaviors, enabling research on multi-task learning across motion and affective behavior recognition. Benchmark results demonstrate that the strategic sensor placement is complementary across modalities, with distinct performance characteristics for different behavioral categories.
\end{abstract}



\maketitle

\section{Introduction}
Robust recognition of hand gestures, daily activities, and affective behavior is essential for applications of human-computer interaction, ambient intelligence, or ubiquitous computing~\cite{picard2000affective, cicirelli2019internet}. 

Despite this demand, current behavioral sensing datasets remain fragmented along two dimensions.
First, recent multimodal RF datasets have begun to cover human actions at several kinematic scales within a single collection. XRF55~\cite{wang2024xrf55}, for example, spans 55 action classes from fine-grained human--object interactions to whole-body motions, and MM-Fi~\cite{yang2023mm} supports pose and activity analysis. 
However, these collections target discrete actions of short duration and do not extend to affective behavioral states that last over minutes. 
Second, affective computing datasets capture states of longer duration, but they rely predominantly on cameras, microphones, or physiological instrumentation~\cite{busso2008iemocap,mollahosseini2017affectnet,koelstra2011deap}, which are difficult to deploy in privacy-sensitive everyday spaces. 
For these reasons, learned representations rarely transfer across behavioral scales, and longer duration affective context cannot be grounded in physical movements of shorter duration that have caused it~\cite{thukral2025cross}. 
In contrast to all prior work, RF-Behavior couples micro-kinematics, macro-kinematics, and affective behavioral expressions in one synchronized, RF-centric, multi-environment collection.

Specifically, RF-Behavior jointly captures hand gestures, full-body activities, and affective behavioral expressions on same and different participants, in the same and different environments, with an heterogeneous RF, IMU, and motion-capture sensor stack.
Our affective labels represent physical behavioral expressions (e.g., variations in movement speed, gait, or posture) associated with affective states (rather than internal psychological emotions).

Activity recognition datasets commonly feature on-body IMU or RGB~\cite{xu2025exploring}. 
Examples are, the UCI HAR~\cite{anguita2013public}, WISDM~\cite{kwapisz2011activity}, and PAMAP2~\cite{reiss2012introducing} datasets (smartphone accelerometer and gyroscope), and of course the OPPORTUNITY dataset~\cite{roggen2010collecting}, which combined multiple body-worn sensors for activities of daily living. 
Furthermore, Video-based datasets like Kinetics~\cite{kay2017kinetics} and ActivityNet~\cite{caba2015activitynet} provide rich visual information from environmental mounted camera systems for macro-activities.
On the other hand, affective computing datasets predominantly rely on facial expressions, speech, and physiological signals. 
Examples are DEAP~\cite{koelstra2011deap} (EEG/physiology and video) or AffectNet~\cite{mollahosseini2017affectnet} (facial images). 
Also, speech-based datasets like IEMOCAP~\cite{busso2008iemocap} can provide rich affective cues.
A limitation of speech-based datasets are noisy environments or when subjects are silent, so that some multimodal datasets, such as RECOLA~\cite{ringeval2013introducing}, add audio-visual modalities. 

While video and audio provide rich cues for the recognition of activities, gestures and affective states, an installation of these sensors is not possible in many public, private or also professional environments due to privacy regulations and concerns. 
 
We therefore present a comprehensive multimodal dataset~\textit{RF-Behavior} for the recognition of activity, gesture and affective states, which includes privacy-preserving, non-intrusive affective sensing scenarios where states are inferred purely from macroscopic behavioral expressions.
The various instrumentations include up to 13~millimeter-wave radars (8 ground-mounted and 5 ceiling-mounted) for multi-angle RF, 6 to 8 RFID tags (body mounted), LoRa-based recognition, body-worn IMUs, and time-series motion data from up to 24 infrared cameras. 
Data were collected from 61~participants across controlled lab settings and real-world environments (living room and industrial factory). 
Each participant performed 21~hand gestures, 10~activities, and 6~affective behaviors. 
This allows research in multimodal fusion and robust recognition across viewpoints and may serve as a multimodal benchmark for the emerging field of joint communication and sensing in the domain of ambient intelligence.

RF-Behavior may be useful for (1)~\emph{RF-sensing researchers}: the first public corpus with thirteen synchronized and coordinate-aligned mmWave radars on ground and ceiling (e.g. to study viewpoint robustness, angle-invariant recognition, sensor selection, and multi-radar fusion); (2)~\emph{Multimodal machine-learning researchers}: time-aligned modalities of deliberately heterogeneous resolution, ranging from dense radar point clouds to a single-link LoRa amplitude stream (e.g. to support research on sensor fusion, cross-modal distillation, and multi-task or foundation-model pretraining across behavioral scales); (3)~\emph{Affective-computing researchers}: a corpus that links macroscopic kinematics to induced and self-reported affective behavior expressions, with ambient RF and motion-capture ground truth recorded simultaneously; (4)~\emph{Application developers} interested to assess how RF recognition degrades outside of lab settings in a controlled laboratory and two realistic environments (living room and industrial factory) e.g for eldercare, workplace well-being, smart environments, and industrial safety.

Our contributions are
\begin{enumerate}
    \item a unified behavioral dataset that covers gestures, daily activities and affective behaviors from lab and real-world settings (living room and industrial factory), enabling research on multi-task learning across localized motion and global affective behavior.
    \item a complementary dataset recorded simultaneously with multi-angle RF: thirteen synchronized radars (8~ground, 5~ceiling) provide dense angular diversity. Ceiling-mounted radars offer a potential path toward angle-invariant motion recognition. 
    \item a comprehensive multi-task baseline evaluating gesture, activity, and affective behavior recognition. This benchmark establishes robust algorithmic reference points for the community. 
    \item a dataset that allows behavior recognition without capturing identifiable facial features or RGB imagery, addressing growing privacy concerns in ubiquitous sensing applications.    
\end{enumerate}
The multi-sensor setup simulates realistic indoor spaces. At the same time, a consistent protocol across 61 participants ensures repeatability, balanced class coverage, and statistically meaningful evaluation across tasks.

\begin{table*}
\centering
\caption{Comparison to existing gesture, activity, and affective behavior datasets/frameworks from recent years.}
\label{tab:dataset_comparison}
\resizebox{\textwidth}{!}{
\begin{tabular}{l|l|c|c|c|c|l}
\hline
\textbf{Dataset/Framework} & \textbf{Type} & \textbf{Modalities} & \textbf{Subjects} & \textbf{Classes} & \textbf{Privacy} & \textbf{Year} \\
\hline
\hline
\multicolumn{7}{c}{\textit{Gesture Recognition Datasets}} \\
\hline
Pantomime~\cite{palipana2021pantomime} & Hand Gesture & Radar & 10 & 21 & High & 2021 \\
HaGRID~\cite{kapitanov2024hagrid} & Body Gesture & RGB Vision & 38K & 18 & Low & 2024 \\
8-radar Dataset~\cite{salami2024angle} & Hand Gesture & Radar & 15 & 21 & High & 2024 \\
Env-RFID~\cite{golipoor2024environment} & Gesture/Track & RFID & 10 & N/A & High & 2024 \\
iRadar~\cite{yang2025iradar} & Hand Gesture & Radar, Wearable IMU & 15 & 10 & High & 2025 \\
mmPencil~\cite{guo2025mmpencil} & Hand Gesture & Radar & 20 & 26 & High & 2025 \\
\hline
\multicolumn{7}{c}{\textit{Activity Recognition Datasets}} \\
\hline
UCI HAR~\cite{anguita2013public} & Activity & Wearable IMU & 30 & 6 & High & 2013 \\
MM-Fi~\cite{yang2023mm} & Activity/Pose & RGB, Depth, LiDAR, Radar, Wi-Fi & 40 & 27 & Low & 2023\\ 
XRF55~\cite{wang2024xrf55} & Activity & Wi-Fi, Radar, RFID, RGB-D & 39 & 55 & Low & 2024\\ 
MMVR~\cite{rahman2024mmvr} & Activity/Pose & Radar & 25 & N/A & High & 2024 \\
CrossHAR~\cite{hong2024crosshar} & Activity & Wearable IMU & N/A & N/A & High & 2024 \\
LLM-HAR~\cite{wei2025one} & Activity & Wearable IMU & N/A & 18 & High & 2025 \\
HAR Foundation~\cite{qiu2025towards} & Activity & Multi-Sensor (IMU, RF) & N/A & N/A & High & 2025 \\
\hline
\multicolumn{7}{c}{\textit{Affective Behavior / Sentiment Recognition Datasets}} \\
\hline
IEMOCAP~\cite{busso2008iemocap} & Emotion & Audio, Video, Motion & 10 & 10 & Low & 2008 \\
AffectNet~\cite{mollahosseini2017affectnet} & Emotion & RGB Images & 440K & 8 & Low & 2017 \\
Detecting Emotion~\cite{gebhardt2024detecting} & Emotion & Wearable Physio, Smartphone & 29 & 8 & Medium & 2024 \\
Stress ID~\cite{neigel2024using} & Stress/Affect & Consumer Wearable & 103 & N/A & High & 2024 \\
AuraVox~\cite{imran2025auravox} & Emotion & mmWave Radar, Audio & 30 & 6 & High & 2025 \\
AFFEC~\cite{sekiavandi2025advancing} & Emotion & EEG, Eye Tracking, GSR, Video & 71 & 6 & Low & 2025 \\
\hline
\hline
\textbf{RF-Behavior (This work)} & \textbf{All Types} & \textbf{Radar, RFID, LoRa, IMU, IR} & \textbf{61} & \textbf{37 (21+10+6)} & \textbf{High} & \textbf{2026} \\
\hline
\end{tabular}
}
\end{table*}

\section{Related work}
We briefly summarize the related work in the domains of gesture recognition, human activity recognition and affective behavior analysis in recent years.

\subsection{Gesture Recognition}
Gesture recognition has been extensively studied across multiple sensing modalities, including vision, inertial, and radio frequency (RF). 
Vision-based datasets, such as HaGRID~\cite{kapitanov2024hagrid}, provide large-scale RGB or RGB-D recordings for hand and body gesture understanding. The strength of these datasets is however also a limitation: vision-centric approaches are not feasible in many public, private and workspace settings. Furthermore, the performance of optical sensors, degrades sharply under poor illumination or occlusion conditions.

Alternative setups may feature radar, RFID or inertial sensors for gesture recognition. 
Millimeter-wave (mmWave) radar, in particular, enables contactless sensing by capturing fine-grained micro-Doppler signatures. mmPencil~\cite{guo2025mmpencil} utilizes millimeter-wave radar for high-precision, person-independent gesture classification and 3D in-air handwriting. Similarly, the iRadar framework~\cite{yang2025iradar} provides extensive resources for cross-modal IMU-to-radar gesture synthesis and sparse point-cloud gesture recognition, respectively. 
While these datasets offer exceptional spatial and temporal resolution for micro-kinematics, a limitation is that they strictly isolate hand gestures from the participant's broader physical and affective context, so that hierarchical, cross-task learning (required to connect brief gestures to long-term behavioral states) is not possible. 
By merging data from multiple radars, recent studies improve spatial density and field of view~\cite{salami2024angle}.
As a further evolution, RF-Behavior achieves accurately synchronized, multi-radar, multi-modal measurements with coordinate alignment by integrating data from 13~mmWave radars and applying time synchronization and coordinate transformation to produce a unified, denser, and information-rich 3D representation of human motion. 
Five radars are fixed on the ceiling, while the remaining radars are placed on the ground surrounding the subject. 
They provide diverse data from all directions. 
Researchers may freely choose data from desired radar channels to fit different experimental setups or tasks.

Body-mounted backscatter tags may further be employed to support the perception of gestures, daily activities and affective behaviors.
In our work, we integrate RFID-type backscattering tags since setups with these standardization-compliant tags are easy to reproduce by other researchers~\cite{golipoor2024environment, golipoor2024rfid,Graph2026,AoA2026, LowPower2026}.

\subsection{Human activity recognition}
Human activity recognition (HAR) has evolved significantly over the past decades, progressing from accelerometer-based classification to complex multimodal sensing systems. 
Pioneering datasets such as WISDM~\cite{kwapisz2011activity} utilized smartphone accelerometers to recognize six basic activities. 
The UCI HAR dataset~\cite{anguita2013public} expanded this approach by incorporating both accelerometer and gyroscope, establishing a widely-used benchmark for wearable sensor-based activity recognition. 

To address the limitations of on-body sensing (subject compliance, battery life, and the requirement to constantly carry or wear devices), researchers have explored ambient and contact-free sensing modalities, such as multimodal RF. 
Recent works, such as CrossHAR~\cite{hong2024crosshar} and LLM-assisted HAR models~\cite{wei2025one}, emphasize the critical need for large-scale, heterogeneous sensor data to train models capable of generalizing across different users, environments, and devices. 
The development of HAR Foundation Models is also rapid~\cite{qiu2025towards}. 
It requires massive, multimodal datasets to align semantic contexts with unconstrained sensor patterns. 
To support these advanced learning strategies, foundational RF datasets have emerged, including MM-Fi~\cite{yang2023mm}, XRF55~\cite{wang2024xrf55} and MMVR~\cite{rahman2024mmvr}. 
These state-of-the-art multimodal datasets already provide broad coverage of discrete human actions. 
XRF55~\cite{wang2024xrf55} spans 55 classes from fine-grained interactions to whole-body motions across four RF and vision modalities -- 
optimized for high-precision 3D human pose estimation, spatial tracking, and discrete, short-duration human activity recognition. 

The RF-Behavior dataset is complementary in three aspects: (1) it adds an affective behavior dimension with states lasting over minutes; 
(2) it provides marker-based infrared motion capture as metric ground truth for the RF streams in the lab setting; 
(3) it utilizes a synchronized thirteen-radar array with released global coordinates.
Specifically, the Dataset shifts the paradigm from structural pose geometry to continuous, unconstrained macroscopic kinematics, enabling the study of daily activities over longer temporal horizons.

\subsection{Affective Behavior Analysis}
Affective state and behavior recognition have been extensively studied across multiple modalities, including facial expressions, speech, physiological measurements, and text analysis. 
Early research focused primarily on facial expressions, building upon Ekman's theory of universal affective states~\cite{ekman1992argument}. 

Foundational datasets (e.g., IEMOCAP~\cite{busso2008iemocap}, AffectNet~\cite{mollahosseini2017affectnet}) provide significant advances conditioned on clear facial visibility or (intrusive) physiological wiring. 

Recent research has also explored contact-free approaches and utilized wearables and ambient sensors. 
The authors in~\cite{gebhardt2024detecting} have demonstrated predictive models to detect emotional states using behavioral dynamics during passive media consumption. 
Similarly, in~\cite{neigel2024using}, wearables are utilized to unobtrusively identify periods of cyclical stress in real-world environments. 
In the ambient sensing domain, works like AuraVox~\cite{imran2025auravox} have utilized mmWave radar to capture emotion-related facial micro-Doppler signatures, while multi-modal datasets like AFFEC~\cite{sekiavandi2025advancing} fuse physiological and kinematic streams to map complex cognitive load.

A critical limitation of existing affective datasets is their focus on brief, discrete moments or internal micro signals (e.g., heart rate, micro expressions). 
Most datasets capture states lasting seconds to a few minutes, often in response to controlled stimuli, so that longer duration temporal dynamics are not captured. 
Additionally, public datasets rely heavily on facial video, or scripted displays, which may not reflect authentic macroscopic expressions (e.g., changes in gait, movement speed, or posture) or constraints of realistic living and working spaces. 
Hence, in contrast to the RF-Behavior dataset, the integration of RF sensing modalities to continuously capture global affective behaviors remains underexplored.

\section{The RF-Behavior Dataset}\label{MocapDataSet}
\begin{figure}
\centering
\includegraphics[width=10cm]{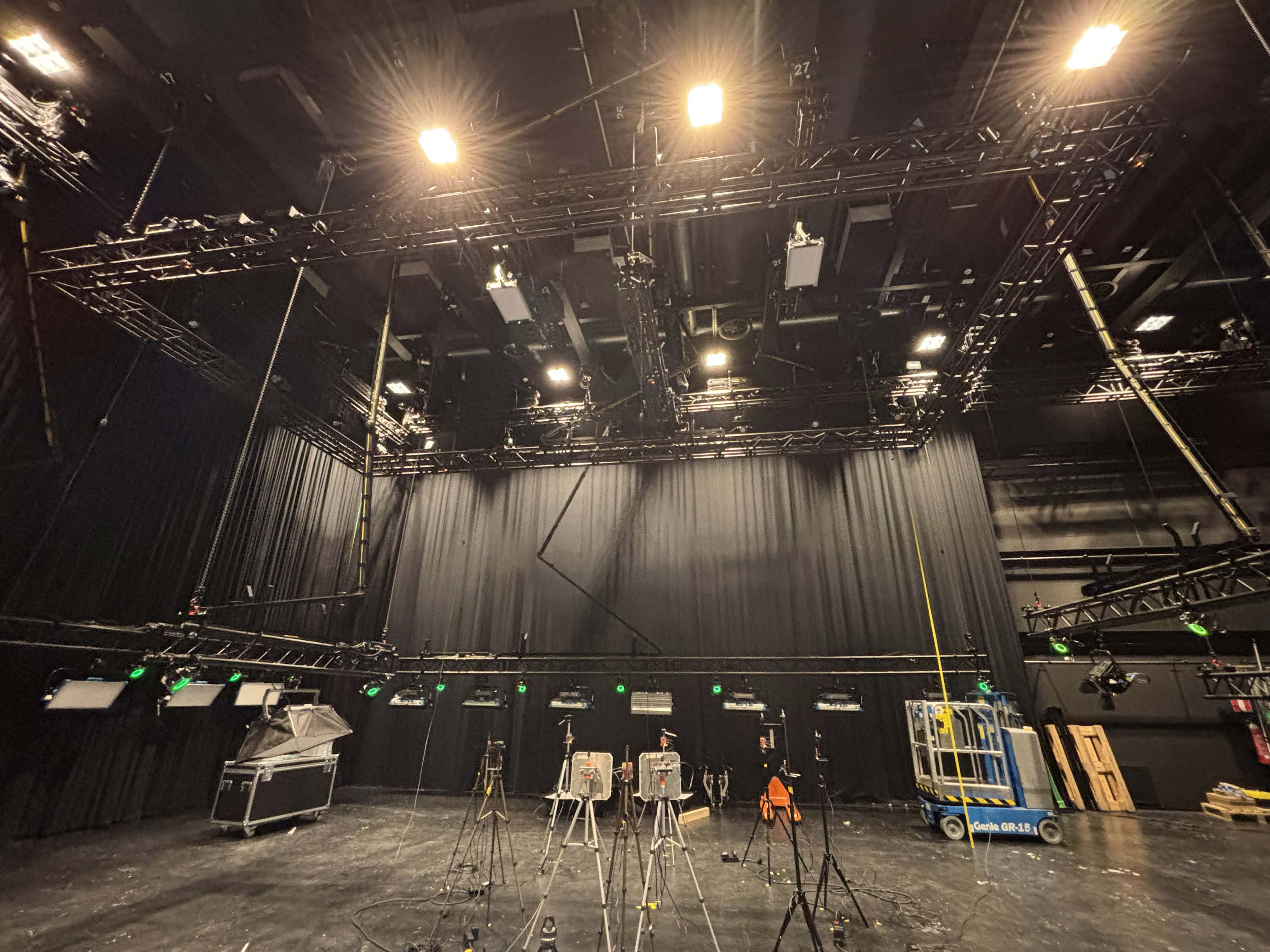}
\caption{The digital studio for data collection.}
\label{fig:studio}
\end{figure}

The RF-Behavior dataset comprises 21~hand gestures, 10~daily activities, and 6~affective behaviors, collected from a total of 61~subjects (see information on the participants in appendix~\ref{sec:appendix_particpantinfo}). 
We divided the collection into four campaigns: 
Campaigns~1 and~4 target on hand gestures, campaign~2 captures full-body activities, and campaign~3 records macroscopic affective behaviors. 
To thoroughly evaluate cross-environment generalization, campaigns 1--3 were conducted in three diverse settings: 
a controlled Digital Studio (lab environment, see Fig.~\ref{fig:studio}), a Living Room, and an Industrial Workplace. 
Campaign~4 was held exclusively in an Industrial Workplace.

The dataset contains simultaneous measurements from multiple sensors, including Radio-Frequency (RF) data from radars, backscatter-type devices (here: RFID), LoRa, motion data from three Inertial Measurement Unit (IMU) sensors, and 24 infrared cameras capturing motion information. 
Data from the IMUs and infrared cameras serve as the ground truth. The radar operates in the 77-81 GHz frequency range, the backscatter tags at 865 MHz, and the LoRa at 915 MHz.

Sensor configurations vary across campaigns 1 to 3 due to environmental constraints. 
The 24 infrared cameras featured within the digital laboratory are not deployed in real-world environments for practical constraints. 
Similar deployment constraints apply also to the backscatter tags. 
IMU data can be collected by body-worn devices (e.g. smartwatches). 
LoRa is rarely utilized in application scenarios involving gesture or activity recognition.

The data collection protocol and sensor placements (both on-body and environmental) remain identical between actual instrumentations in real environments and those used in the laboratory setting. 
The laboratory dataset provides the most comprehensive sensor modalities to support the exploration in activity and affective behavior recognition, while the living room and industrial datasets reflect practical constraints in real-world implementation (cf. Table~\ref{tab:env_sensor}).

\begin{table}
\centering
\caption{Sensor usage for different data collection environments.}
\label{tab:env_sensor}
\begin{tabular}{cccccccc}
\toprule
  & \textbf{Radar} & \textbf{RFID} & \textbf{LoRa} & \textbf{IMU} & \textbf{Infrared Camera} & \textbf{ECG}$^{*}$ & \textbf{EEG}$^{*}$ \\
\midrule
Digital Studio & \checkmark  & \checkmark  & \checkmark  & \checkmark  & \checkmark & - & - \\
Living Room  &  \checkmark & -  & - & \checkmark & - & \checkmark & \checkmark \\
industrial factory & \checkmark  &  -  & - & \checkmark & - & - & - \\

\bottomrule
\multicolumn{8}{l}{\footnotesize $^{*}$physiological reference subset only (Section~\ref{sec:physio_subset}).}
\end{tabular}
\end{table}

\subsection{Design Rationale}\label{sec:design_rationale}
\subsubsection{Sensor selection}\label{sec:SensorSelection} 
The sensing stack was selected to span three complementary operating points of the RF-sensing design space, rather than to enumerate every available RF technology. 
For instance, WiFi is another technology that had received significant attention for environmental sensing in the literature, however, it's sensing principle is similar to LoRa-type sensing, but constrained to shorter range. 
Likewise, UWB, which may be featured alongside future WiFi chipsets, and indeed prospectively replace WiFi sensing, shares several similarities with radar.
In the literature, datasets that include UWB and WiFi CSI are e.g., MM-Fi~\cite{yang2023mm} or XRF55~\cite{wang2024xrf55}. 

RF-Behavior complements these existing datasets by providing data from a densely synchronized multi-radar array, LoRa-based behavior sensing, as well as body-worn backscatter. 
In the RF-Behavior dataset, mmWave FMCW radar (77--81\,GHz) represents the active, high-resolution extreme and produces 3D point clouds of the moving body. 
Passive RFID backscatter (865\,MHz) represents battery-free, on-body sensing with precise limb-level association. 
LoRa (915\,MHz, chirp spread spectrum) represents the communication-centric, sub-GHz, single-link extreme. 
The three modalities cover active, passive, and backscatter-type sensing in three well separated frequency bands. 
The data representation ranges from point clouds down to a single scalar channel. 
Body-worn IMUs and marker-based infrared motion capture provides kinematic ground truth against which every RF stream can be evaluated.
The scope of the RF stack is a choice that is based on depth over breadth. Adding more radio modalities would increase synchronization complexity, interference risks, and the recording burden for participants. We therefore selected three complementary RF modalities and collected detailed measurements from each.
Wi-Fi CSI and UWB could provide useful extensions in future work. The released synchronization framework and coordinate information can support such extensions.
 
\subsubsection{Behavior selection}
The three behavior sets were chosen to maximize comparability with existing work while covering three distinct kinematic scales. 
The 21 gestures adopt the publicly released gesture vocabulary of Pantomime~\cite{palipana2021pantomime}. 
This set has since been used as an evaluation benchmark by several radar gesture recognition studies (e.g., \cite{salami2022tesla,salami2024angle}).
Our data is therefore directly comparable with this line of work rather than introducing yet another gesture vocabulary. 
The 10 activities combine the core Activities of Daily Living that anchor standard HAR benchmarks (walking, running, sitting, lying, and stair ascent/descent, as it has been used in the OPPORTUNITY dataset~\cite{chavarriaga2013opportunity}, 
UCI-HAR~\cite{anguita2013public}, as well as WISDM~\cite{weiss2019wisdm}) with four sports actions. 
The sports actions extend the dynamic range of the dataset toward fast, high-amplitude, and less constrained motion. 
The six affective behavior categories are derived from the quadrants of Russell's circumplex model of affect~\cite{russell1980circumplex} and one attentional state pair, as detailed in Section~\ref{sec:affectiveBehaviors}.

\subsubsection{Environment setup}
A single fixed geometry cannot optimally observe both micro- and macro-kinematics. 
Fine hand gestures require sensors densely focused on a small area, whereas locomotion and sports actions need room for natural full-body movement. 
We therefore held the sensing stack, the ceiling radar array, the on-body sensor scheme, the synchronization procedure, and the global coordinate frame fixed across campaigns 1--3, and adapted only the ground-plane geometry to the spatial envelope of each behavior class; 
the resulting per-campaign layouts are described in Section~\ref{sec:sensors}. 
Joint multi-scale monitoring is thus enabled through a shared synchronized sensing infrastructure, a common metric reference frame with released radar coordinates, and a common participant pool (required for cross-campaign representation learning).

\subsection{Sensors}\label{sec:sensors}
\subsubsection{Radar}
Thirteen mmWave Frequency-modulated Continuous Wave (FMCW) radars (TI IWR1443) are employed. 
The radar devices operate over a 4~GHz bandwidth from 77~GHz to 81~GHz and emit linearly frequency-modulated chirps to determine object range and velocity by interpreting received reflected signals relative to the transmitted signals. 
The onboard 3~transmit and 4~receive antennas allow angle estimation. 
FMCW-MIMO technology is used to generate range–azimuth–elevation maps and convert them into 3D point-cloud data representing reflections from surrounding objects. 
The resulting point clouds, captured frame by frame in x–y–z coordinates, form spatio-temporal sequences for analysis. 
When saving the data, the current timestamp, x, y, z, and density are stored together.
Each radar has its own coordinate system, so that coordinate transformation and time synchronization are conducted to merge all data.

The global coordinate system is sketched in Fig.~\ref{fig:setup_s1}~(b). 
The direction the radar is facing is its x-axis. 
Ceiling-mounted radars face the floor, while radars on the ground face the study-participants. 
To unify all observations into a global reference frame, a rigid transformation consisting of a rotation and a translation is applied. 
Specifically, for a point $\mathbf{p}_i^{(k)} \in \mathbb{R}^3$ captured by the $k$-th radar at frame $i$, its global coordinate $\mathbf{p}_i^{(k,\text{global})}$ is computed as:
\begin{equation}
\mathbf{p}_i^{(k,\text{global})} = \mathbf{R}_k \mathbf{p}_i^{(k)} + \mathbf{t}_k
\end{equation}
where $\mathbf{R}_k \in \mathbb{R}^{3 \times 3}$ is the rotation matrix and $\mathbf{t}_k \in \mathbb{R}^3$ is the translation vector (also the global coordinates) of the $k$-th radar. 
For ceiling-mounted radars, a fixed coordinate transformation matrix is used to align the local radar coordinate system with the global reference frame:
\begin{equation}
\mathbf{R}_k ={T} =
\begin{bmatrix}
0 & 0 & 1 \\
0 & 1 & 0 \\
-1 & 0 & 0
\end{bmatrix}
\end{equation}

For radars on the ground, instead, the rotation is defined as a yaw rotation around the vertical axis:
\begin{equation}
\mathbf{R}_k =
\begin{bmatrix}
\cos \theta_k & -\sin \theta_k & 0 \\
\sin \theta_k & \cos \theta_k & 0 \\
0 & 0 & 1
\end{bmatrix}
\end{equation}
where $\theta_k$ is determined based on the radar placement orientation. Specifically, it is computed from the radar translation vector $\mathbf{t}_k = [x_k, y_k, z_k]^T$ as
\begin{equation}
\theta_k = -\arctan2(y_k, -x_k)
\end{equation}

Although all radars are triggered simultaneously, their recorded timestamps are not perfectly aligned due to position and processing delays (e.g. the same motion may be captured by radar~5 and radar~12 at timestamps $0.15, 0.18, 0.21, \dots$, and $0.20, 0.23, 0.26, \dots$ respectively). 
We construct a global timeline using the minimum and maximum timestamps across all sensors:
\begin{equation}
t_{\min} = \min_k \{ t_0^{(k)} \}, \quad
t_{\max} = \max_k \{ t_{N_k}^{(k)} \}
\end{equation}

A fixed-length temporal sequence of $N = 16$ frames is constructed for each sample. 
The temporal interval is then defined as:
\begin{equation}
\Delta t = \frac{t_{\max} - t_{\min}}{N - 1}
\end{equation}

The corresponding temporal grid is:
\begin{equation}
t_n = t_{\min} + n \cdot \Delta t, \quad n = 0,1,\dots,N-1
\end{equation}

For each radar $k$ and each target timestamp $t_n$, the closest frame is selected by:
\begin{equation}
\hat{i} = \arg\min_i \left| t_i^{(k)} - t_n \right|
\end{equation}

The corresponding point cloud $\mathbf{P}_{\hat{i}}^{(k)}$ is used as the synchronized observation at time $t_n$. 
This fixed-length temporal alignment ensures that all radar sequences are uniformly sampled into 16 frames, enabling consistent input to the learning model while preserving the original temporal structure.

After time synchronization, the multi-radar data are organized into a unified frame-based format. 
Each frame contains point clouds from all radars at the aligned timestamp, represented as tuples of frame ID, radar ID, and point attributes:

\begin{center}
\begin{tabular}{c c c c c c}
Frame ID & Radar ID & $x$ & $y$ & $z$ & Intensity \\
1 & 0 & $\cdots$ & $\cdots$ & $\cdots$ & $\cdots$ \\
1 & 1 & $\cdots$ & $\cdots$ & $\cdots$ & $\cdots$ \\
1 & $\cdots$ & $\cdots$ & $\cdots$ & $\cdots$ & $\cdots$ \\
2 & $\cdots$ & $\cdots$ & $\cdots$ & $\cdots$ & $\cdots$ \\
$\cdots$ & $\cdots$ & $\cdots$ & $\cdots$ & $\cdots$ & $\cdots$ \\
\end{tabular}
\end{center}

\subsubsection{RFID}
We use Alien AZ 9662 passive RFID tags, an Impinj Speedway R420 RFID reader, and a circularly polarized Vulcan RFID PAR90209H antenna with a circular polarization gain of $9$~dBiC as well as elevation and azimuth beamwidth of $70^{\circ}$.
A laptop running the Impinj ItemTest software is used to control the RFID reader. 
The RFID system functions at a frequency of 865 MHz with a transmission power of 30 dBm.
The tag’s integrated circuit (IC) modulates its reply to the reader’s continuous wave signal by varying the impedance on its antenna. 
This change causes the tag to either reflect or absorb portions of the incoming signal, generating reflection coefficient $\Gamma(t) \in {0,1}$.
The reflection coefficient is $\Gamma = \frac{Z_\text{Ant}-Z_\text{Load}(t)}{Z_\text{Ant}+Z_\text{Load}(t)}$, where $Z_\text{Ant}$ and $Z_\text{Load}(t)$ denote the antenna impedance (around $50\Omega$) and the load impedance controlled by the tag's IC.
When $Z_\text{Load} = Z_\text{Ant}$, the reflection coefficient becomes $\Gamma = 0$, meaning no signal is reflected. Conversely, when $Z_\text{Load} = 0$, the reflection coefficient reaches $\Gamma = 1$, and the reader’s signal is fully reflected. In practice, due to imperfect load matching, $\Gamma$ typically varies between values that are approximately $0$ and $1$.
We denote the basedband received signal at the reader by 
\begin{align}\label{eq:ym}
    y (t) &= \sum_{i=1}^{N_p} g_i G_a\sqrt{P_t}s(t) + \nu(t) ;\ \ \ \ y (t) \in \mathbb{C}.
\end{align}
where $g_i\triangleq \frac {\lambda}{4\pi d_i}e^{\jmath \theta_i} $ denotes the channel gain of the $i$th round-trip path with $\theta_i\sim \mathcal{U}[0,2\pi)$, wavelength $\lambda$, path distance $d_i$, the number of multipath components $N_p$, antenna gain $G_a$, reader transmit power $P_t$ and the transmitted signal $s(t)$.
$\nu(t)$ represents the additive thermal noise at the reader, modeled as a zero-mean Gaussian random variable with variance $\sigma^2$.
The data obtained from the tag includes Electronic Product Code (EPC), a timestamp, the Received Signal Strength (RSS), and the signal phase.

\subsubsection{LoRa}
LoRa employs CSS modulation to achieve long-range communication. 
A Commercial Semtech SX1276 RF module configured by an Arduino Uno board is utilized as the LoRa node for transmitting LoRa signals via a bipolar antenna. 
The LoRa node sends a series of linear chirps, each characterized by its duration $T$ and frequency bandwidth $B$. 
CSS modulates the data by varying the starting frequency and the slope of the chirp. 
Fig.~\ref{fig:lora3a} illustrates the signal spectrum of a typical LoRa packet. 
The packet begins with a preamble containing full up-chirps. 
Subsequently, two down-chirps are transmitted to represent the start frame delimiter (SFD). 
After the SFD, chirps with different starting frequencies are used to encode the data to be transmitted. The encoded data bits within a chirp are defined by the Spreading Factor (SF). 
LoRa supports seven different spread factors, ranging from 6 to 12. 
Each chirp can encode data using $2^{SF}$ different starting frequencies. By default, the LoRa node is configured to transmit signals with a carrier frequency of 865.5 MHz with a bandwidth of 125 kHz. 
To interpret the LoRa signals, a USRP N200 co-located with the LoGa node is configured as a LoRa gateway to receive the reflected LoRa signals. 
The received LoRa signals are down-converted to the baseband and then sampled by ADC with 500~kHz sampling rate. 
The amplitude of the baseband signal is exported for sensing purposes. 
LoRa is included in the dataset specifically to capture macroscopic movement and room-level presence.

In addition to the raw baseband recordings, we release three extra features per trial. The amplitude of the baseband signal is low-pass filtered with a 6th-order Butterworth filter (1\,kHz cut-off) and decimated to 2\,kHz, then filtered again (6th-order Butterworth, 100\,Hz cut-off) and decimated to 200\,Hz, yielding the \emph{amplitude} feature 
which captures motion-induced variations of the channel while suppressing the chirp structure and receiver noise. From this signal we derive the \emph{differential} feature
, the first-order difference that removes the static path component and emphasizes motion, and the sliding-window \emph{variance} feature (0.05\,s windows), a motion-energy descriptor. 

\subsubsection{IMU}
We used three \textit{Movesense} sensors~\footnote{https://www.movesense.com/docs/} for IMU data collection. 
\textit{Movesense} is a battery-powered device that incorporates low power sensor components with Bluetooth Low Energy (BLE) enabled Micro Controller Unit (MCU). 
It is based on Nordic Semiconductor's nRF52 BLE chip fitted with 9-axis motion sensor (3-axis accelerometer, 3-axis gyroscope and 3-axis magnetometer).  
We used the sampling frequency of 52Hz for the gesture collection and 104Hz for the rest of the campaigns. 
Three \textit{Movesense} sensors were placed on various parts of the human body in different campaigns. 

\subsubsection{Infrared Camera}
The digital studio is equipped with 24 infrared cameras for motion tracking. 
The camera-based system uses multiple synchronized cameras to capture motion from different angles by tracking reflective markers placed on the human body. 
Cameras are strategically positioned around the studio to cover the entire recording area and to minimize occlusions. 
Data from all cameras are fused to reconstruct a three-dimensional trajectory for each marker via triangulation, after which the system computes the spatial positions to digitally reconstruct the motion. 
The sampling rate was set to 100 Hz for all recording campaigns. 

\subsection{Environmental Setup}\label{sec:env_setup}
\begin{figure}
  \centering
  \begin{subfigure}[t]{0.45\textwidth}
    \centering
    \includegraphics[width=\textwidth]{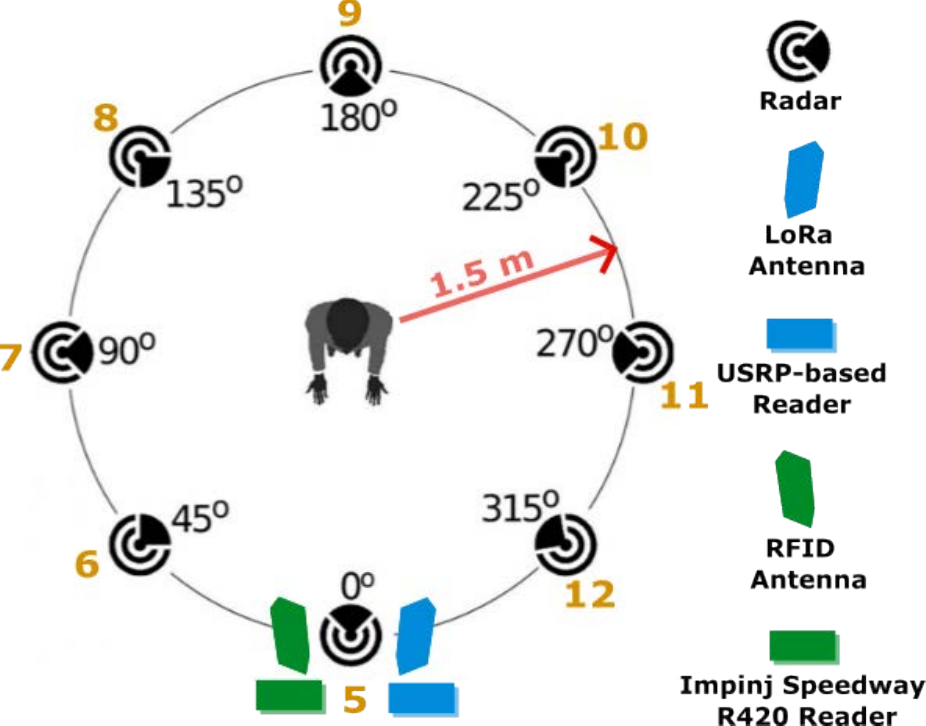} 
    \caption{Setup on the ground.}
  \end{subfigure}
  \hfill
  \begin{subfigure}[t]{0.45\textwidth}
    \centering
    \includegraphics[width=\textwidth]{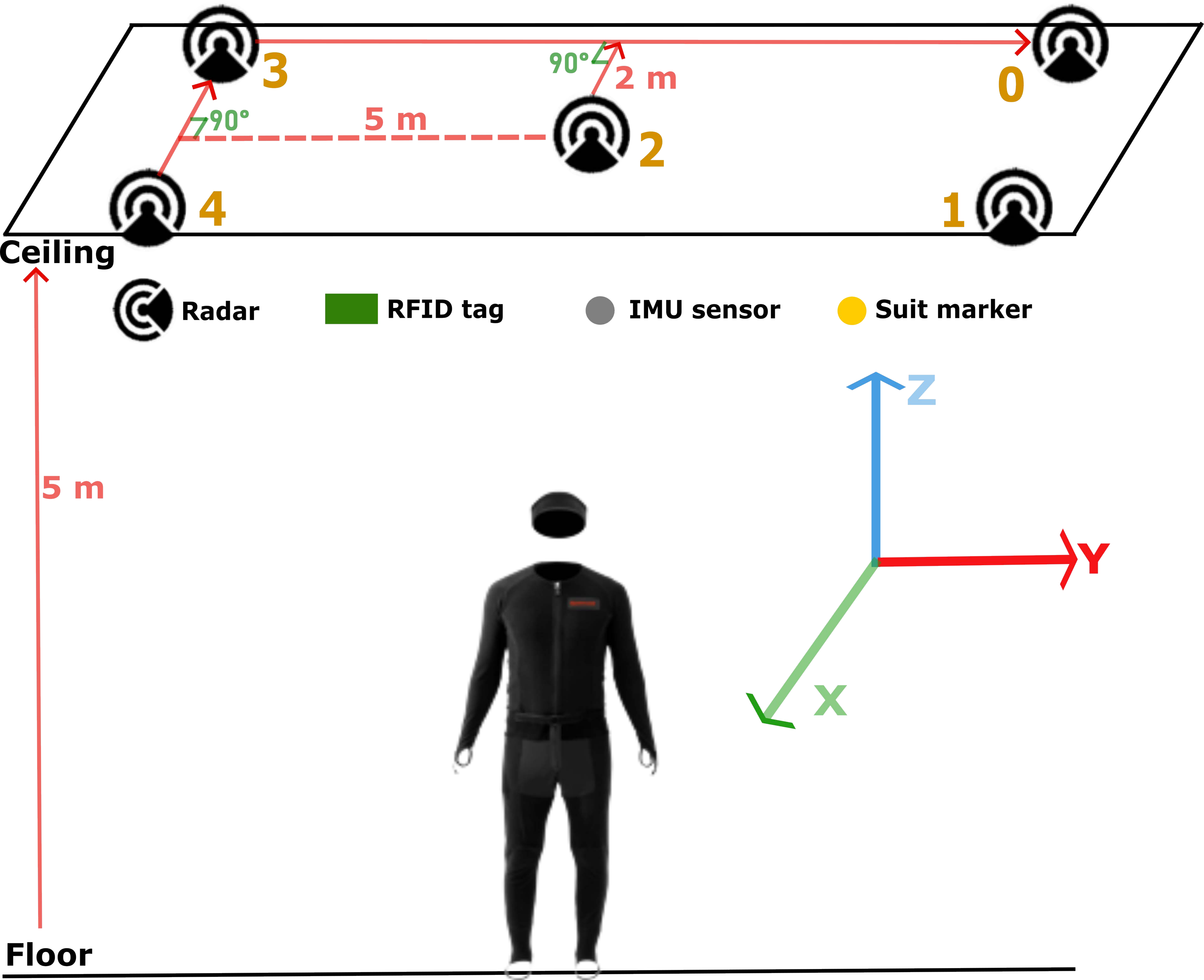} 
    \caption{Setup on the ceiling.}
  \end{subfigure}
  \caption{Setup for campaign~1: gesture data collection.}
  \label{fig:setup_s1}
\end{figure}

\subsubsection{Campaigns 1 \& 4 (Gestures)}
Fig.~\ref{fig:setup_s1} illustrates the environment setup, which comprises two sections: ground and ceiling. 
The number next to the radar symbol indicates the index of the radar. During each data collection experiment, a subject is positioned on the ground, at the center of a circular configuration with a radius of 1.5 meters. Surrounding the subject, there are eight radars placed at specific angular coordinates: 0°, 45°, 90°, 135°, 180°, 225°, 270°, and 315°. These radars detect movement and velocity by measuring reflected radio waves. At the 0° position, an RFID antenna was installed and connected to an Impinj Speedway R420 reader to receive backscattered signals from passive RFID tags attached to the subject. 

On the ceiling, four radars were positioned in each corner, spaced 4 meters apart on one side and 10 meters apart diagonally on the opposite side, providing comprehensive coverage of movement within the area. 
An additional radar was placed in the center of the ceiling, directly above the subject. 
Furthermore, 24~infrared cameras were arranged around the space at various angles and heights to ensure unobstructed views as the subject performs gestures from a standing position (see Fig.~\ref{fig:camera_setup} in the Appendix).

After preliminary experiments, we decided to lower the height to 3 meters and reposition the four radars at the corners of the ceiling: 1.5 meters apart on one side and $2\sqrt{3}$ meters apart diagonally on the opposite side. 
These adjustments (illustrated in Fig.~\ref{fig:setup_s1_v2} in the Appendix) were made to improve the detection of hand movements and to enhance the overall accuracy of motion information collected.

The sensors are concentrated on the upper part of the body, as most gestures involve upper-body movements. 
This strategic placement ensures accurate data collection for gesture recognition, focusing on areas with significant motion activity. 
As the subject performs gestures, the movement is captured by sensors placed on the body as well as devices positioned on the ground and ceiling. 

\subsubsection{Campaign~2 (Human activities)}
\begin{figure}
\centering
\includegraphics[width=12cm]{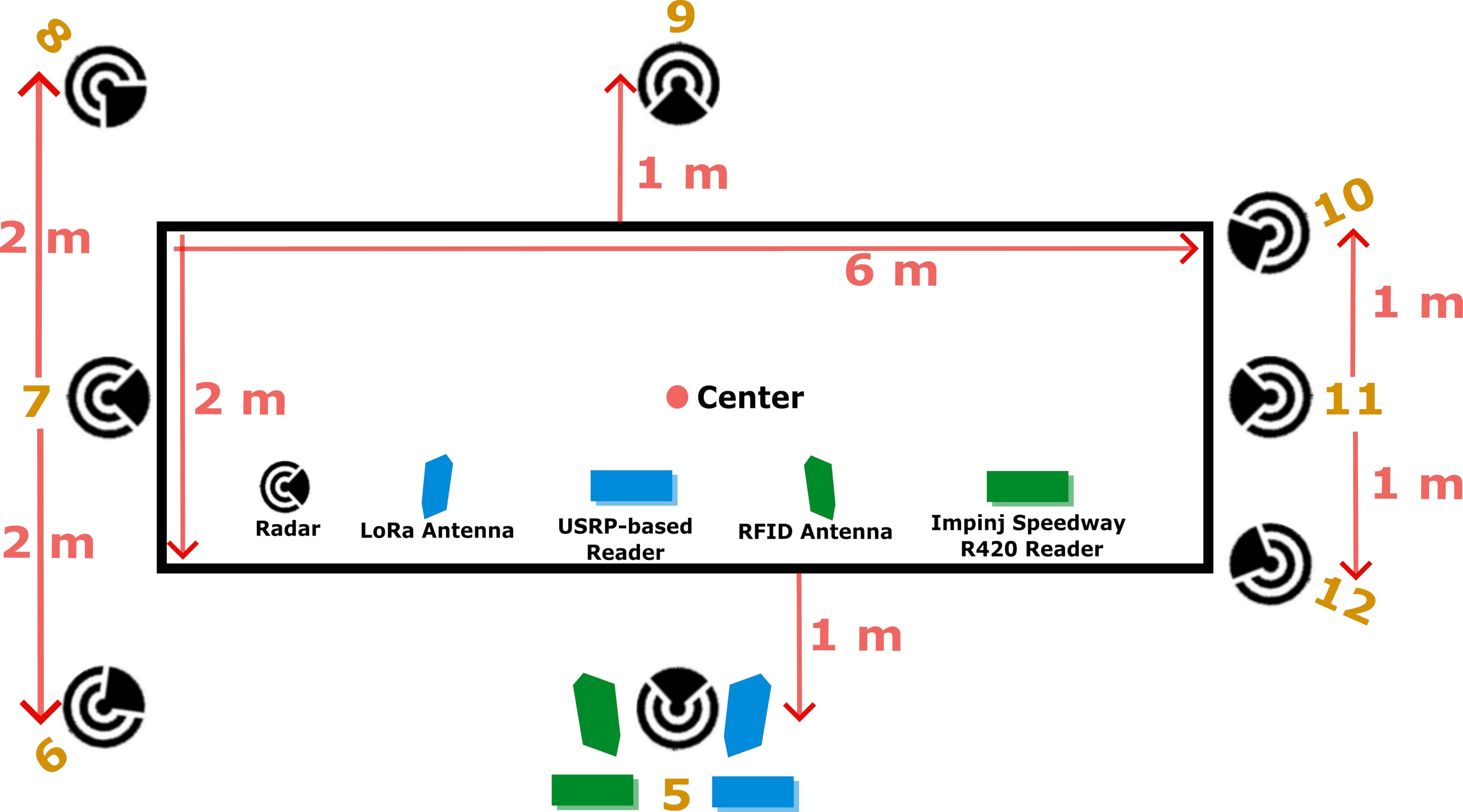}
\caption{Setup on the ground for campaign~2: human activity collection.}
\label{fig:setup_s2s3}
\end{figure}

For the human activity collection in campaign~2, we maintained the ceiling setting and modified the experimental setup on the ground by enlarging the recording area to 6m~x~2m (see Fig.~\ref{fig:setup_s2s3}). 
Radar devices were placed to ensure comprehensive coverage: devices with indices 6,~7, and~8 were aligned along one of the long sides, devices 10,~11, and~12 were on the opposite side, and devices~5 and~9 were centered on the shorter ends. Within this area, the RFID and LoRa antennas were also placed to capture multi-modal data during the activities.

The adjustment of the setup compared to the previous campaign was primarily due to the fundamentally different requirements of the target activities versus gestures, particularly the need for more space (the 1.5-meter radius of the circular setup is too restrictive for full-body dynamic activities). 
Motions like walking, running, playing badminton, and kicking a soccer ball require significant room for participants to move naturally, and the 6m~x~2m rectangular area provides this necessary space. 
The surrounding radar placement allows for the capture of activities from multiple angles, which is especially beneficial for tasks like walking and running. 

As indicated by the red arrows and text in the figure, the distances between the radar nodes were asymmetrical. 
Notably, the spacing between radars 8, 7, and 6 is greater than the spacing between radars 10, 11, and 12, which is on purpose. For sports-related activities, participants were instructed to face the side with radars 6, 7, and 8 when throwing a basketball, playing badminton/floorball, or kicking a football. 
To ensure the safety of the equipment and provide adequate space for these actions, we increased the distance between the devices on that side. 
The sports-related classes have a preferred body orientation toward radars 6, 7, and 8, so orientation diversity for these classes is provided primarily by the ceiling-mounted radars, whose top-down view is orientation-complete. 
The released per-radar channels allow researchers to quantify this effect directly.

Furthermore, to ensure the highest positional accuracy, we addressed a discrepancy observed between the desired and the actual physical placements of the devices. 
We precisely localized the exact 3D coordinates of each radar device using an infrared camera system in the studio. 
For this process, the center (see Fig.~\ref{fig:setup_s2s3}) of the experimental area was established as the origin.
Establishing this unified coordinate system is useful to synchronize data from the different radars. 
The measured 3D coordinates of all radar devices are listed in Table~\ref{tab:radar_coordinates} in the Appendix.

\subsubsection{Campaign~3 (Affective Behaviors)}\label{sec:affectiveBehaviors}
\begin{figure}
\centering
\includegraphics[width=8cm]{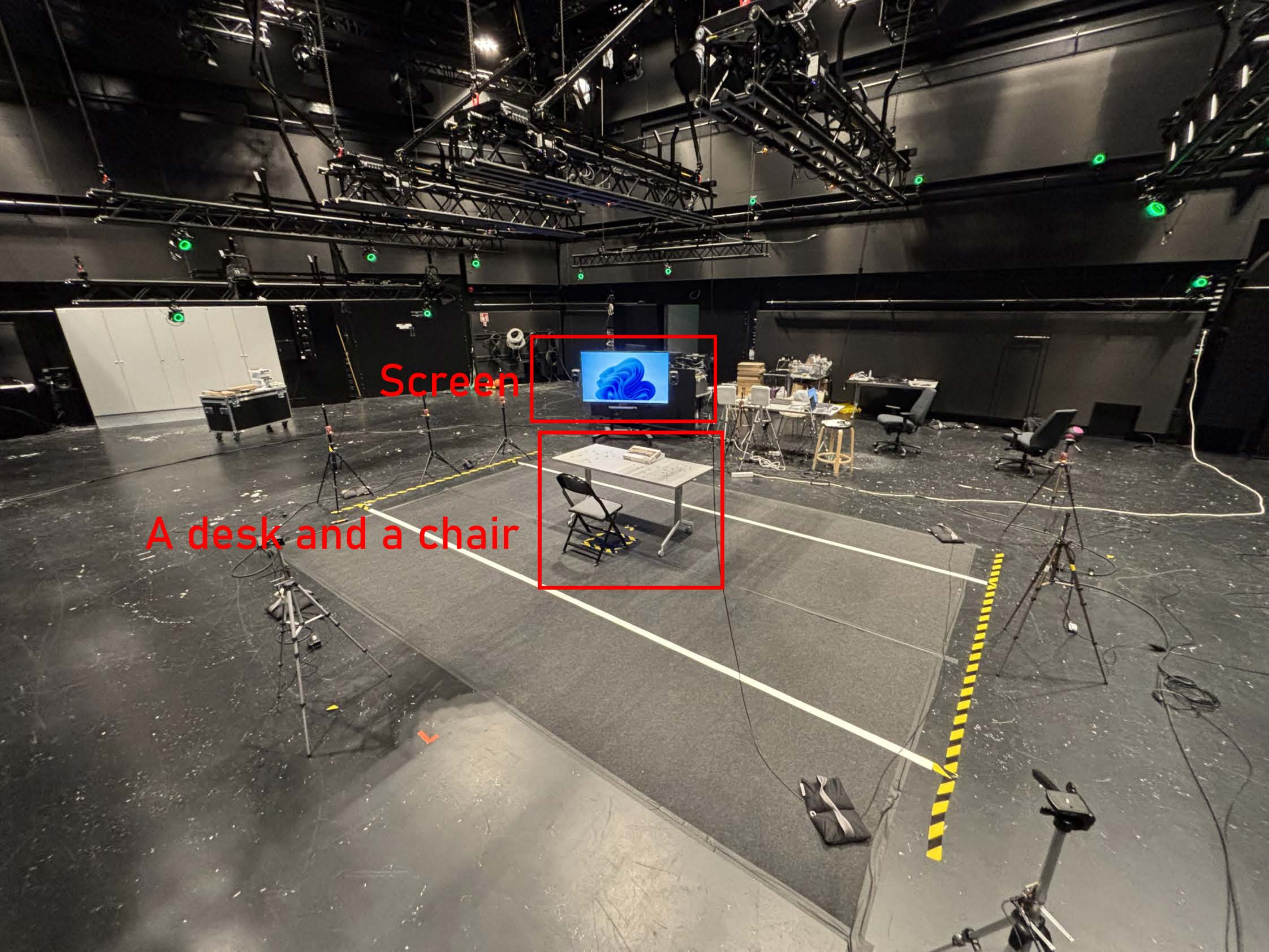}
\caption{Ground setup updated for campaign~3.}
\label{fig:setup_s3}
\end{figure}

The environmental setup for the affective behavior data collection was nearly identical to that of the human activity collection, preserving the device settings on the ceiling and around the 6m x 2m rectangular ground area. 
This campaign requires a specific setup with a desk and chair centered in the recording area. 
A screen was also placed in front of the participant outside the recording area, as shown in Fig.~\ref{fig:setup_s3}. 
The data collection was split into two parts: while the screen was present for both, the desk and chair from the first part were removed for the second part. 
Details of the data collection protocol are given in Section~\ref{sec:c3_protocol}.

\subsubsection{Campaign~4 (Gestures)}
\begin{figure}
\centering
\includegraphics[width=8cm]{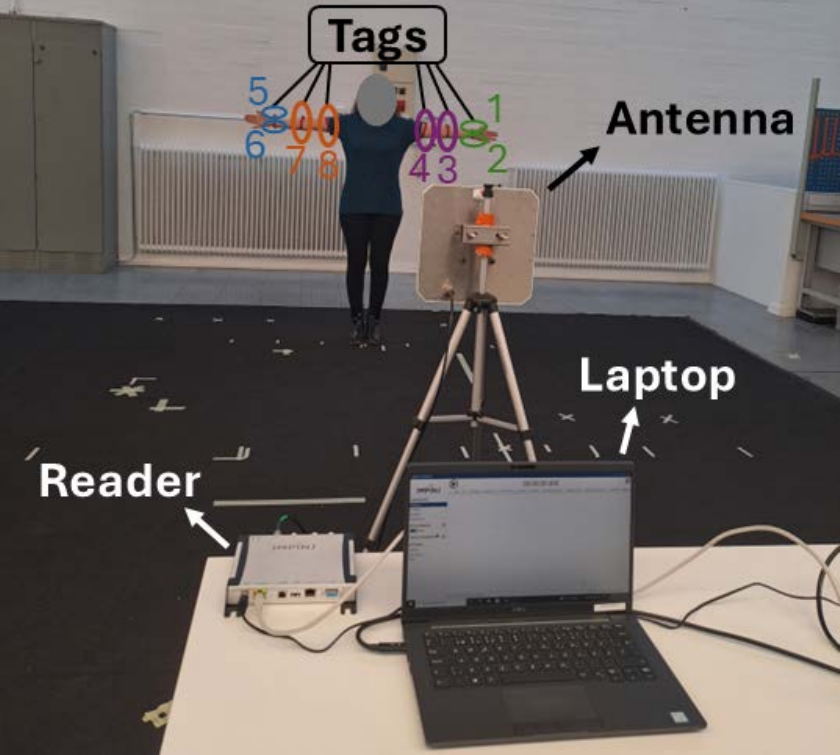}
\caption{Experimental setup. Eight passive RFID tags were affixed to the participant. Phase and RSS data were recorded using an Impinj Speedway R420 reader with a circularly polarized antenna as the participants performed gestures at distances of 1.5 m and 3 m from the antenna.}
\label{fig:c4_RFIDSetup}
\end{figure}

The same equipment described in Section~\ref{MocapDataSet} was employed in an industrial laboratory setting (see Fig.~\ref{fig:c4_RFIDSetup}), involving a different group of participants (totaling 17) along with one additional RFID tag placed on each side of the wrists for data collection.

\subsection{Protocol}
\subsubsection{Campaign~1 (Gestures)}
We collected data from subjects performing 21 gestures, with each gesture repeated 8 times.
This gesture set of gestures was originally proposed in~\cite{palipana2021pantomime}, and we have utilized the same set but within a different environmental setup. Specifically, here are the names for each gesture: (a) lateral-raise, (b) push-down, (c) lift, (d) pull, (e) push, (f) lateral-to-front, (g) swipe-right, (h) swipe-left, (i) throw (j) arms-swing, (k) two-hand-throw, (l) two-hand-push, (m) two-hand-pull, (n) two-hand-lateral-raise, (o) left-arm-circle, (p) right-arm-circle, (q) two-hand-outward-circles, (r) two-hand-inward-circles, (s) two-hand-ateral-to-front, (t) circle-clockwise, (u) circle-counter-clockwise.
The set is divided into two groups, namely easy and complex, according to their execution difficulty. Gestures (a)-(i) are considered to be the easy set, consisting of single-hand motions that are straightforward to execute and recall. The remaining gestures (j)-(u) belong to the complex set, which includes bimanual, linear, and circular movements.

\subsubsection{Campaign~2 (Human activities)}
\begin{figure}
  \centering
  \begin{subfigure}[t]{0.48\textwidth}
    \centering
    \includegraphics[width=\textwidth]{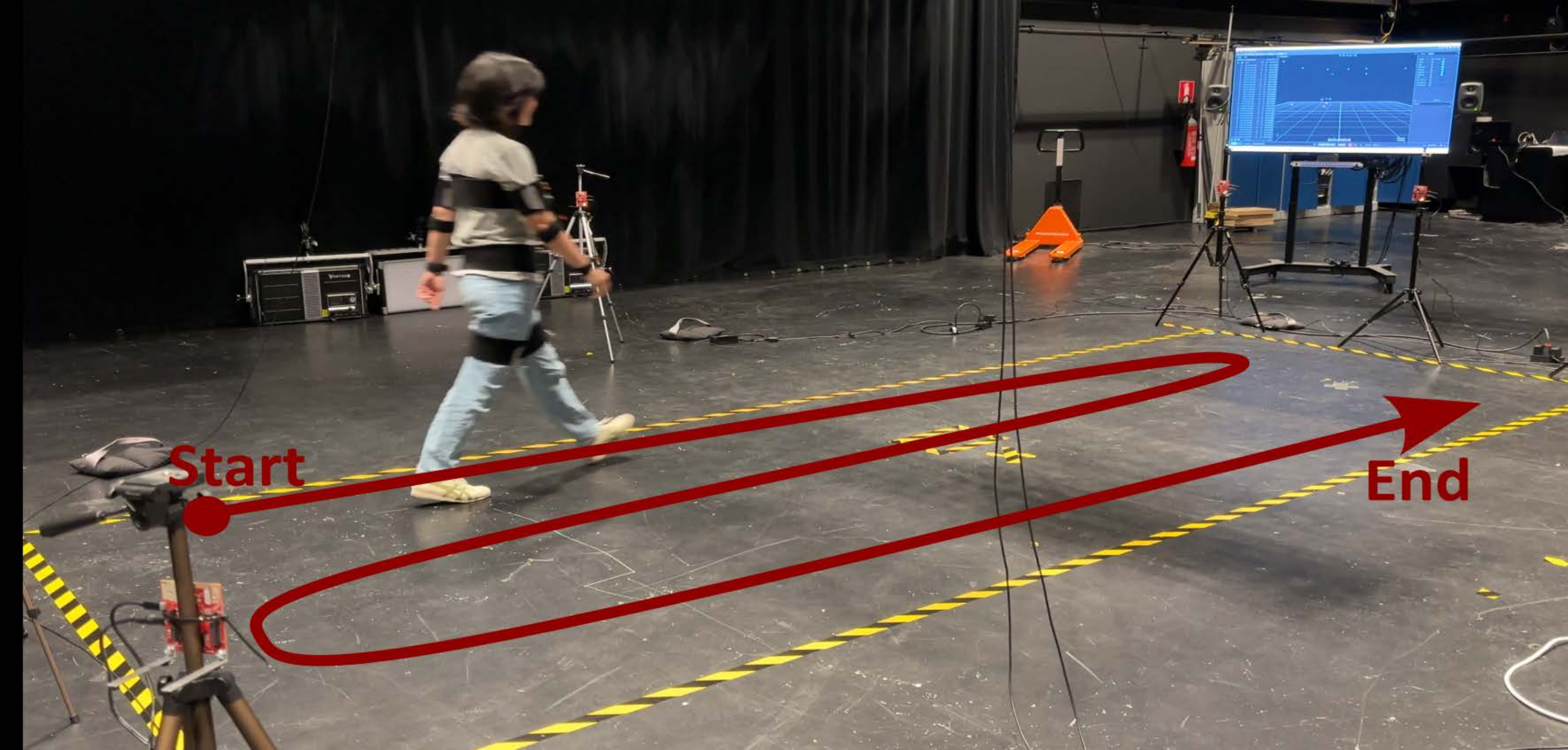} 
    \caption{Walking}\label{fig:s2_walking}
  \end{subfigure}
  \hfill
  \begin{subfigure}[t]{0.48\textwidth}
    \centering
    \includegraphics[width=\textwidth]{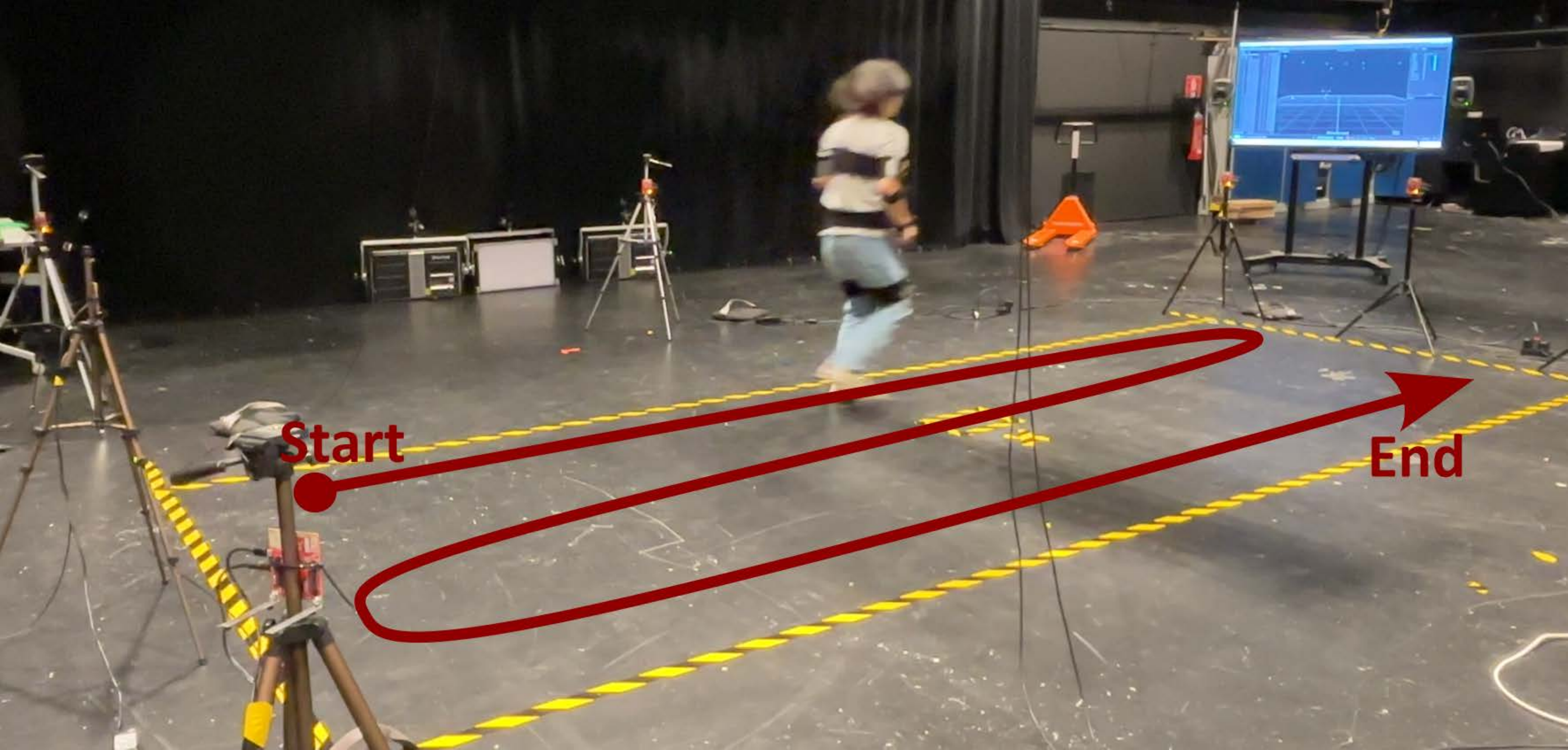} 
    \caption{Running}\label{fig:s2_running}
  \end{subfigure}
    \centering
  \begin{subfigure}[t]{0.48\textwidth}
    \centering
    \includegraphics[width=\textwidth]{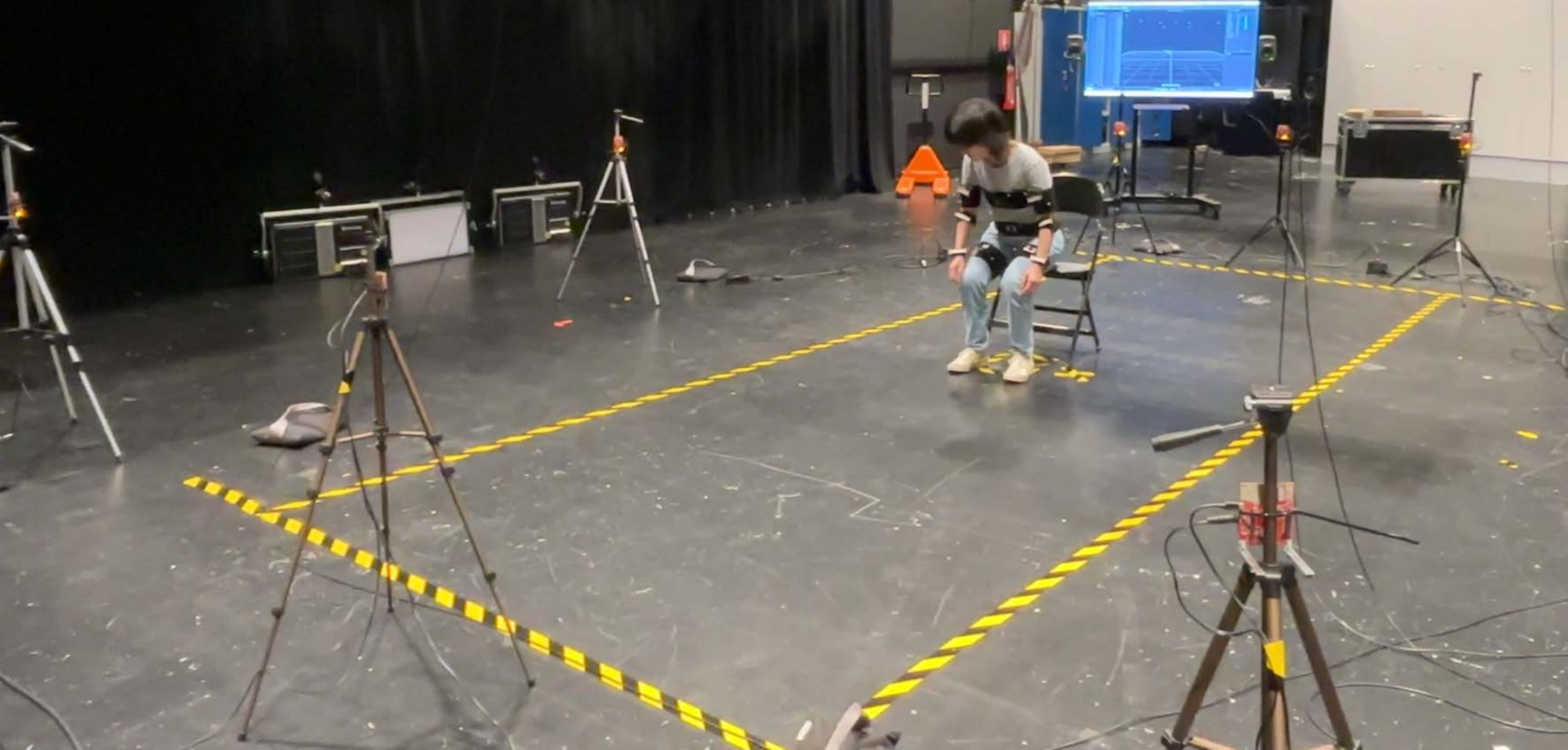} 
    \caption{Sitting}
  \end{subfigure}
  \hfill
  \begin{subfigure}[t]{0.48\textwidth}
    \centering
    \includegraphics[width=\textwidth]{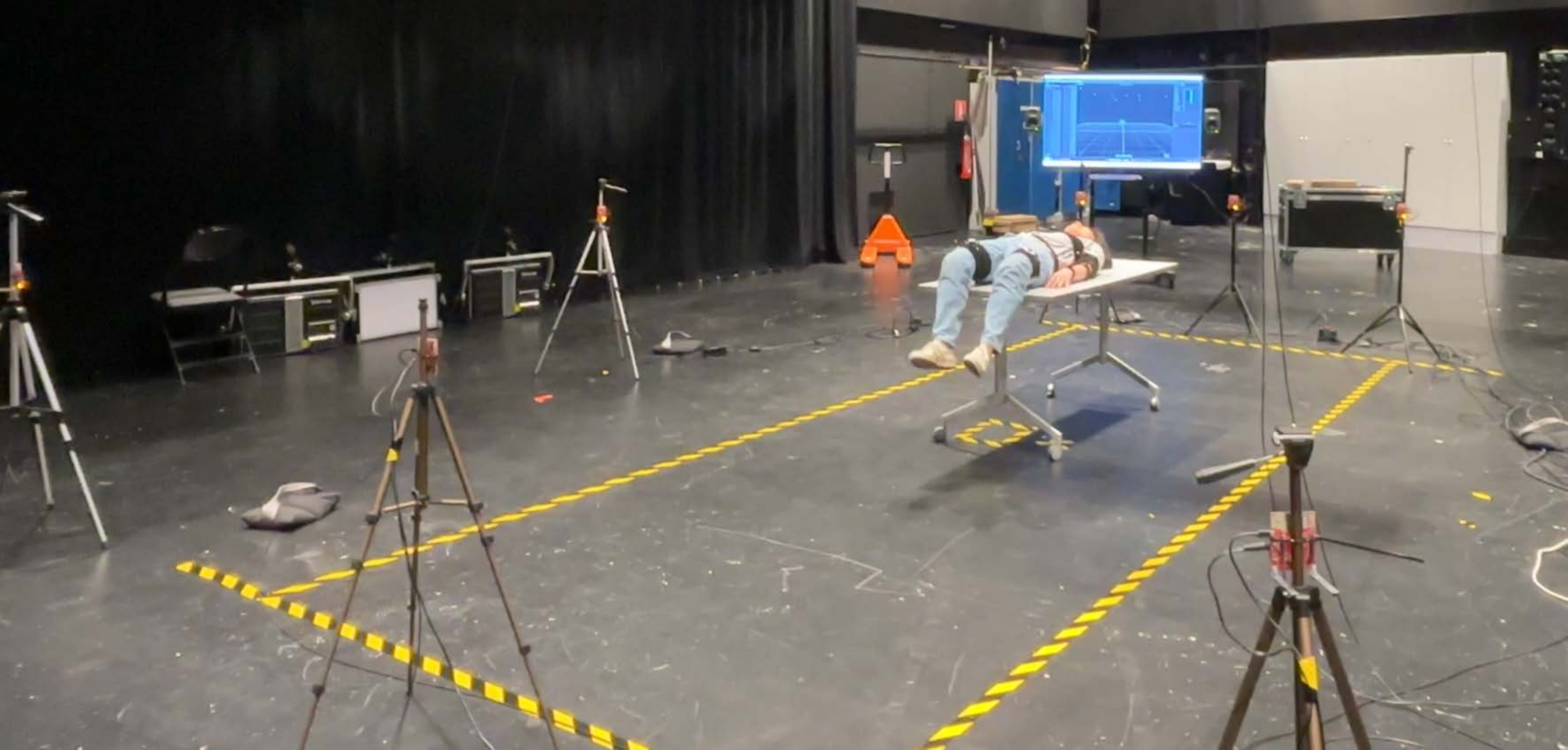} 
    \caption{Lying}
  \end{subfigure}
    \centering
  \begin{subfigure}[t]{0.48\textwidth}
    \centering
    \includegraphics[width=\textwidth]{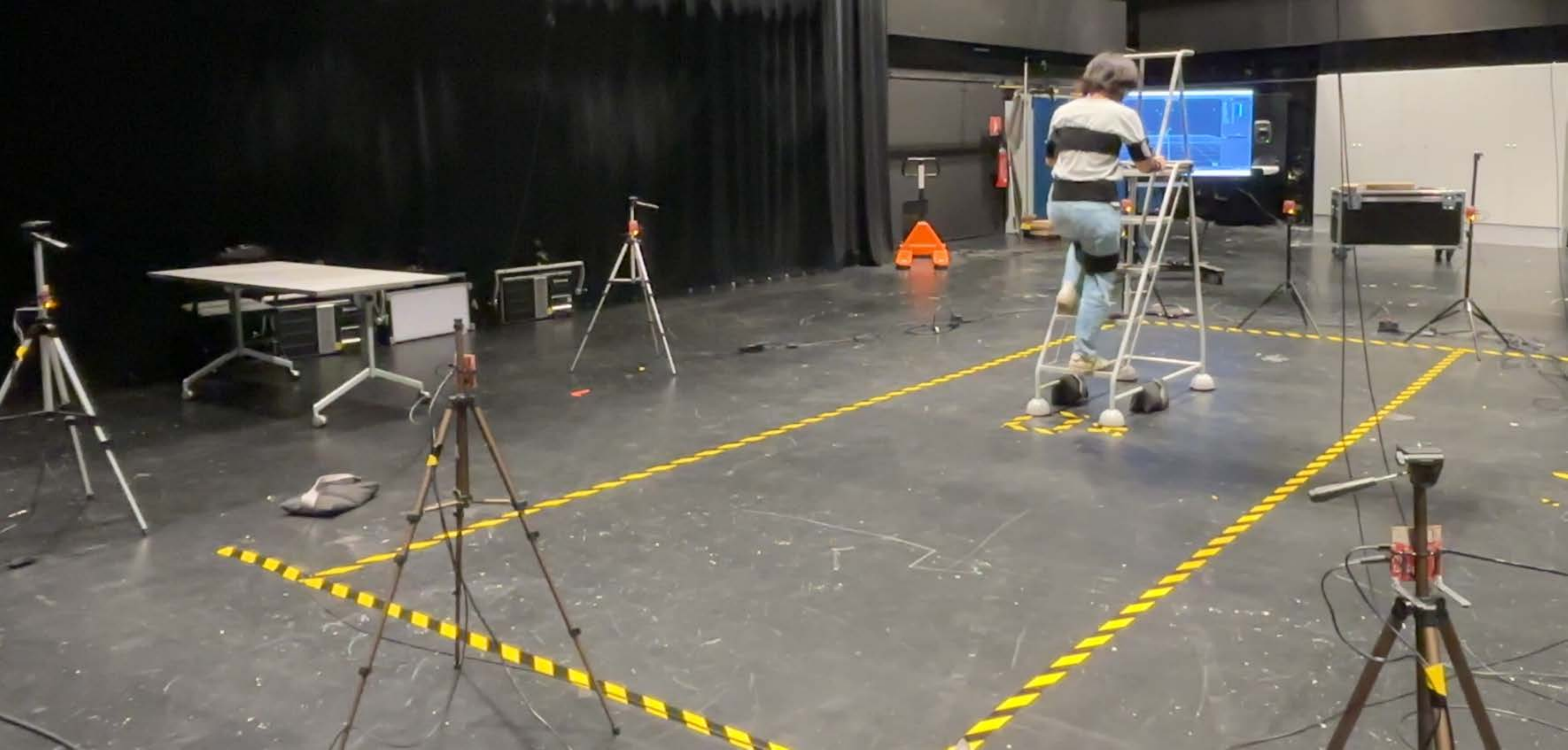} 
    \caption{Ascending stairs}
  \end{subfigure}
  \hfill
  \begin{subfigure}[t]{0.48\textwidth}
    \centering
    \includegraphics[width=\textwidth]{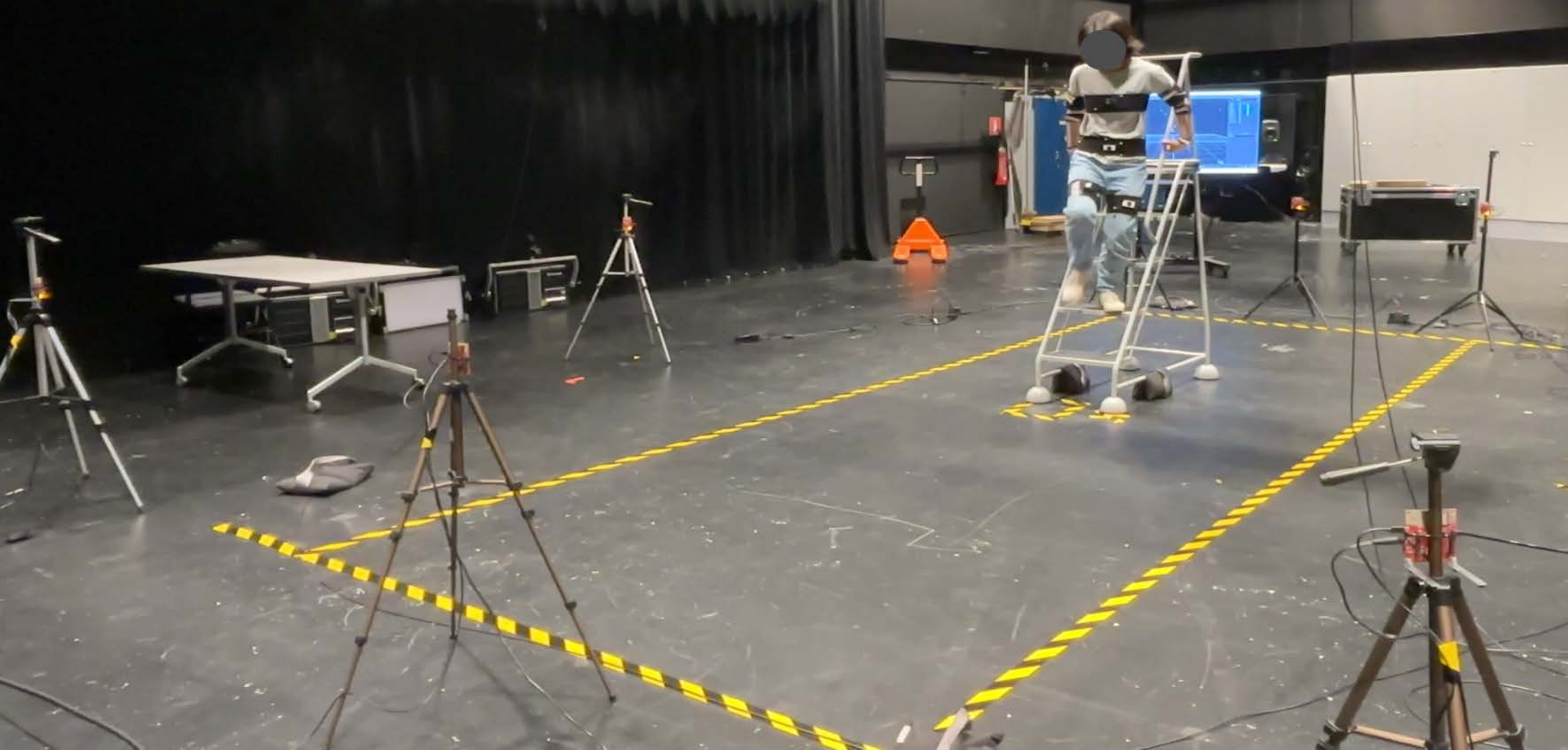} 
    \caption{Descending stairs}
  \end{subfigure}
    \centering
  \begin{subfigure}[t]{0.48\textwidth}
    \centering
    \includegraphics[width=\textwidth, height=3.65cm]{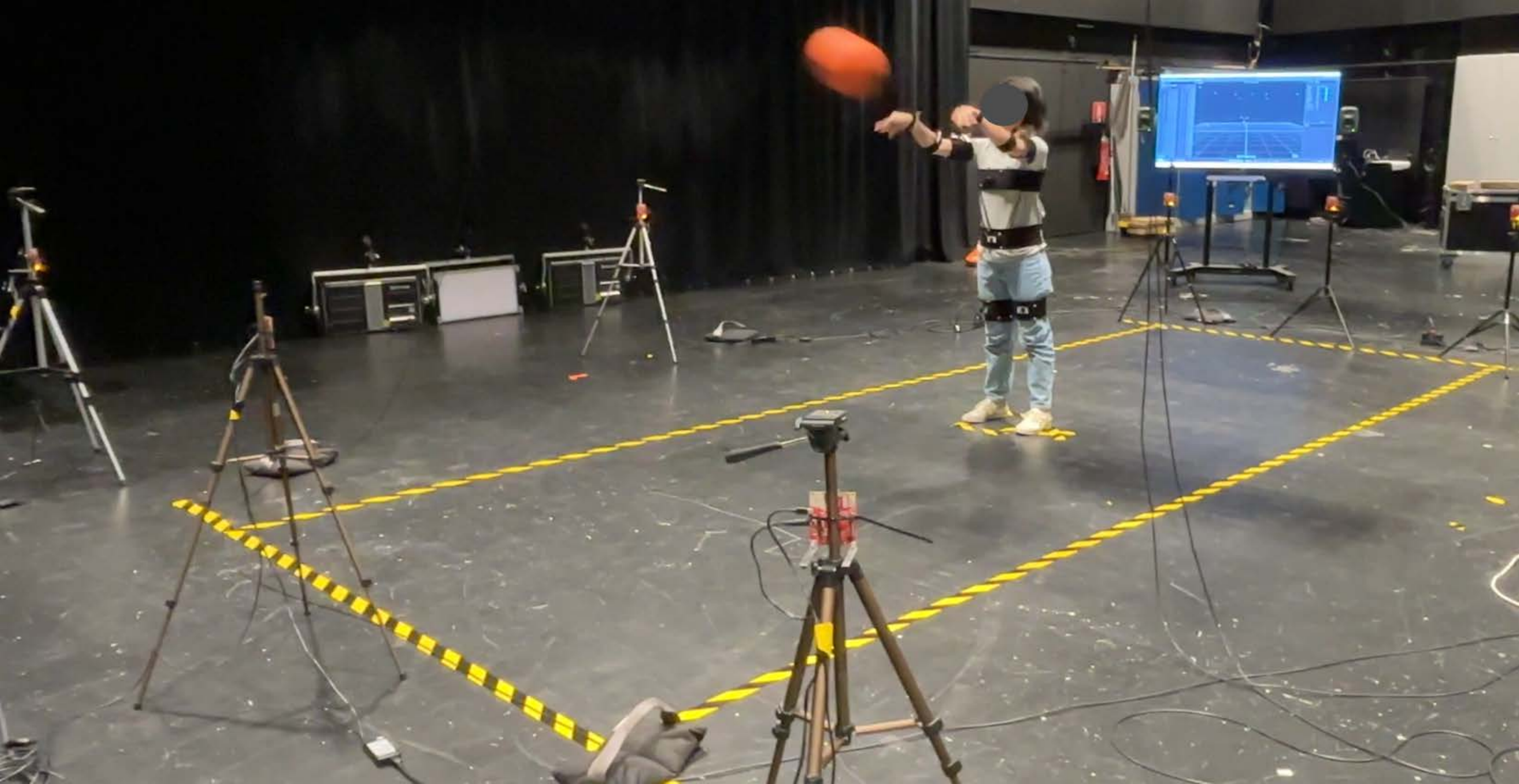} 
    \caption{Passing a basketball}
  \end{subfigure}
  \hfill
  \begin{subfigure}[t]{0.48\textwidth}
    \centering
    \includegraphics[width=\textwidth]{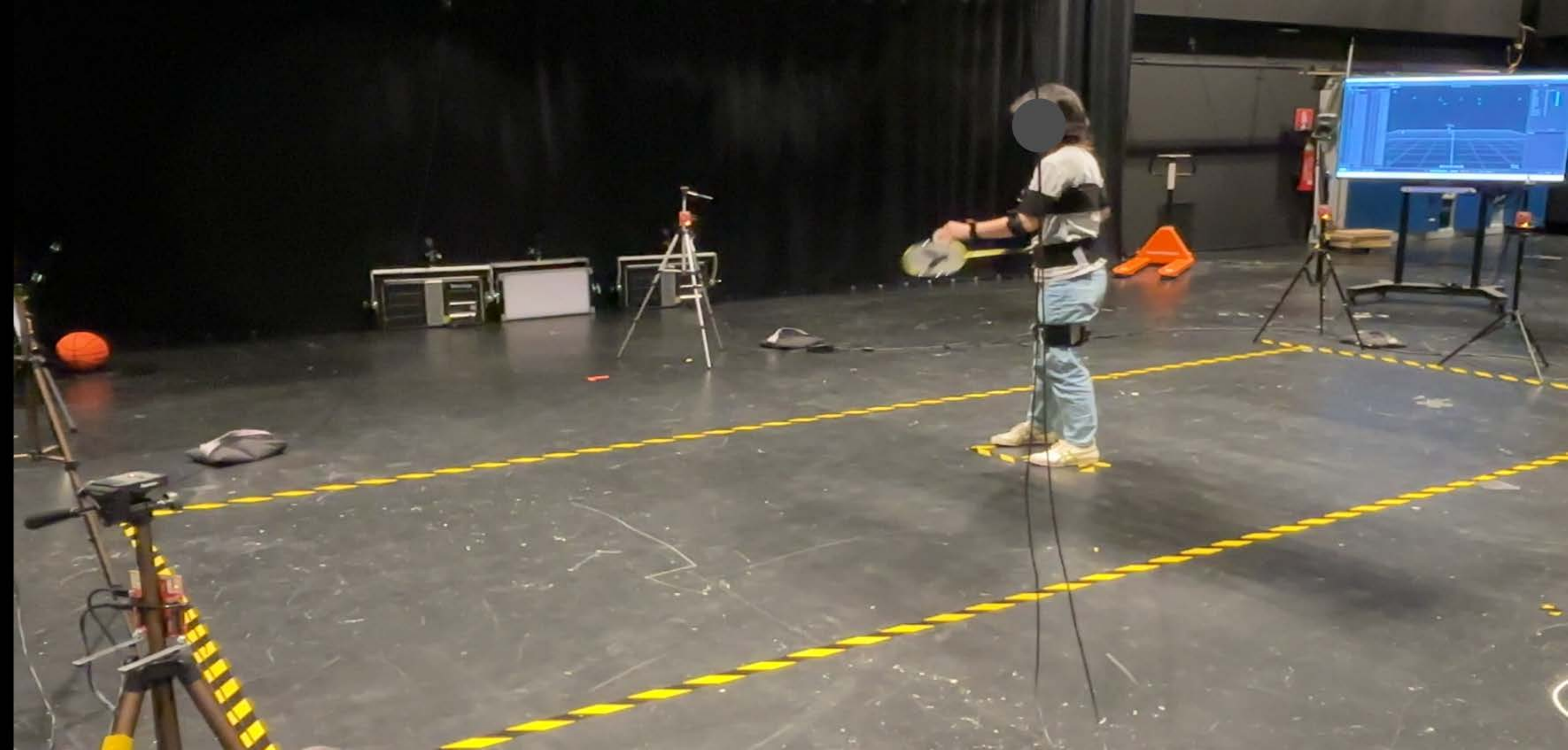} 
    \caption{Playing badminton}
  \end{subfigure}
    \centering
  \begin{subfigure}[t]{0.48\textwidth}
    \centering
    \includegraphics[width=\textwidth]{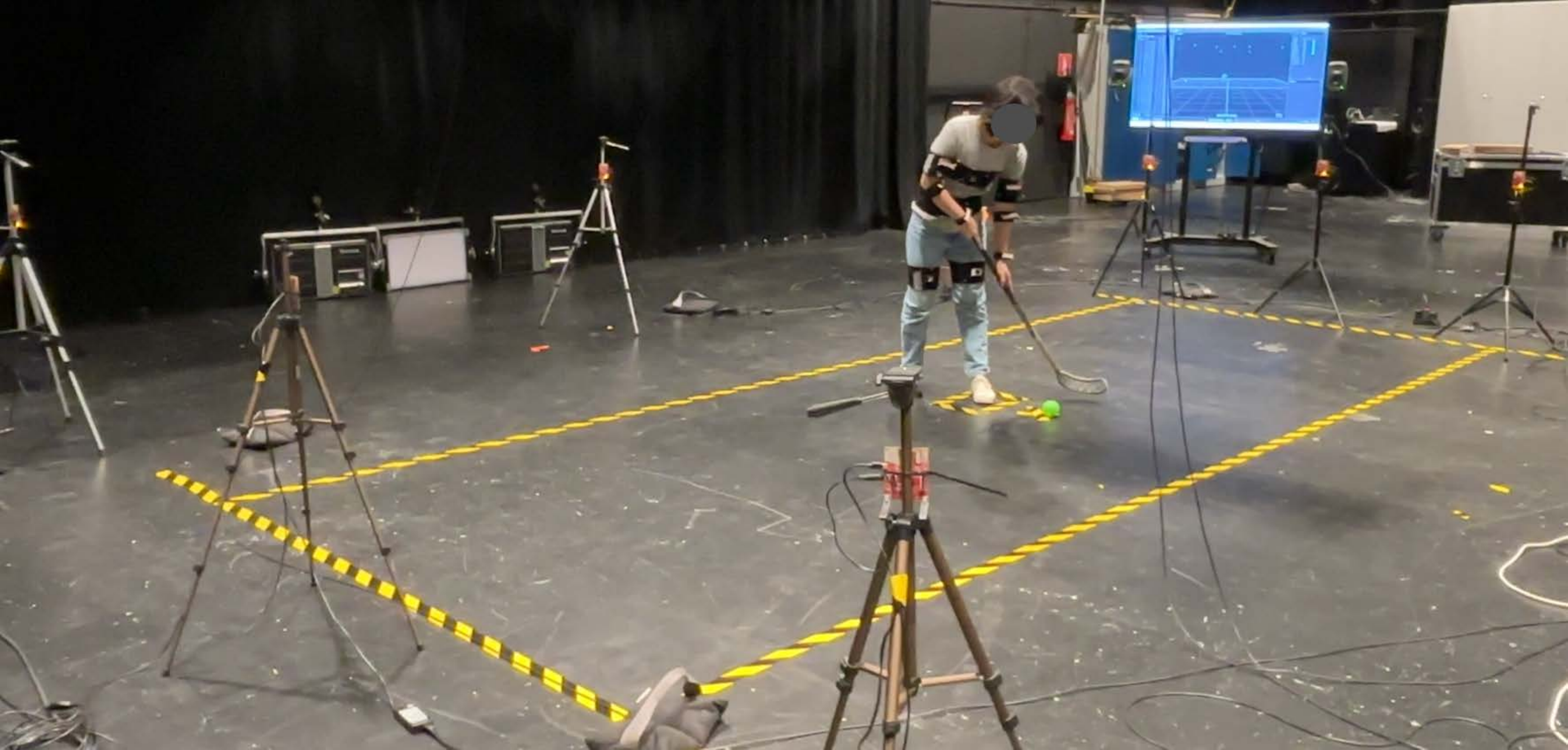} 
    \caption{Playing floorball}
  \end{subfigure}
  \hfill
  \begin{subfigure}[t]{0.48\textwidth}
    \centering
    \includegraphics[width=\textwidth]{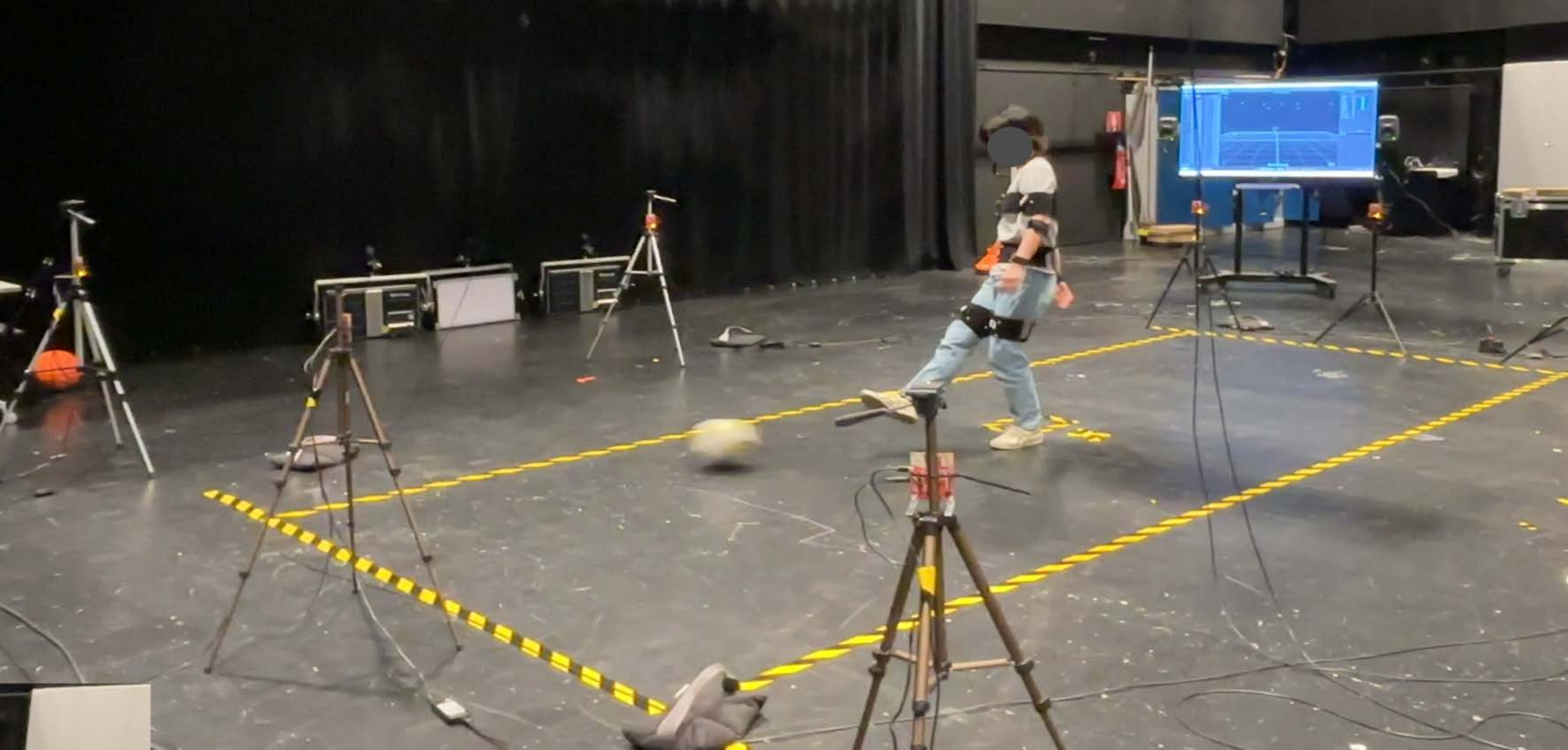} 
    \caption{Kicking a football}
  \end{subfigure}
  \caption{Human activities set for campaign~2.}
  \label{fig:s2_sets}
\end{figure}

We collected data from subjects for ten activities (walking, running, sitting, lying, ascending stairs, descending stairs, passing a basketball, playing badminton, playing floorball, and kicking a football; see Fig.~\ref{fig:s2_sets}). Each gesture was repeated 8 times by each subject.

Our selection of activities combines fundamental Activities of Daily Living (ADLs) with more complex sports-related movements. As the most frequent and essential actions in daily life, these ADLs serve as standard benchmarks found in commonly used HAR datasets, including \textit{Opportunity}~\cite{chavarriaga2013opportunity}, UCI-HAR~\cite{anguita2013public}, and WISDM~\cite{weiss2019wisdm}. Specifically, the execution protocols for the activities were adapted according to the recording environment. In the laboratory setting, the primary objective was to acquire high-fidelity motion data. Therefore, for walking and running activities, participants were instructed to follow an 'S'-shaped trajectory. This ensured that the sensors distributed throughout the environment could optimally capture the required data. Conversely, in the living room and industrial settings, participants were not constrained to a strictly predefined trajectory. Instead, they received general instructions to move around the space, thereby generating more naturalistic and ecologically valid data. Similar adjustments also apply to other activities, such as lying on the sofa or ascending stairs in an unconstrained manner.

To push the boundaries beyond these foundational tasks, we incorporated sports activities that introduce a higher degree of complexity through high-intensity, variable movements. Training a model on these dynamic actions forces it to learn more sophisticated and generalizable features, thereby enhancing the dataset's value for emerging applications in sports science and digital coaching.

To manage this complexity and create a focused dataset for analysis, we adopted a specific data collection strategy for the sports activities. Rather than recording long, continuous sequences of gameplay (e.g., a full basketball sequence involving dribbling, crossover, and layup), we captured several repetitions of a single, fundamental action from multiple participants. For instance, we focused on the distinct motion of passing a basketball, a frequently used and representative movement.  

\subsubsection{Campaign~3 (Affective behaviors)}\label{sec:c3_protocol}
For the affective behavior data collection, we recorded the physical expressions of six target states: focus, distraction, stress, relaxation, depression, and excitement. 
To ground this label space in established affect research, we anchor it in Russell's circumplex model of affect~\cite{russell1980circumplex}, which organizes affective states along the two dimensions of valence (pleasant--unpleasant) and arousal (activated--deactivated)~\cite{posner2005circumplex}. 
Four of the six labels instantiate the four quadrants of this space: \emph{excitement} (positive valence, high arousal), \emph{stress} (negative valence, high arousal), \emph{depression} (negative valence, low arousal), and \emph{relaxation} (positive valence, low arousal). 
In this work, ``depression'' denotes a transient, induced low-energy affective behavior expression in the sense of this quadrant, characterized by psychomotor slowing, reduced movement amplitude, and slumped posture (It does not refer to a clinical disorder). 
Under this model, the depression--excitement pairing used in the elicitation protocol is a \emph{diagonal} contrast: the two states differ in both valence (negative versus positive) and arousal (low versus high) and therefore occupy opposite quadrants of the circumplex.
The pairing does not claim that depression and excitement are opposite poles of a single underlying dimension. Whether affective opposites form strictly bipolar dimensions is itself debated in the affect literature~\cite{russell1999bipolarity}.
We contrast the two diagonal quadrants because a condition pair that differs in both dimensions maximizes the expected kinematic separation between the recorded expressions, from psychomotor slowing and reduced movement amplitude at one end to fast, high-amplitude, animated movement at the other, and therefore yields the clearest behavioral signal for sensing research.
The same logic applies to the stress--relaxation pair of Part~I, which contrasts the two remaining quadrants.
The remaining pair, \emph{focus--distraction}, is considered as an attentional state pair grounded in the task-engagement literature~\cite{matthews2002fundamental}. 
It is included since attentional state is among the most application-relevant conditions for workplace, learning, and living scenarios, and because it is expressed through the same macroscopic kinematic channel of task-oriented stillness versus restlessness that the sensors observe. 
Table~\ref{tab:affective_mapping} summarizes the mapping.

\begin{table}[t]
\caption{Grounding of the six affective behavior categories. Four categories instantiate the quadrants of Russell's circumplex model of affect~\cite{russell1980circumplex}; focus/distraction form an attentional state pair from the task-engagement literature~\cite{matthews2002fundamental}. Labels denote observable behavioral expressions, verified by per-phase self-assessment.}
\label{tab:affective_mapping}
\centering
\small
\begin{tabular}{lllll}
\toprule
Category & Dimensional grounding & Elicitation & Typical observable expression \\
\midrule
Excitement & Circumplex: valence $+$, arousal $+$ & Video priming + role-play & Fast, high-amplitude, animated movement \\
Stress & Circumplex: valence $-$, arousal $+$ & Time pressure + countdown & Hurried, tense, jerky manipulation \\
Relaxation & Circumplex: valence $+$, arousal $-$ & Pressure removal & Slow, smooth, unhurried movement \\
Depression & Circumplex: valence $-$, arousal $-$ & Video priming + role-play & Psychomotor slowing, slumped posture \\
Focus & Attentional engagement (high) & Undisturbed puzzle task & Task-oriented stillness, regular reaching \\
Distraction & Attentional engagement (low) & Layered interruptions & Head/torso reorientation, irregular pauses \\
\bottomrule
\end{tabular}
\end{table}

Note that the labels denote \emph{observable behavioral expressions} of these states, since internal emotional or cognitive states can not be measured directly via environmental RF sensors. 
The sensors capture the macroscopic physical expressions that are symptomatic of the induced conditions, such as variations in pacing, movement speed and smoothness, or irregular posture. 
The induction itself is verified on the experiential level through per-phase self-assessments (see below). 
Physiological signals such as ECG or EEG reflect intrinsic states directly and therefore provide a valuable explainable reference, and contactless RF-based sensing of important signs is itself an established research direction~\cite{liu2020wireless}.
In the core RF-Behavior campaigns, we did not use physiological instruments. Body-worn electrodes could restrict the natural movements needed to capture macroscopic behavioral expressions. Besides, our labels describe observable expressions rather than internal states.
To connect both levels, the dataset includes a physiological reference subset, in which ECG and EEG were recorded simultaneously with radar and IMU during the affective behavior protocol (Section~\ref{sec:physio_subset}).
We follow established methodology by providing controlled elicitation combined with self-reports~\cite{busso2008iemocap,gross1995emotion}.

There are three main considerations for including affective behavior in this motion-centric dataset. 
First, ambient well-being monitoring in eldercare, single-occupancy homes, and workplaces targets precisely the states covered here, such as sustained stress, low-energy (depression-like) behavior, and distraction. Such deployments can typically use neither cameras (privacy) nor body-worn physiological sensors (compliance), which leaves macroscopic movement as the only continuously observable signal. Second, whether affective state modulates whole-body kinematics strongly enough to be detected by ambient RF is an open question that cannot be studied without a corpus coupling the two. RF-Behavior provides this testbed with motion-capture-grade ground truth. Third, since gestures, activities, and affective behaviors are recorded with the same synchronized infrastructure and share a participant pool, the dataset enables the study of whether representations learned from short-horizon motion transfer to long-horizon behavioral states.

\begin{figure}
\centering
\includegraphics[width=10cm]{fig/s3_p1.pdf}
\caption{Experimental setup for Part I affective behavior data collection: assemble a jigsaw puzzle.}
\label{fig:s3_p1}
\end{figure}

Although data was collected across multiple environments, the protocol for emotion data collection remained consistent. 
To ensure the capture of genuine responses, each participant took part in the third campaign only once, regardless of the environment. 
In contrast, for the remaining campaigns, participants were allowed to participate multiple times across different environmental settings.

We split the data collection into two parts. 
In Part I, we collected a continuous recording for the first two pairs (focus-distraction; stress-relaxation), approximately 10 to 12 minutes per participant without interruptions between the different stages. 
Part II focuses on the recording of depression-excitement. 
In this campaign, we have 23 participants. 

\paragraph{Part I}: The participant was seated at a large desk, approximately 2 meters wide, with a chair. 
The primary task was to assemble a jigsaw puzzle (see Fig.~\ref{fig:s3_p1}). 
The puzzle pieces were intentionally scattered across the wide desk surface. 
This design choice was made to encourage participants to make large, deliberate, and easily detectable movements with their arms and upper body when searching for and reaching for pieces, providing richer data for the motion sensors. 
The experimental setup was adjusted slightly for different environments. 
For example, in the living room setting, participants were seated on a sofa with puzzle pieces placed on a coffee table.

The 10-12 minute campaign was structured into four consecutive phases, each designed to induce a specific target state.
\begin{enumerate}
    \item Focus Phase (2 minutes): This initial phase served as a baseline. 
    The participant was instructed to begin assembling the jigsaw puzzle at their own pace, without any external pressures or interruptions. 
    This was designed to capture a state of natural concentration on the task.
    \item Distraction Phase (3-5 minutes): To trigger a state of distraction, a series of escalating interruptions was introduced while the participant continued the puzzle task. 
    The interruptions were layered in three levels: \textbf{Level 1}: Auditory distractions were introduced, including random background noises and a phone ringtone played every 30 seconds; \textbf{Level 2}: Verbal interruptions were added on top of the auditory distractions; \textbf{Level 3}: While the auditory and verbal distractions continued, three data recorders began engaging in sports activities within the experiment room to create a physical distraction.
    \item Stress Phase (3 minutes): The participant was verbally informed, "You have now only 3 minutes to finish it." This was reinforced with two additional stressors: 
    a large countdown timer was displayed on a screen in front of the participant; verbal countdowns were announced, including an oral reminder for the final 30 seconds.
    \item Relaxation Phase (2 minutes): In the final phase, the stressors were removed. 
    The participant was explicitly told that the time pressure was gone and that they could continue building the puzzle at their own speed.
\end{enumerate}

\paragraph{Part II}: The second part of the data collection aimed to record behaviors associated with depression and excitement through a role-playing methodology. To avoid inducing genuine distress, participants viewed video stimuli to help them understand and simulate a target state, which they were then asked to embody during the recording.

To induce affective behavior, we employed a two-stage protocol combining audio-visual priming and improvised role-playing. 
For the depression state, participants watched a short video that personified the emotion. 
For the contrasting excitement state, they viewed a high-energy video, with the option to choose a clip they personally found motivating (e.g., rock climbing). 
Film-based elicitation is a standardized method to induce specific affective states~\cite{gross1995emotion}. 
After each video, participants responded to neutral prompts, such as "Describe an experience with sports," by role-playing the target state through their speech, tone of voice, and body language.
This role-playing approach ensures the generation of behavioral signs without relying on rigid scripts (cf. e.g.~\cite{busso2008iemocap}).

Note that the semantic content of the participants' stories was not subject to analysis; 
the focus was exclusively on the expressive, non-verbal cues in their speech and body language. 
Furthermore, participants were encouraged to speak in their preferred language to ensure they could express themselves naturally and comfortably.

\paragraph{Post-Task Assessment (Ground Truth Collection)} Immediately after the recording task was completed, participants were asked to provide a self-assessment of their experience.
The participants rated the intensity of each of the six target states (focus, distraction, stress, relaxation, depression, and excitement) they experienced during the relevant phases of the experiment. 
This self-reported data can serve as the detailed ground truth label for the sensor data collected during each phase. 
Ratings used a 6-point Likert scale from 0 (not in the mood) to 5 (quite in the mood) (cf. Table~\ref{tab:s3_selfassessment} in the \textbf{Appendix} for further details).

\paragraph{Physiological Reference Subset}\label{sec:physio_subset}
To complement the behavioral recordings with physiological reference signals, we collected an additional subset in the living-room environment.
Seven new participants (in addition to the 61 core participants) completed the campaign~3 affective protocol while ECG, EEG, IMU, and radar were recorded simultaneously.
ECG was recorded using a Shimmer3 ExG unit at a sampling rate of 512\,Hz. Four disposable electrodes (RA, LA, RL, and LL) were used. RA and LA were placed below the right and left clavicles, respectively, and RL and LL on the right and left lower abdomen, respectively. EEG was recorded using a NeuroSky MindWave Mobile headset, with one EEG channel at a sampling rate of 512 Hz.  The active electrode was positioned over the left frontal forehead, with the reference and ground contacts attached to the left ear. Raw EEG samples were recorded in ADC units. The device-derived signal-quality index, eight EEG band-power measures (delta, theta, low and high alpha, low and high beta, and low and mid gamma), and proprietary eSense Attention and Meditation scores (0–100) were recorded at approximately 1 Hz.
The streams were recorded with approximately coordinated start and stop times.
The physiological recordings serve as a reference that links the induced conditions to internal responses, for example heart-rate and heart-rate-variability changes during different phases, and they complement the per-phase self-assessments.
The affective behavior labels remain defined as observable behavioral expressions.
With 7 participants, it supports per-subject reference analyses and sanity checks of the induction, not population-level conclusions. Demographic details and example recordings of this subset are documented in the dataset repository.

\subsubsection{Campaign~4 (Gestures)}\label{sec:rfid_8tags}
A total of 7,140 gesture instances were recorded from 17~volunteers (10~male and 7~female) with heights ranging from 155~cm to 185~cm. 
Each participant executed 21 hand gestures (20 repetitions). 
All gestures were performed at two different distances from the antenna: 1.5~and 3~meters. 
The duration of individual gestures differed depending on both the complexity of the movement and the participant’s execution speed.

Examples of collected data are provided in the \textbf{Appendix} (see~\ref{sec:apx_datasample}).

\subsection{On-body Sensor Setup}
\begin{figure}
  \centering
  \begin{subfigure}[t]{0.32\textwidth}
    \centering
    \includegraphics[width=0.6\textwidth, height=5cm]{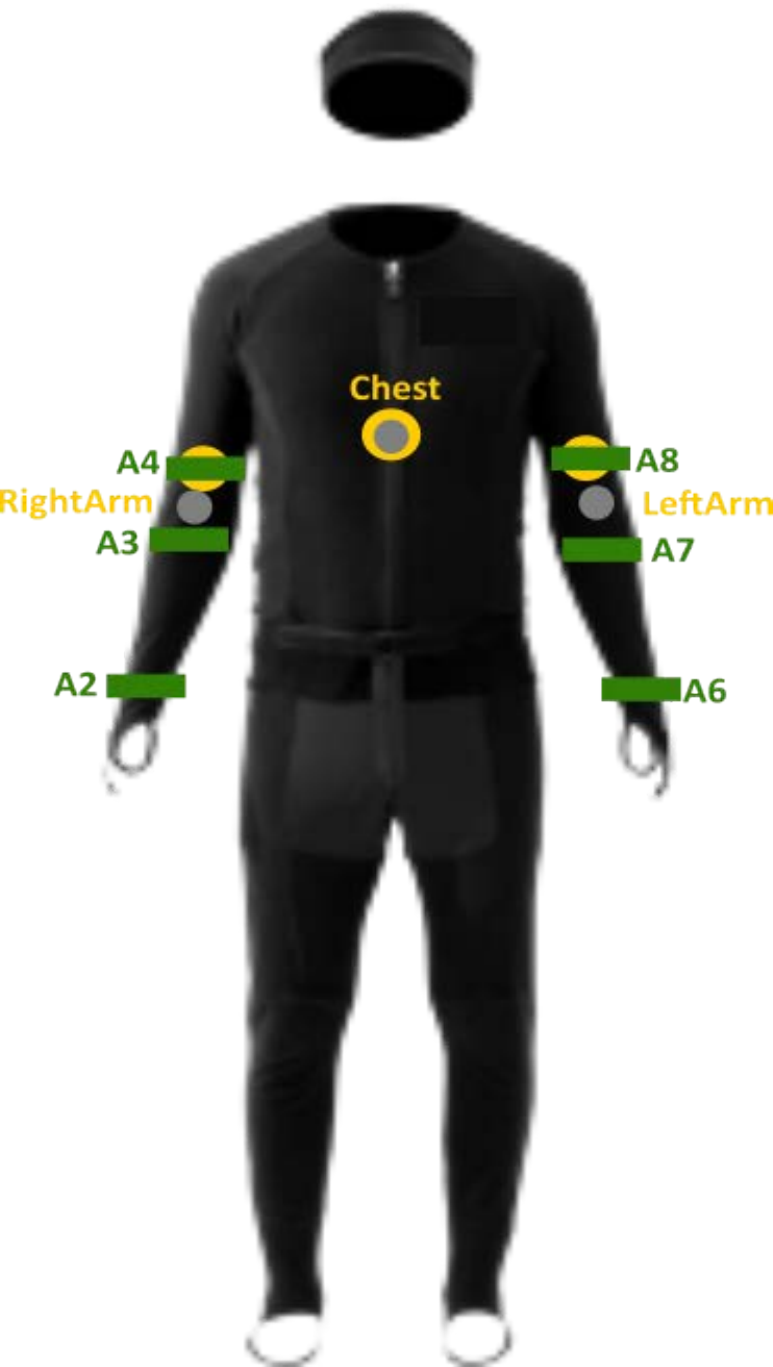}
    \caption{Campaign~1: gestures.}
    \label{fig:s1_sensor_onbody}
  \end{subfigure}
  \hfill
  \begin{subfigure}[t]{0.32\textwidth}
    \centering
    \includegraphics[width=0.6\textwidth, height=5cm]{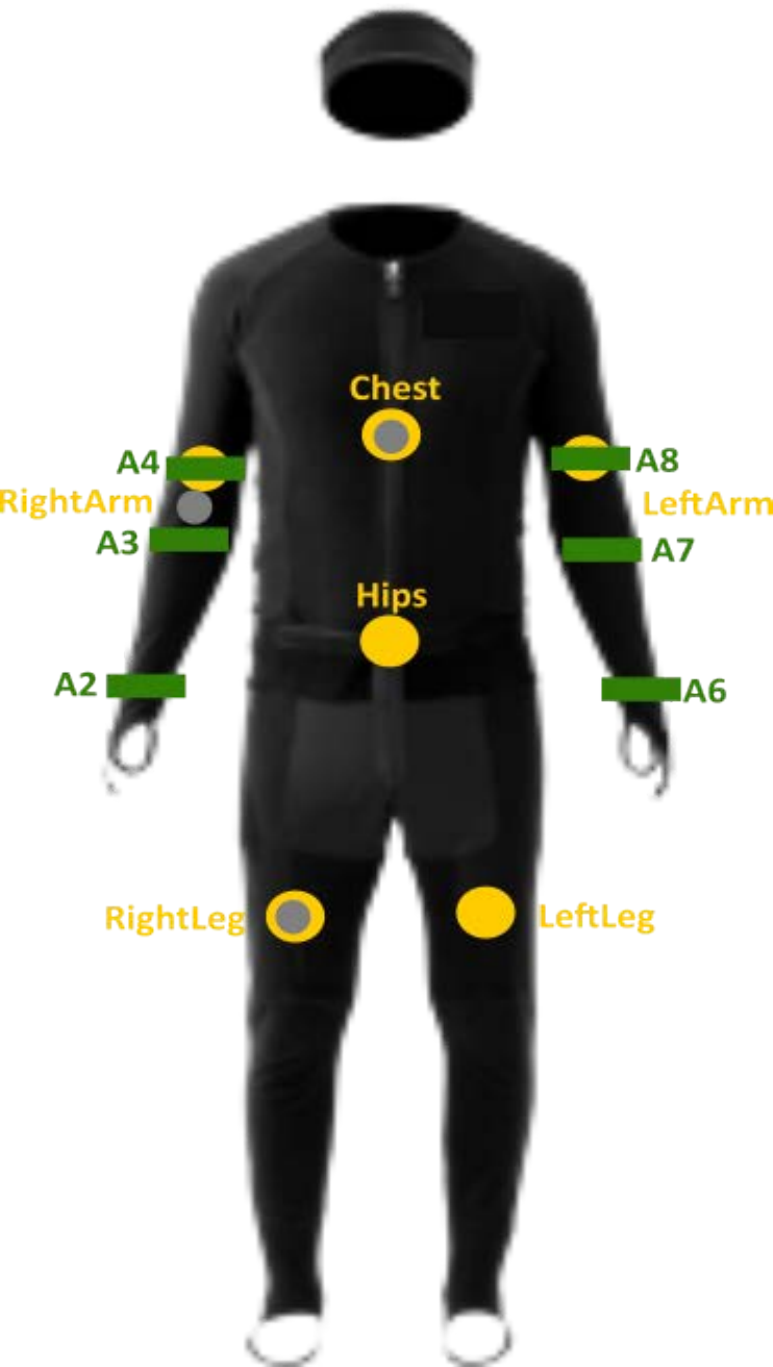}
    \caption{Campaign~2: human activities.}
    \label{fig:s2_sensor_onbody}
  \end{subfigure}
  \hfill
  \begin{subfigure}[t]{0.32\textwidth}
    \centering
    \includegraphics[width=0.6\textwidth, height=5cm]{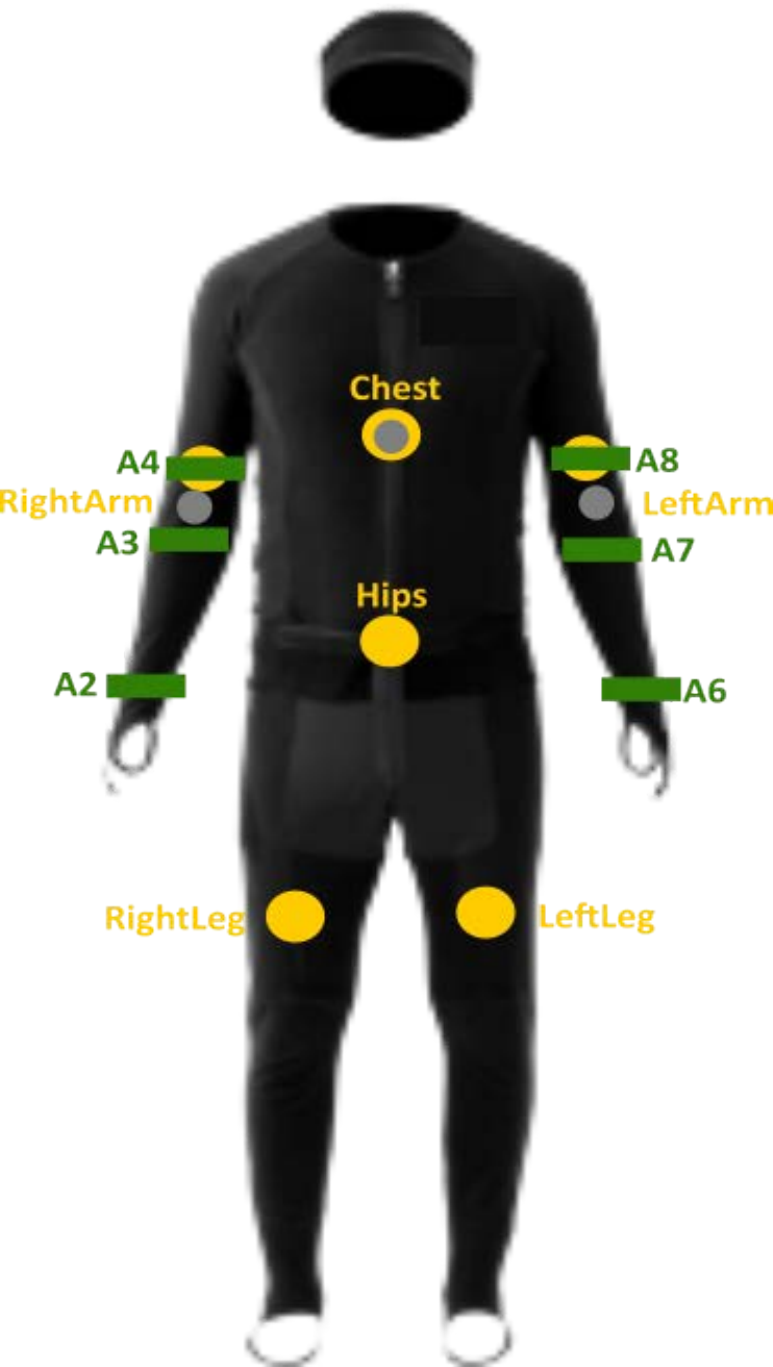}
    \caption{Campaign~3: affective behaviors.}
    \label{fig:s3_sensor_onbody}
  \end{subfigure}
  \caption{On-body sensor placement for the three lab campaigns.}
  \label{fig:onbody_placement}
\end{figure}

In addition to the sensors placed in the environment, different sensors were also mounted at different body locations of the subject (see Fig.~\ref{fig:onbody_placement}). 
We considered three types of sensors attached to the human body:
\begin{enumerate}
    \item RFID tags (shown in green): These tags are used for tracking and are placed on each arm --- on the wrist (A2, A6), below the elbow (A3, A7), and above the elbow on the front side (A4, A8). Specifically, for campaign~4 (Gestures), we have two additional tags (A1 and A5) placed on the wrists of both hands (see Fig.~\ref{fig:c4_RFIDSetup}). 
    \item Infrared camera markers (shown in yellow): These markers are positioned to facilitate motion capture. 
    The placement of the infrared camera markers \textbf{varies} \textbf{across} different campaigns.
    \item IMU sensors (shown in gray): 
    These sensors provide data on orientation, acceleration, and angular velocity. The placement of the IMU sensors \textbf{varies} \textbf{across} different campaigns.
\end{enumerate}
As mentioned above, in the lab setting, we have all three types of sensors placed on the human body. 
For the living room and industrial factory settings, we used only the IMU sensors.

\begin{table}
  \centering
  \caption{Sensor placement}
  \label{tab:sensor_placement}
  \begin{tabular}{l *{4}{c}} 
    \toprule
    & \multicolumn{4}{c}{Sensor placement} \\ 
    \cmidrule(lr){2-5} 
    & \textbf{Campaign~1} & \textbf{Campaign~2} & \textbf{Campaign~3} & \textbf{Campaign~4}  \\
    \midrule
    RFID Tag - Left arm          & \checkmark & \checkmark & \checkmark & \checkmark \\
    RFID Tag - Right arm         & \checkmark & \checkmark & \checkmark & \checkmark \\
    Radar - Floor (8)            & \checkmark & \checkmark & \checkmark & - \\
    Radar - Ceiling (5)          & \checkmark & \checkmark & \checkmark & - \\
    IMU - Left arm               & \checkmark & - & \checkmark & - \\
    IMU - Right arm              & \checkmark & \checkmark & \checkmark & - \\
    IMU - Chest                  & \checkmark & \checkmark & \checkmark & - \\
    IMU - Right leg              & - & \checkmark & - & - \\
    Camera markers - Left arm    & \checkmark & \checkmark & \checkmark & - \\
    Camera markers - Right arm   & \checkmark & \checkmark & \checkmark & - \\
    Camera markers - Chest       & \checkmark & \checkmark & \checkmark & - \\
    Camera markers - Left leg    & - & \checkmark & \makecell{\checkmark} & - \\ 
    Camera markers - Right leg   & - & \checkmark & \makecell{\checkmark} & - \\
    Camera markers - Hips        & - & \checkmark & \makecell{\checkmark} & - \\
    \bottomrule
  \end{tabular}
\end{table}

Table~\ref{tab:sensor_placement} details the sensor placement for each campaign. 
Across all experiments, eight backscatter-type tags were consistently used. 
The placement of IMU sensors varied by campaign. 
For campaign~1 (gesture collection) and campaign~3 (affective behavior collection), sensors were placed exclusively on the upper body (chest, left and right arms), as these experiments were designed to capture upper-body motion. 
In contrast, for campaign~2, IMU data were collected from the right arm, chest, and right leg.

Camera markers were placed to help the studio's infrared camera system to capture movements from desired body locations. 
For the gesture data in campaign~1, markers were limited to the upper body. 
For the other campaigns, they were placed on the arms, chest, hip, and legs. 
Note that, during the first part of campaign~3, participants were seated at a desk to complete a puzzle. 
The desk obstructed the cameras' view. 
Therefore, the marker data from this segment is unusable for analysis.

\section{Benchmarks and Baseline Results}
To demonstrate the utility and multi-scale nature of the RF-Behavior dataset, we establish baseline benchmarks tailored to the physical characteristics of each task. 
We adopt a task-driven evaluation strategy to ensure that the baselines reflect realistic deployment scenarios and highlight the specific strengths of each modality. 
The benchmarking is split into two primary directions: an upper-bound baseline in a controlled setting, and a cross-domain robustness evaluation in extended environments. 

Thus, we present three types of tasks in the paper: 
\textbf{1) Fine-grained gesture recognition: } 
hand and arm gestures are highly localized, short-range interactions. 
For this task, we benchmark radar, RFID, LoRa, and IMU independently. 
RFID provides a performance upper bound for precise, device-bound limb tracking. 
Radar establishes a markerless, volumetric tracking baseline. 
LoRa and IMU delineate what a single communication link and a body-worn inertial sensor can achieve on the same task. 
2)~\textbf{Macro activity recognition}. 
Macro-activities (e.g., walking, running) involve significant spatial displacement and are highly sensitive to environmental multipath. 
We evaluate radar across all three environments, complement it with LoRa and IMU baselines, and add a radar and IMU fusion study for cross-environment transfer. 
3)~\textbf{Affective behavior recognition}. The physical expression of affective states involves both overall posture changes and minor localized movements. 
We evaluate radar, IMU, their statistical-feature fusion, and LoRa. Where applicable, the same model ladder (statistical features with a Random Forest, followed by deep models of increasing capacity) and the same protocol names (within-subject, unseen users, cross-environment) are used across modalities, so that the contribution of each sensing modality can be compared directly.

Table~\ref{tab:results_overview} summarizes the strongest unseen-subject result per modality and task. 
Based on these results, we notice that no single modality leads on all three tasks. 
Wherever fusion is compared against single modalities under matched conditions, it improves over the best of them (parenthesized gains).
\begin{table}[t]
\caption{Overview of the baseline results: best unseen-subject accuracy (\%) of the strongest model per modality and task (--: not evaluated). Models and protocols are detailed in the referenced subsections. 
Affective values are the best across the three environments (Tab.~\ref{tab:affective_baseline}). Parenthesized values in the fusion row give the improvement over the best single modality under the matched evaluation setting (Secs.~\ref{sec:gesture_fusion}, \ref{sec:activity_fusion} and \ref{sec:emotion_fusion}). The fusion studies run in the real-world and cross-environment settings, where single-modality performance degrades, so fusion values are not directly comparable with the lab-based radar and RFID rows.
}
\label{tab:results_overview}
\centering\small
\begin{tabular}{lccc}
\toprule
Modality & Gestures (21 cl.) & Activities (10 cl.) & Affective (6 cl.) \\
\midrule
Radar (multi-radar) & 99.2 & 95.8 & 82.2 \\
RFID (on-body tags) & 98.1 & -- & -- \\
LoRa (single link) & 30.3 & 66.7 & 40.7 \\
IMU (3 body-worn) & 57.2 & 84.8 & 82.2 \\
radar and IMU fusion & 75.9 ($+$9.3) & 95.0$^{\dagger}$ ($+$50.0) & 80.6 ($+$4.8) \\
\bottomrule
\end{tabular}\\[2pt]
{\footnotesize $^{\dagger}$cross-environment with six adaptation subjects (Tab.~\ref{tab:sensor_fusion_cross_env}); direct radar-only transfer drops to 47.4 (Section~\ref{sec:cross_env_radar}).}
\end{table}

\subsection{Fine-grained gesture recognition}
We benchmark four modalities on the 21 gestures: 
(1)~radar (ground, ceiling, and their combination); 
(2)~RFID; 
(3)~LoRa; and 
(4)~IMU. 
Radar, RFID, and LoRa are evaluated on the lab recordings. 
The IMU baseline uses the two real-world settings. 
We additionally evaluate radar and IMU fusion in the real-world settings. 
Accuracy, macro-averaged F1-score, and (for radar and RFID) AUC are used for evaluation.

\subsubsection{Radar-based Gesture Recognition}\label{sec:RadarBased}
All reported baselines are trained and tested on the radar point clouds using the same preprocessing, data split, optimization, and downstream pipeline settings. 

\paragraph{Preprocessing pipeline: }
Radar point clouds contain background clutter, multipath reflections, and frame-to-frame variability in point density. 
A unified preprocessing protocol is applied to every recorded gesture instance before training any model.
\textbf{(1) Clutter suppression: }
For each gesture instance, all frames are first merged into a single aggregated point cloud.
Using DBSCAN with $\epsilon = 1$ and a minimum cluster size of 3, we segment the cloud and retain the largest high-density cluster, presumed to be the performer, while discarding minor clusters as noise; \textbf{(2) Temporal normalization: }
To ensure consistent temporal structure for model input, each gesture instance is re-segmented into a fixed number of frames (e.g., 2, 4, or 8);
\textbf{(3) Point set normalization: }
We therefore Resample the original point-cloud for each frame to have the same target number of points.

\paragraph{Benchmark models: } We include three representative models. 
\textbf{PointNet++}~\cite{qi2017pointnet++} is a widely used hierarchical network for point cloud processing. It progressively aggregates local neighborhoods to learn both fine-grained and global spatial features directly from unordered points. 
\textbf{Pantomime}~\cite{palipana2021pantomime} combines a PointNet++-style spatial encoder with recurrent Long Short-Term Memory (LSTM) layers. PointNet++ operates on each frame to learn hierarchical geometric descriptors from sparse 3D points, and the LSTM stack captures the temporal dynamics across frames. \textbf{Tesla}~\cite{salami2022tesla} is a compact gesture recognition architecture tailored for mmWave radar. It extracts spatial features frame-by-frame from the radar point cloud and then models their temporal evolution over the gesture sequence. 
All experiments are implemented in PyTorch and run on an NVIDIA GeForce RTX~\num{4080} GPU. Same configuration applied to all experiments.

\begin{figure}
  \centering
  \begin{subfigure}[t]{0.32\textwidth}
    \centering
    \includegraphics[width=\textwidth]{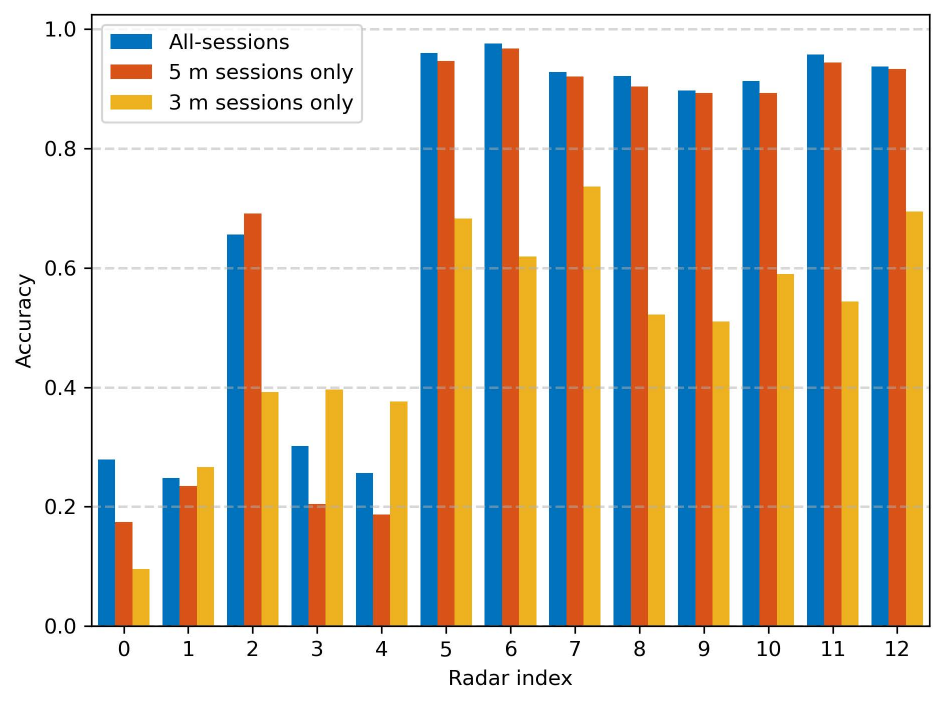} 
    \caption{PointNet++}
  \end{subfigure}
  \begin{subfigure}[t]{0.32\textwidth}
    \centering
    \includegraphics[width=\textwidth]{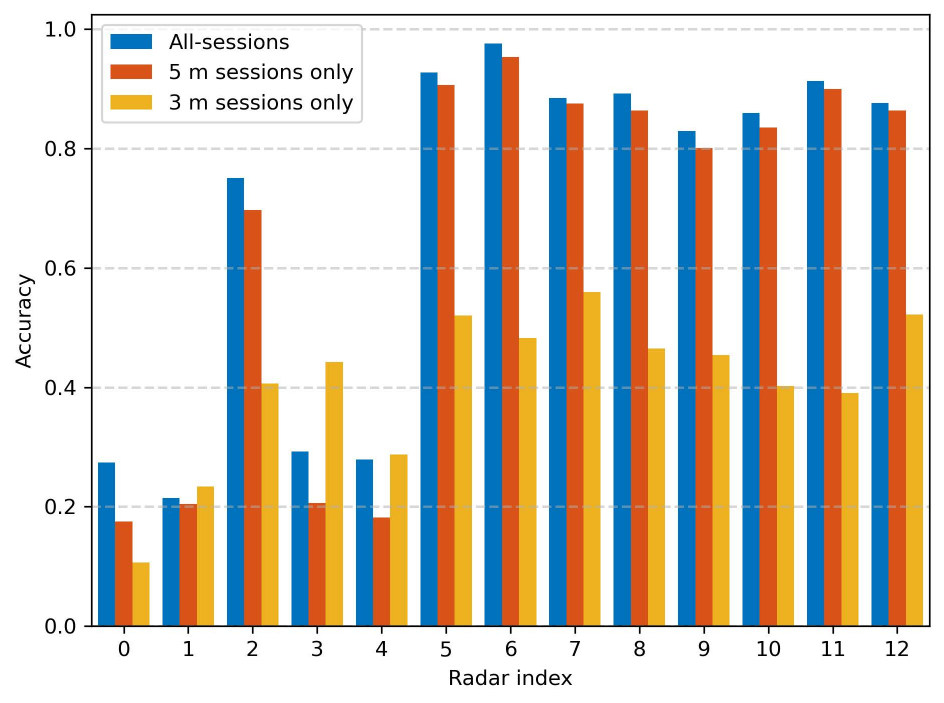} 
    \caption{Pantomime}
  \end{subfigure}
    \begin{subfigure}[t]{0.32\textwidth}
    \centering
    \includegraphics[width=\textwidth]{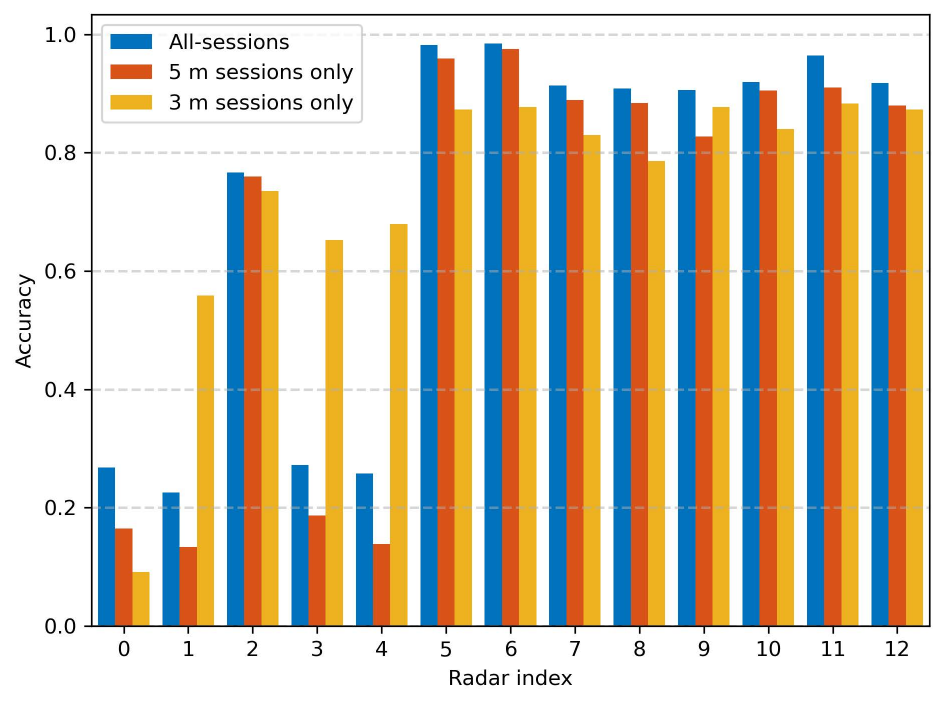} 
    \caption{Tesla}\label{fig:single_radar_tesla}
  \end{subfigure}
  \caption{Gesture recognition accuracy on benchmark models. From Left $\rightarrow$ Right: Pointnet++, Pantomime, Tesla. The X-axis indicates the Radar Index (see Fig.~\ref{fig:setup_s1}) while the Y-axis shows the accuracy.}
  \label{fig:single_radar_acc}
\end{figure}

We split the gesture data into three groups: 
all samples, 5-meter samples and 3-meter samples. 
As mentioned in Section~\ref{sec:env_setup}, we adjusted the height of the radar placement on the ceiling from 5 meters to 3 meters after initial data collection. 
We first evaluated the gesture recognition performance using a single radar.

Fig.~\ref{fig:single_radar_acc} shows the results for gesture recognition using data from a single radar with different models (From left to right: Pointnet++, Pantomime and Tesla). 
The X-axis indicates the radar index (see Fig.~\ref{fig:setup_s1}) while the Y-axis shows the accuracy. 
Indices 0-4 represent radars mounted on the ceiling, while other numbers indicate those on the ground. 
For the radars on the ceiling, across all models, radar~2 provides the highest accuracy for data from the `All' and `5-meter' groups, while the accuracy for the `3-meter' group remains relatively low, similar to results from the other radars mounted on the ceiling. 
A probable explanation for this phenomenon is that radar~2 is positioned directly above the subject, capturing more information while the subject performs various gestures. 
As the other ceiling-mounted radars are relatively distant from the subject, they may not capture a sufficient amount of information. 
For the radars placed on the ground, PointNet++ and Pantomime models show poor performance with data from the `3-meter' group, which might be due to the limited amount of subjects in the group. 
In contrast, the Tesla model exhibits superior performance with the same group, even outperforming other models for radars placed on the ceiling. 
To reflect different deployment constraints, we define three evaluation settings.
\begin{description}
    \item[Ceiling-only: ]  We use only the 5 ceiling-mounted radars. This provides a predominantly top-down view with reduced self-occlusion from the arms and torso. This setting is representative of ceiling-installed infrastructure in smart-room or smart-home scenarios.
    \item [Ground-only: ] We aggregate only the point clouds from the 8 ground radars.
    It also matches common gesture-sensing setups where devices are placed around the subject at approximately chest/arm height.
    \item [Ceiling+Ground: ] We merge all available radars (8 ground + 5 ceiling) after time synchronization and coordinate alignment. This produces denser, more complete 3D point clouds with improved coverage of subtle hand/arm motion as well as full upper-body kinematics. 
\end{description}

Tab.~\ref{tab:radar_fusion_acc} presents the results of three different evaluation settings. 
The Ground-only setting achieves the best performance in both the '3-meter' and '5-meter' groups, while combining data from ceiling- and ground-mounted radars yields slightly better results in the 'All' group setting. 
Although using ceiling-only data results in the lowest average accuracy, the performance remains acceptable, indicating the potential to replace ground-mounted radars with ceiling-mounted ones (a more user-friendly setting) in smart environments.

\begin{table}[]
\caption{Gesture recognition accuracy (\%) on different evaluation settings with Tesla model.}
\label{tab:radar_fusion_acc}
\begin{tabular}{lccc}
\toprule
\multicolumn{1}{l}{\textbf{Data Group}} & \multicolumn{1}{l}{\textbf{Ceiling-only}} & \multicolumn{1}{l}{\textbf{Ground-only}} & \multicolumn{1}{l}{\textbf{Ceiling+Ground}} \\ \hline
3-meter campaigns & 85.0 & 99.1 & 99.0 \\
5-meter campaigns & 78.5 & 99.4 & 98.6 \\
All campaigns & 83.8 & 99.0 & 99.2 \\
\bottomrule
\end{tabular}
\end{table}

\begin{figure}
\centering
\includegraphics[width=8cm]{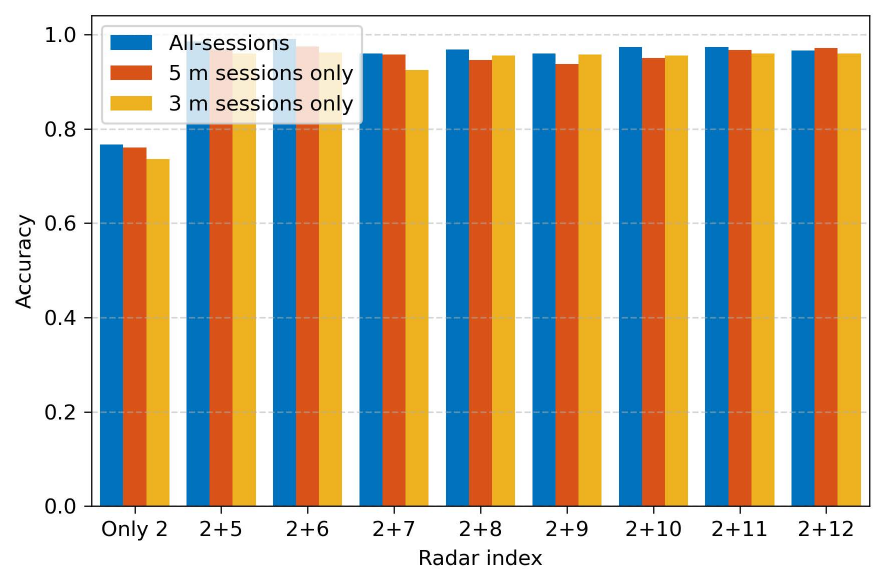}
\caption{Gesture recognition accuracy based on the combination of Radar 2 (from ceiling, directly above the subject) and radars on the ground (Indexes 5-12) using Tesla model.}
\label{fig:radar_2_plus_ground}
\end{figure}

To further evaluate the contribution of gesture recognition enabled by ceiling-mounted radars, we combined radar~2 with the radars positioned on the ground. 
The results indicate that integrating data from radar~2 enhances the performance of ground radars, particularly those not placed directly in front of the subject (see Fig.~\ref{fig:radar_2_plus_ground}). 
Obvious improvements have been made for all data groups when comparing the result with a single radar setting (see Fig.~\ref{fig:single_radar_tesla}) for radars with indices 6-12. 
Ceiling-mounted radars offer additional advantages such as fewer physical obstructions, broader coverage, and greater flexibility in placement, all of which contribute to more reliable recognition and a more holistic view of the interaction space.

A performance gap remains between ceiling-mounted and frontal configurations. 
Since radars measure radial velocity, a top-down perspective compresses the subject's vertical kinematic features, making small gestures and vertical body movements harder to classify. 
We argue that this algorithmic trade-off is justified by deployment practicality. 
Ceiling-mounted radars avoid the line-of-sight (LoS) blockages caused by furniture and environmental clutter, making them robust for continuous, unconstrained behavioral monitoring in real-world spaces.

\subsubsection{RFID-based Gesture Recognition}
This section presents the RFID-based gesture recognition reference model and benchmark models evaluated using the data from campaign~4. 
We utilize RSS and phase information reflected by the tags. 
The raw data from the reader is first processed. 
Specifically, this process includes data formatting, phase unwrapping, normalization, and managing missing data points. 
\paragraph{Preprocessing and Learning Framework: }
Two types of missing data are identified. 
The first occurs when a tag temporarily disconnects during gesture execution, while the second arises when a tag is not detected at all for a given execution. 
Adaptive interpolation is employed to handle temporary disconnections, whereas within-class imputation combined with proximity-based imputation is applied to tackle completely missing tag data.
Each recorded gesture consists of 8~dataframes, with each dataframe corresponding to a unique tag (identified by its EPC). 
Each dataframe contains 4~columns: timestamp, RSS, phase, and the tag's EPC.

For classification, we adopt the graph-based convolutional neural network proposed in~\cite{salami2022tesla}. 
A temporal K-nearest neighbors (K-NN) approach is used to construct a graph, where each EPC is represented as a node, and edges connect nodes based on the similarity of their RSS and phase values across successive timestamps. 
An edge is created between EPCs $i$ and $j$ if the prior reading of EPC $i$ at time $t-1$, characterized by the RSS and phase pair $(\alpha_{i}^{t-1}, \phi_{i}^{t-1})$, falls within the $K$ nearest neighborhood defined by the current reading of EPC $j$ at time $t$, $(\alpha_{j}^{t}, \phi_{j}^{t})$.
In this way, each EPC is linked to its most similar counterparts at the following timestamp, capturing temporal correlations and enabling effective feature propagation across the graph.

\paragraph{Benchmark models: }
A Random Forest Classifier model developed by~\cite{zhang2022real} is used as a benchmark, testing its performance with three feature configurations: features extracted from the phase signal alone (SP); a combination of wavelet coefficients and statistics of phase (SWP); and statistical features from both the phase and RSS signals (SPR).
Other established models are also utilized as benchmarks: 
Early Fusion~\cite{calatrava2023light}, Late Fusion~\cite{golipoor2024rfid}.
Metrics including accuracy, precision, recall, and F1-score derived from the benchmarks and the reference model are presented in Table~\ref{tab:RFIDBenchmarks}.

\begin{table}
\centering
\caption{Evaluation scores (\%) for RFID-based gesture recognition.}
\label{tab:RFIDBenchmarks}
\begin{tabular}{lcccc}
\toprule
\textbf{Method} & \textbf{Accuracy} & \textbf{Precision} & \textbf{Recall} & \textbf{F1-score} \\
\midrule
Random Forest with SP & 83.6 & 83.7 & 83.6 & 83.4 \\
 Random Forest with SWP & 86.3 & 86.5 & 86.2 & 86.1 \\
 Random Forest with SPR & 95.3 & 95.4 & 95.3 & 95.2 \\
Early Fusion & 83.6 & 84.3 & 83.6 & 83.3 \\
Late Fusion & 87.1 & 88.3 & 87.1 & 86.9 \\
The reference model & 98.1 & 98.2 & 98.1 & 98.1 \\
\bottomrule
\end{tabular}
\end{table}

\subsubsection{LoRa-based Gesture Recognition}\label{sec:lora_gesture}
We benchmark LoRa-only gesture recognition on the 21~gestures of campaign~1. 

\paragraph{Preprocessing:} Rather than operating on the raw baseband IQ samples, the baseline is trained on the extracted features released with the dataset: 
the low-pass-filtered amplitude (200\,Hz), its first-order differential (200\,Hz), and the sliding-window variance (20\,Hz). 
Each trial is represented by these three channels (amplitude, differential, and log-variance upsampled to 200\,Hz), z-normalized per trial. 
Every trial is linearly resampled to a fixed number of time steps (512), while the trial duration is retained as an explicit additional feature.

\paragraph{Benchmark models:} We evaluate a set of models of increasing capacity: 
(1)~a Random Forest (500 trees) on an 87-dimensional statistical descriptor (amplitude/differential statistics, 1\,Hz-band spectral energies up to 10\,Hz, log-variance statistics, and 16-segment temporal profiles), mirroring the statistical-feature baselines used for RFID; 
(2)~a 2D CNN over the log-magnitude STFT of the three channels (window 256, hop 16, keeping the 0--25\,Hz band that contains motion-induced amplitude modulation); 
(3)~a 1D CNN over the resampled time series; 
(4)~a CNN-BiLSTM (convolutional front-end, two bidirectional LSTM layers, attention pooling) and (5)~a fusion of the 2D and 1D CNN. 

We report two settings (see Table~\ref{tab:lora_baseline_C1}): \emph{within-subject} (stratified random trial split, 70/10/20 train/val/test; the classifier additionally receives a learned subject embedding) and \emph{cross-subject} (20\% of subjects are held out for testing, 10\% for validation, unseen-subject conditions; no subject embedding). 
Accuracy improves consistently with model capacity, from 38.9\% (Random Forest) to 58.9\% within-subject for the spectrogram CNN. Under unseen-subject conditions the best model reaches 30.3\%. 
An interesting finding is that the easy (single-hand) and complex (bimanual/circular) gesture subsets perform comparably (61.6\% vs.\ 56.6\% within-subject). Unlike for radar, bimanual motions perturb the single LoRa link strongly, compensating for their kinematic complexity. 

\begin{table*}[t]
\caption{LoRa-only baseline results on campaign~1 (21 gestures).}
\label{tab:lora_baseline_C1}
\centering\small
\setlength{\tabcolsep}{3.5pt}
\begin{tabular}{lcccccccc}
\toprule
\multirow{2}{*}{Model} & \multicolumn{2}{c}{Accuracy (\%)} & \multicolumn{2}{c}{Precision (\%)} & \multicolumn{2}{c}{Recall (\%)} & \multicolumn{2}{c}{F1-score (\%)} \\
\cmidrule(lr){2-3}\cmidrule(lr){4-5}\cmidrule(lr){6-7}\cmidrule(lr){8-9}
 & Within & Cross & Within & Cross & Within & Cross & Within & Cross \\
\midrule
Random Forest (stat.\ features) & 38.9 & 22.4 & 39.0 & 21.9 & 38.7 & 22.6 & 37.5 & 20.9 \\
CNN-2D (spectrogram) & \textbf{58.9} & \textbf{30.3} & \textbf{58.9} & \textbf{31.8} & \textbf{58.7} & \textbf{30.5} & \textbf{58.2} & \textbf{28.8} \\
CNN-1D (temporal) & 49.0 & 26.3 & 49.0 & 26.5 & 48.8 & 26.5 & 48.3 & 25.0 \\
CNN-BiLSTM & 50.8 & 29.2 & 51.5 & 30.3 & 50.7 & 29.6 & 50.3 & 28.2 \\
CNN-2D+1D fusion & 54.5 & 30.2 & 55.4 & 30.0 & 54.4 & 30.1 & 53.6 & 28.0 \\
\bottomrule
\end{tabular}
\end{table*}

\subsubsection{IMU-based Gesture Recognition}\label{sec:imu_gesture}
\paragraph{Sensors and preprocessing:} 
We benchmark the body-worn inertial modality using three Movesense IMUs. 
For each IMU sensor, we use the accelerometer and gyroscope axes (six channels per sensor). 
We use the three upper-body sensors (chest, left arm, right arm), giving an 18-channel signal. 
Each trial is linearly resampled to a fixed length of 128 time steps and z-normalized per channel, and trials with missing or empty IMU streams are excluded.

\paragraph{Benchmark models:} We evaluate 
(1)~a Random Forest (500 trees) on a statistical descriptor comprising per-channel time-domain statistics (moments, quartiles, RMS, zero-crossing rate, and first-difference statistics), spectral descriptors (centroid, entropy, and dominant frequency), and per-sensor acceleration and gyroscope magnitude statistics; 
(2)~a compact 1D convolutional encoder (three temporal convolution blocks with global average pooling); and (3)~a CNN-BiLSTM with attention pooling. The IMU streams recorded in the lab setting are affected by a technical logging issue, so all results use the two real-world settings. We report three protocols: within-subject, cross-subject (unseen users), and cross-environment (trained in the living room, tested in the industrial factory). 
All results are means over three seeds (Tab.~\ref{tab:imu_ladder_C1}).

The Random Forest on statistical features is the strongest IMU model in every protocol. 
It reaches 91.7\% within-subject accuracy, 57.2\% under unseen-subject conditions, and 64.1\% cross-environment (macro-F1 91.4, 55.4, and 63.8). 
The deep models reach 74.4\% (CNN) and 71.1\% (BiLSTM) within-subject and 46.9--49.1\% in the two transfer settings. 
This pattern echoes the affective benchmark: with hundreds of trials per class, compact statistical descriptors outperform deep sequence models. Within-subject IMU accuracy clearly exceeds the single-link LoRa baseline on the same task (91.7\% vs.\ 58.9\%, Section~\ref{sec:lora_gesture}), as expected from a body-mounted sensor. 
The pronounced drop towards unseen subjects indicates that the three upper-body IMUs capture gross arm motion but not the fine wrist articulation that distinguishes several of the 21 gestures, which is where radar's spatial resolution is complementary. 
Accuracy is higher in the living room than in the industrial factory (73.9\% vs.\ 68.3\% within-subject for the CNN), and the cross-environment results stay close to the cross-subject level. 
For gestures, the inter-subject gap therefore dominates the inter-environment gap.

\begin{table*}[t]
\caption{IMU-only baseline results on campaign~1. Within: stratified trial split; Cross(-subject): unseen users; Cross-env: trained in the living room and tested in the industrial factory. Best value per column in bold.}
\label{tab:imu_ladder_C1}
\centering\small
\setlength{\tabcolsep}{3.0pt}
\resizebox{\linewidth}{!}{%
\begin{tabular}{lcccccccccccc}
\toprule
\multirow{2}{*}{Model} & \multicolumn{3}{c}{Accuracy (\%)} & \multicolumn{3}{c}{Precision (\%)} & \multicolumn{3}{c}{Recall (\%)} & \multicolumn{3}{c}{F1-score (\%)} \\
\cmidrule(lr){2-4}\cmidrule(lr){5-7}\cmidrule(lr){8-10}\cmidrule(lr){11-13}
 & Within & Cross & Cross-env & Within & Cross & Cross-env & Within & Cross & Cross-env & Within & Cross & Cross-env \\
\midrule
Random Forest (stat.\ features) & \textbf{91.7} & \textbf{57.2} & \textbf{64.1} & \textbf{91.8} & \textbf{61.2} & \textbf{71.3} & \textbf{91.5} & \textbf{57.3} & \textbf{64.1} & \textbf{91.4} & \textbf{55.4} & \textbf{63.8} \\
CNN-1D (temporal) & 74.4 & 49.1 & 47.0 & 75.7 & 55.8 & 52.2 & 73.9 & 49.0 & 46.7 & 73.9 & 49.7 & 47.2 \\
CNN-BiLSTM & 71.1 & 48.7 & 46.9 & 70.7 & 52.2 & 48.7 & 70.5 & 48.2 & 46.7 & 70.1 & 48.9 & 46.5 \\
\bottomrule
\end{tabular}%
}
\end{table*}

\subsubsection{Sensor fusion for gesture recognition}\label{sec:gesture_fusion}
We evaluate the fusion of radar and IMU data for fine-grained gestures in the two real-world settings, where both modalities were recorded simultaneously. 
The radar branch is the Tesla backbone operating on point clouds of the central floor radar (16 frames, 16 points per frame). 
We use Tesla because it is designed for sparse point clouds and was the strongest single-radar model. 
The IMU branch is the 1D convolutional encoder of Section~\ref{sec:imu_gesture}. 
We report late fusion, which combines the two posterior distributions with a validation-selected weight, and feature fusion, which concatenates the two embeddings and trains a fusion head. 
The models are trained separately, and the three protocols follow Section~\ref{sec:imu_gesture}. 
Results are given in Table~\ref{tab:gesture_fusion}.

\begin{table}[t]
\caption{Radar and IMU for gesture recognition in the two real-world settings (Accuracy in \%; Tesla radar backbone on the central floor radar; protocols as in Section~4.1.4).}
\label{tab:gesture_fusion}
\centering\small
\begin{tabular}{lcccc}
\toprule
Protocol & Radar-only & IMU-only & radar and IMU (Feature fusion) & radar and IMU (Late fusion)   \\
\midrule
Within-subject & 73.0 & 74.5 & 86.2 & \textbf{89.4} \\
Unseen subjects & 66.6 & 49.0 &   68.6 & \textbf{75.9} \\
Cross-environment & 61.2 & 47.1 & 67.9 & \textbf{69.4}\\
\bottomrule
\end{tabular}
\end{table}

Late fusion improves over the best single modality in all three protocols, reaching 89.4\% within-subject accuracy (best single modality: IMU, 74.5\%), 75.9\% under unseen-subject conditions (radar-only 66.6\%), and 69.4\% cross-environment (radar-only 61.2\%). The complementarity is asymmetric across protocols. Within subjects, the IMU branch is stronger and radar adds spatial detail; under subject and environment shift, the radar branch is more stable and the IMU contributes complementary motion dynamics.

\subsection{Macro activity recognition}
We establish baseline models for activity recognition using radar data collected in all environments and LoRa data collected in lab environment. 

Models are first trained on data from the lab setting. To evaluate the generalization capability of the multi-radar system in realistic scenarios, we test the trained models on (1) unseen subjects within the lab environment, and (2) two unseen environments: a living room and an industrial factory.

The evaluation metrics include Accuracy, Precision, Recall, F1-score, and normalized confusion matrices.

\paragraph{Preprocessing pipeline:} 
The preprocessing pipeline follows a unified protocol combining steps introduced in Section~\ref{sec:SensorSelection} and Section~\ref{sec:RadarBased}. Specifically,
(1) Coordinate transformation and temporal alignment are applied to synchronize multi-radar data into fixed-length sequences (16 frames);
(2) Clutter suppression is performed using DBSCAN to remove noise and retain the primary subject;
(3) Point set normalization is applied by resampling each frame to a fixed number of points (16 points per frame).

\paragraph{Benchmark models:}
We adopt the same baseline models as in Section~\ref{sec:RadarBased}, including PointNet++, Pantomime, and Tesla. 
These models represent three complementary design paradigms: hierarchical spatial feature learning (\textbf{PointNet++}), spatio-temporal modeling with recurrent structures (\textbf{Pantomime}), and lightweight radar-specific temporal modeling (\textbf{Tesla}). 
All experiments are implemented in PyTorch and conducted on an NVIDIA A40 GPU. 
We use the Adam optimizer with early stopping (patience = 20 epochs).

\subsubsection{Multi-Radar Training in Lab}
We evaluate the performance of multi-radar fusion by training Tesla, PointNet++, and Pantomime using data from all 13 radars in the laboratory environment. subjects 1–22 are used for training, while subjects 23–27 are reserved for testing to ensure a fair evaluation under unseen-subject conditions.

The quantitative results are summarized in Table~\ref{tab:activity_multi_radar}. 
Specifically, PointNet++ achieves the best overall performance, with a test accuracy of 95.8\%, precision of 95.9\%, recall of 95.8\%, and F1-score of 95.8\%. Tesla follows closely with all metrics at about 94.5\%, while Pantomime shows comparatively lower performance, with a test accuracy and an F1-score of 91.0\%. Therefore, Pantomime is not used in subsequent activity recognition evaluation.

These results indicate that PointNet++ is particularly effective in leveraging the richer spatial information provided by multi-radar fusion. 
Its hierarchical feature learning mechanism allows it to better capture fine-grained geometric structures from the aggregated point clouds.

\subsubsection{Single-Radar Evaluation in Lab} 
We further evaluate Tesla and PointNet++ under the single-radar setting in the laboratory environment. 
The learning rate is uniformly set to 0.001, and the batch size is fixed at 16 for all experiments. 
Users 1–22 are used as the training set, while subjects 23–27 are reserved for testing.

\begin{figure}
\centering
\includegraphics[width=13cm]{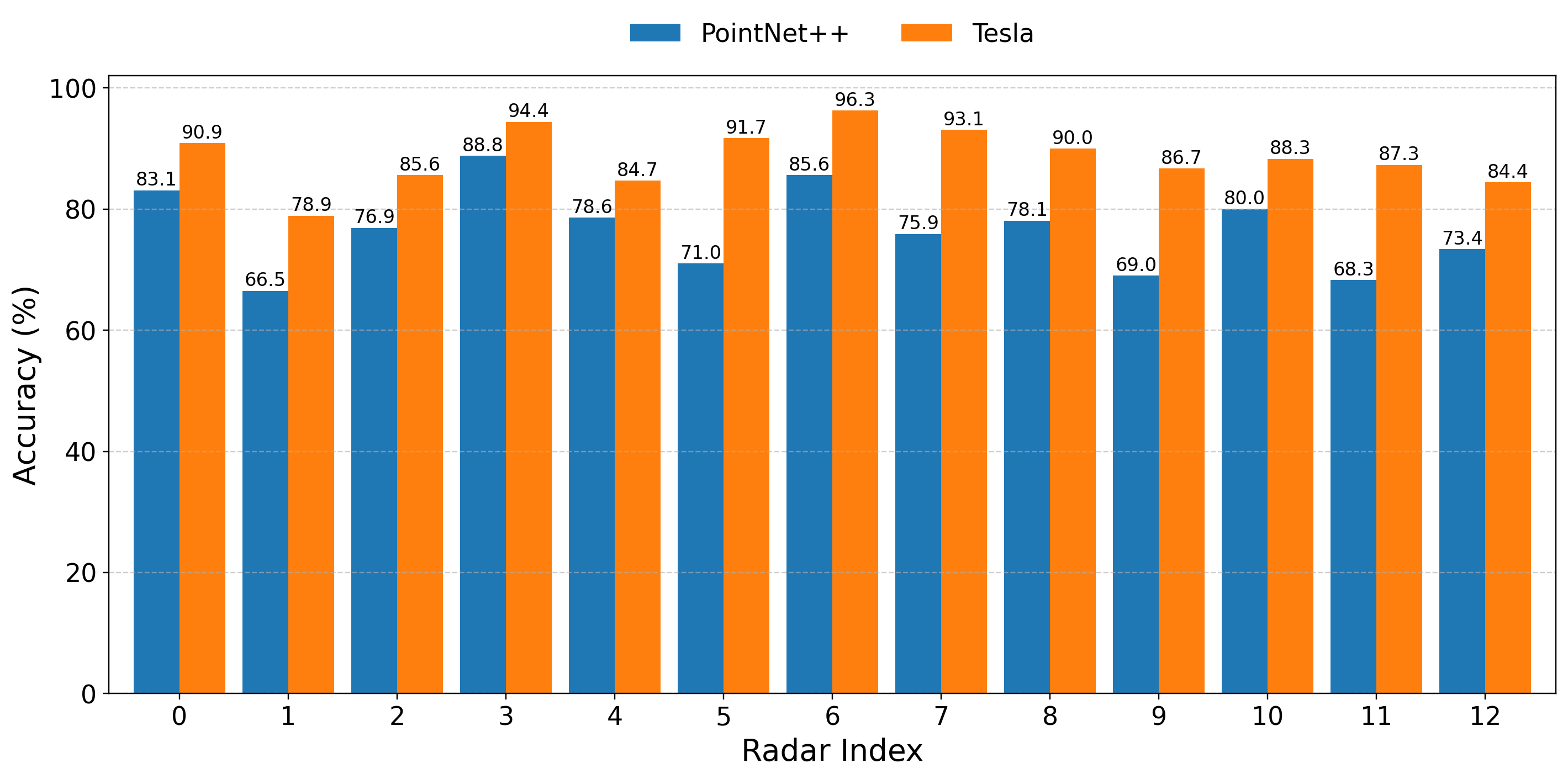}
\caption{Activity recognition accuracy on 2 baseline models. The blue column represents Pointnet++, while the red column represents Tesla. 
The X-axis indicates the Radar Index and the Y-axis shows the test accuracy.}
\label{fig:test_acc_compare_single_radar}
\end{figure}

After training, the test accuracy comparison of the two models is shown in Fig.~\ref{fig:test_acc_compare_single_radar}. 
The x-axis represents the radar index (see Fig.~\ref{fig:setup_s2s3} and Fig.~\ref{fig:c4_RFIDSetup}), and the y-axis denotes the classification accuracy. 
Blue bars correspond to PointNet++, while orange bars represent Tesla. Radar indices 0–4 indicate ceiling-mounted radars, while the remaining indices correspond to ground-mounted radars.

\begin{table}[t]
\centering
\caption{Activity recognition performance in the lab environment using 13 radars.}
\label{tab:activity_multi_radar}
\begin{tabular}{lcccc}
\toprule
Model & Accuracy (\%) & Precision (\%) & Recall (\%) & F1-score (\%) \\
\midrule
Tesla      & 94.5 & 94.7 & 94.5 & 94.5 \\
PointNet++ & 95.8 & 95.9 & 95.8 & 95.8 \\
Pantomime  & 91.0 & 91.1 & 91.0 & 91.0 \\
\bottomrule
\end{tabular}
\end{table}

It can be observed that Tesla consistently outperforms PointNet++ across all radar positions, demonstrating its strong suitability for this task.

For PointNet++, among all radars, Radar~3 (located at the rear-left position on the ceiling) achieves the best performance with an accuracy of 88.8\%, followed by Radar~6 (front-left position on the ground) with an accuracy of 85.6\%. 
In contrast, its performance on Radars~1, 9, and~10 is relatively poor, with accuracies below 70.0\%.

Tesla shows superior overall performance compared to PointNet++. 
The highest accuracy is achieved at Radar~6 (96.3\%), followed by Radar~3 (94.4\%), while the lowest performance is observed at Radar~1 (78.9\%). 

This can be explained by the fact that several activities (e.g., playing basketball, badminton, and lying down) are often oriented toward these radar positions. 
In addition, their viewing angles provide broader and more effective coverage of the recording area. 
Interestingly, despite being positioned directly overhead, Radar~2 does not achieve the best performance. 
Similarly, although Radars~5 and~9 are physically closer to the participant, they do not yield optimal results. 
This suggests that viewing angle and motion coverage are more critical than proximity alone. 
Placing the radar in the optimal position might achieve results as good as those of a multi-radar system. 

The confusion matrices for Radar~3 and Radar~6 using Tesla are shown in Fig.~\ref{fig:cm_activity} in the Appendix. 
For Radar 3, notable confusion occurs between Activities 1 and 2, as well as among Activities~7, 8, and~9. 
For Radar 6, minor confusion is observed among Activities~4, 8, and~10.

\subsubsection{Cross-Environment Generalization}\label{sec:cross_env_radar}
\begin{figure}
  \centering
  \begin{subfigure}[t]{0.45\textwidth}
    \centering
    \includegraphics[width=\textwidth]{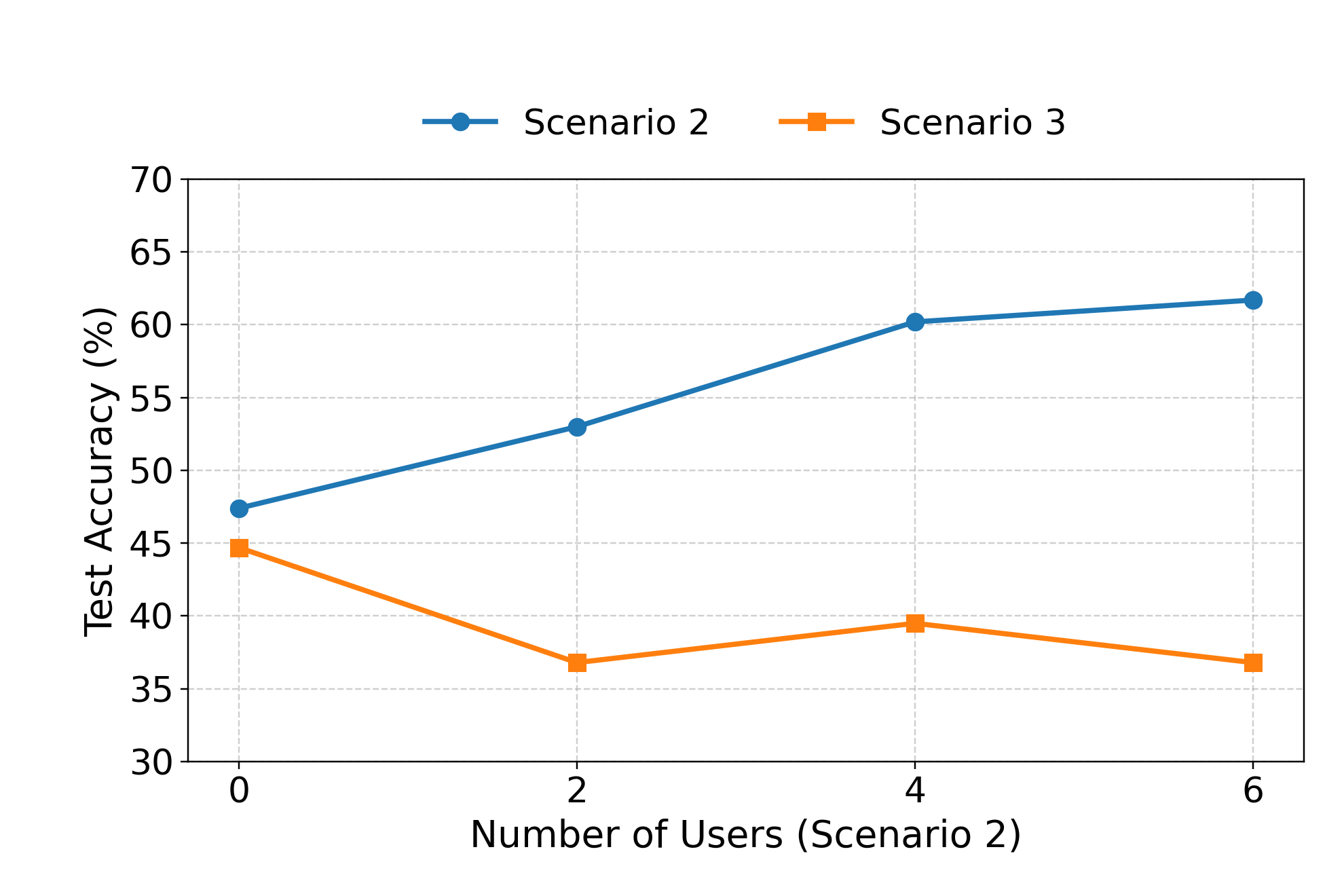} 
    \caption{Test accuracy using Pointnet++ baseline.}
  \end{subfigure}
  \hfill
  \begin{subfigure}[t]{0.45\textwidth}
    \centering
    \includegraphics[width=\textwidth]{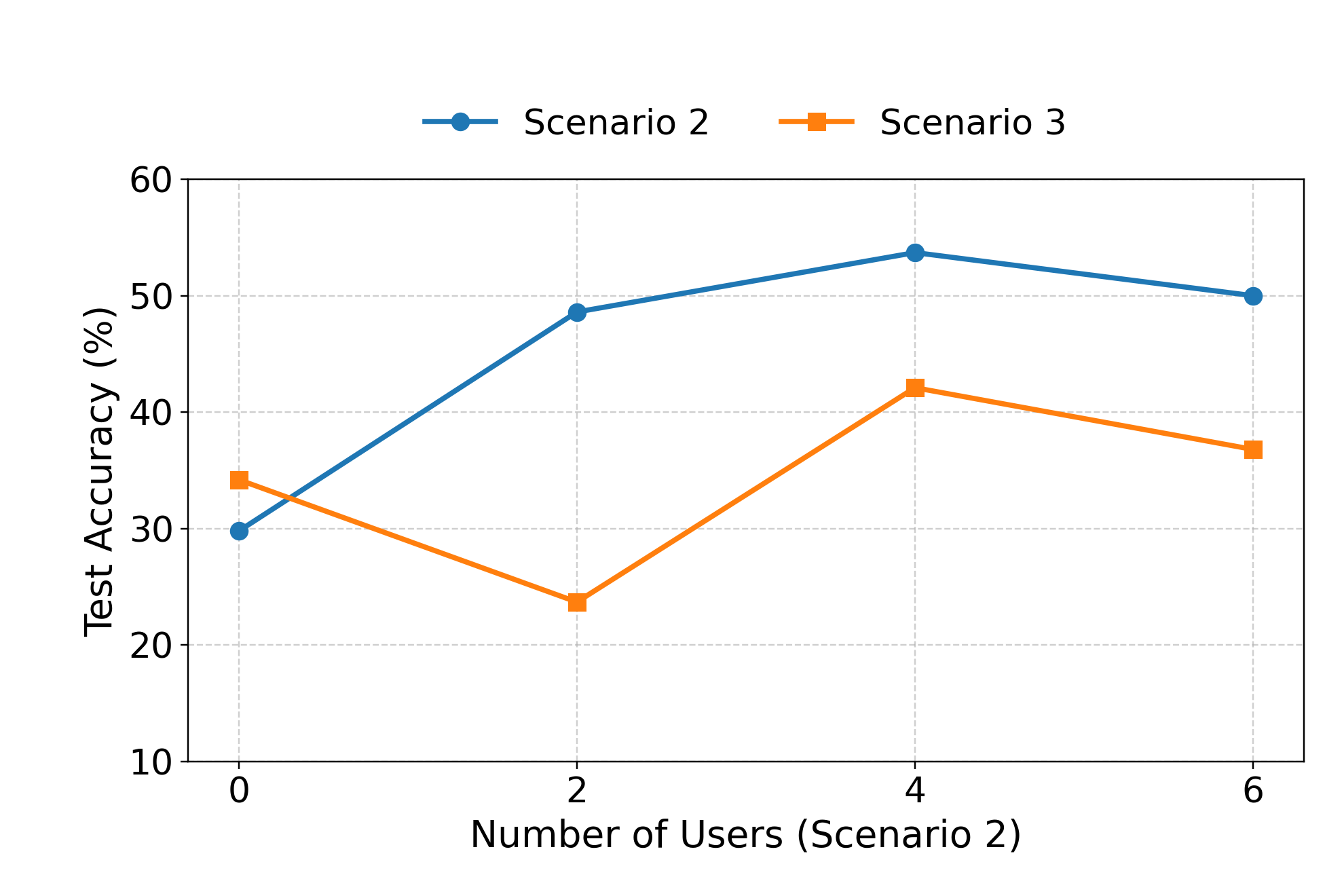} 
    \caption{Test accuracy using Tesla baseline.}
  \end{subfigure}
  \caption{Activity recognition performance on~2 new scenarios (Scenario~2: living room, Scenario~3: industrial factory) based on baseline models using~13 radars. 
  The X-axis represents the number of subjects whose data in scenario~2 are added to the training dataset.}
  \label{fig:new_scenario_activity}
\end{figure}

To evaluate cross-domain generalization, we directly apply the models trained in the lab environment to two unseen environments: the living room (Scenario~2) and the industrial factory (Scenario~\ref{MocapDataSet}), both using 13-radar configurations. The performance drops significantly to 47.4\% and 44.7\%, respectively, based on Pointnet++. The accuracies are also low using the Tesla model: 29.8\% and 34.2\% in two environments, respectively. The results are illustrated in Fig.~\ref{fig:new_scenario_activity}, where the points at $x=0$ represent the accuracies under this setting.

This degradation can be attributed to both environmental and behavioral differences. 
Specifically, the walking/running trajectories differ significantly across environments: 
in the lab setting, the trajectory typically follows an ``S''-shaped path, whereas in other environments it tends to be more circular. 
This variation arises from differences in obstacles and layout constraints. 
The lab environment is relatively open, with minimal clutter around the recording area. 
In contrast, the living room and factory environments contain more obstacles, including tripods used to mount ceiling radars (at a height of $\sim$3\,m), tables, chairs, walls, and various experimental equipment.

Another example is the ``lying'' activity, which corresponds to lying on a table in the lab, on a sofa in the living room, and on the floor in the factory. 
In addition, the ladders used in the living room and factory differ from those in the lab; they are less stable and have narrower steps, which introduces additional motion variations and deformation in the recorded activities. 
There are also slight differences in the direction of shooting a basketball and kicking a football.

To mitigate this issue, we incrementally incorporate samples from the living room environment into the training set. Specifically, we evaluate the performance when including data from~2, 4, and~6 subjects (out of~8 total subjects in the living room), as illustrated in Fig.~\ref{fig:new_scenario_activity}. For the living room (scenario~2), as more samples from the new scenario are introduced, performance gradually improves for both Pointnet++ and Tesla. However, the overall gain remains relatively limited, and the performance still falls short of the desired level. 

The reasons behind this result are similar to those discussed earlier. The performance gap may stem from differences across environments, variations among users, and the limited amount of data available in the new scenarios. In particular, Scenario~3 contains only two users, which further restricts the diversity of training samples.
Even after incorporating data from the new environments, the overall amount of such data remains relatively small. 
As a result, the model is still biased toward patterns learned from the lab environment, making it difficult to fully adapt to unseen scenarios. This imbalance between the lab data and the newly added data limits the effectiveness of cross-environment generalization.
These results suggest that simply adding a small amount of data from a new scenario is insufficient for robust cross-environment generalization. More advanced domain adaptation or multi-environment training strategies may be required to effectively bridge this gap.

\subsubsection{LoRa-based Activity Recognition}\label{sec:lora_activity}
We apply the identical benchmark models, preprocessing, and training protocol as in Section~\ref{sec:lora_gesture} to the ten macro activities of campaign~2.
Since LoRa was deployed only in the laboratory, this evaluation complements the radar cross-environment study with a single-link, lab-setting baseline. Results for both the within-subject and the unseen-subject setting are presented in Tab.~\ref{tab:lora_baseline_C2}.

The fused 2D+1D CNN reaches 75.8\% within-subject accuracy and 66.7\% under unseen-subject conditions (macro-F1 75.7 and 65.9). 
In contrast to fine-grained gestures, the drop towards unseen subjects is small. LoRa amplitude dynamics of whole-body activities generalize across users. Locomotion is recognized almost perfectly (walking 100\%, running 95\%, stair ascent/descent 82--85\%), and errors concentrate among the in-place sports actions (basketball passing, badminton, floorball), whose arm-dominated motions produce similar single-link signatures. This confirms the intended role of LoRa in RF-Behavior as a modality for macroscopic movement, approaching the utility of a dedicated sensing modality for coarse motion while remaining far below the multi-radar baselines for fine-grained discrimination.

\begin{table*}[t]
\caption{LoRa-only baseline results on campaign~2 (10 activities).}
\label{tab:lora_baseline_C2}
\centering\small
\setlength{\tabcolsep}{3.5pt}
\begin{tabular}{lcccccccc}
\toprule
\multirow{2}{*}{Model} & \multicolumn{2}{c}{Accuracy (\%)} & \multicolumn{2}{c}{Precision (\%)} & \multicolumn{2}{c}{Recall (\%)} & \multicolumn{2}{c}{F1-score (\%)} \\
\cmidrule(lr){2-3}\cmidrule(lr){4-5}\cmidrule(lr){6-7}\cmidrule(lr){8-9}
 & Within & Cross & Within & Cross & Within & Cross & Within & Cross \\
\midrule
Random Forest (stat.\ features) & 68.0 & 66.0 & 67.8 & 66.1 & 68.0 & 66.0 & 67.6 & 65.1 \\
CNN-2D (spectrogram) & 72.8 & 62.9 & 74.0 & 64.8 & 72.8 & 62.9 & 72.5 & 61.3 \\
CNN-1D (temporal) & 74.2 & 63.8 & 74.4 & 65.1 & 74.2 & 63.8 & 73.9 & 63.3 \\
CNN-BiLSTM & 72.8 & 64.4 & 74.4 & 66.5 & 72.8 & 64.4 & 72.8 & 64.2 \\
CNN-2D+1D fusion & \textbf{75.8} & \textbf{66.7} & \textbf{77.1} & \textbf{69.6} & \textbf{75.8} & \textbf{66.7} & \textbf{75.7} & \textbf{65.9} \\
\bottomrule
\end{tabular}
\end{table*}

\subsubsection{IMU-based Activity Recognition.}\label{sec:imu_activity}
We apply the identical IMU models and protocols of Section~\ref{sec:imu_gesture} to the ten activities of campaign~2, again using the two real-world settings (Tab.~\ref{tab:imu_ladder_C2}). The body-worn modality is strong on this task. The Random Forest reaches 95.0\% within-subject accuracy, 84.8\% under unseen-subject conditions, and 89.4\% cross-environment (macro-F1 95.0, 84.2, and 88.6). The deep models follow the same ordering (CNN 91.0/75.7/65.3\%, BiLSTM 85.4/71.4/65.1\%). Two observations stand out. First, unlike for gestures, IMU performance transfers well across users, since whole-body activities are expressed similarly across participants at the acceleration level. Second, the statistical descriptor is markedly more environment-robust than the learned representations (89.4\% vs.\ 65.3\% cross-environment). The 1D convolutional encoder evaluated here is also the IMU branch used in the fusion study (Section~\ref{sec:activity_fusion}), so its single-modality results serve as the direct reference for the fusion gains reported there.
Overall, these results provide the wearable reference point against which the ambient RF modalities can be judged.

\begin{table*}[t]
\caption{IMU-only baseline results on campaign~2. Within: stratified trial split; Cross(-subject): unseen users; Cross-env: trained in the living room and tested in the industrial factory. Best value per column in bold.}
\label{tab:imu_ladder_C2}
\centering\small
\setlength{\tabcolsep}{3.0pt}
\resizebox{\linewidth}{!}{%
\begin{tabular}{lcccccccccccc}
\toprule
\multirow{2}{*}{Model} & \multicolumn{3}{c}{Accuracy (\%)} & \multicolumn{3}{c}{Precision (\%)} & \multicolumn{3}{c}{Recall (\%)} & \multicolumn{3}{c}{F1-score (\%)} \\
\cmidrule(lr){2-4}\cmidrule(lr){5-7}\cmidrule(lr){8-10}\cmidrule(lr){11-13}
 & Within & Cross & Cross-env & Within & Cross & Cross-env & Within & Cross & Cross-env & Within & Cross & Cross-env \\
\midrule
Random Forest (stat.\ features) & \textbf{95.0} & \textbf{84.8} & \textbf{89.4} & \textbf{95.3} & \textbf{87.2} & \textbf{89.5} & \textbf{95.0} & \textbf{84.7} & \textbf{89.1} & \textbf{95.0} & \textbf{84.2} & \textbf{88.6} \\
CNN-1D (temporal) & 91.0 & 75.7 & 65.3 & 91.3 & 77.4 & 69.7 & 91.0 & 75.6 & 64.8 & 90.8 & 74.5 & 64.9 \\
CNN-BiLSTM & 85.4 & 71.4 & 65.1 & 86.6 & 73.3 & 68.8 & 85.4 & 71.3 & 64.7 & 84.9 & 70.6 & 65.2 \\
\bottomrule
\end{tabular}%
}
\end{table*}

\subsubsection{Sensor Fusion Cross-Environment Evaluation}\label{sec:activity_fusion}
To evaluate whether multimodal fusion improves cross-environment activity recognition, we conduct radar and IMU fusion experiments under two training protocols. 
In the first protocol, the radar backbone is pre-trained using only the lab-setting radar data for~100 epochs, and then fine-tuned with paired radar and IMU data from the living room (scenario~2) for 30 epochs. In the second protocol, the model is re-trained from scratch for 100 epochs using the lab-setting radar data together with a small number of paired radar and IMU samples from the living room. For both protocols, we vary the number of living room subjects used for adaptation and evaluate the model on the remaining living room subjects and on all industrial factory (Scenario~3) users.

We evaluate three fusion strategies. 
Late fusion combines the posterior probabilities of the radar-only and IMU-only classifiers, where the fusion weight is selected on a held-out validation subset from the adaptation data. Feature fusion concatenates the radar embedding and the IMU embedding before classification. Cross-attention fusion treats radar frame-level features and IMU temporal segments as two input streams and learns cross-modal attention between them.
These three variants allow us to compare a simple decision-level fusion baseline with two trainable representation-level fusion mechanisms.

\begin{figure*}[t]
    \centering
    \includegraphics[width=0.92\linewidth]{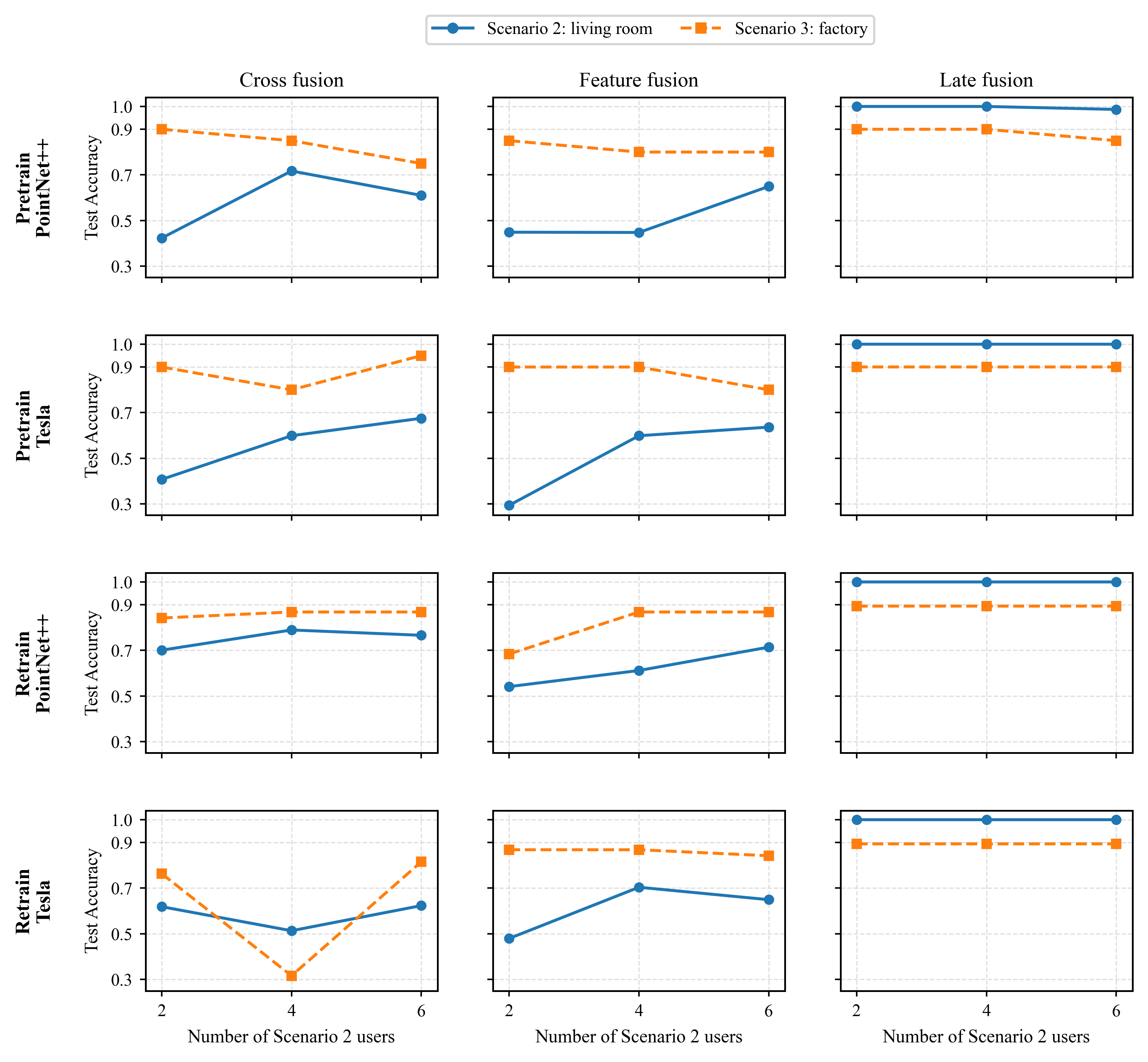}
    \caption{radar and IMU fusion accuracy across cross-environment evaluation settings. The x-axis denotes the number of living-room subjects used for adaptation. Rows compare the fine-tuning and re-training protocols with PointNet++ and Tesla backbones, while columns compare cross-attention, feature-level, and late fusion. Blue and orange curves report test accuracy on living room and industrial factory environments, respectively.}
    \label{fig:sensor_fusion_cross_env}
\end{figure*}

Figure~\ref{fig:sensor_fusion_cross_env} shows that radar and IMU fusion improves cross-environment activity recognition over the radar-only baselines in Figure~\ref{fig:new_scenario_activity}. Late fusion achieves the strongest results in most settings, reaching nearly perfect accuracy on the living room because the IMU branch is highly reliable in the living-room environment. The trainable fusion methods also provide clear gains: under pre-training/fine-tuning, PointNet++ cross-attention reaches 71.7\% on the living room with four adaptation users, while Tesla cross-attention reaches 95.0\% on the industrial factory (Scenario~3) with six adaptation users. Under from-scratch re-training, PointNet++ cross-attention reaches 78.9\% on the living room and 86.8\% on the industrial factory, and PointNet++ feature fusion also reaches 86.8\% on the industrial factory. The corresponding radar-only and radar and IMU fusion accuracies across both environments, backbone models, and fusion strategies are summarized in Tab.~\ref{tab:sensor_fusion_cross_env} in the Appendix. These results show that IMU signals provide complementary motion cues that improve robustness when radar-based activity recognition is transferred from the lab setting to new deployment environments.

The fusion strategies evaluated here (late, feature-level, and cross-attention fusion) are intended as transparent and reproducible reference points rather than as the strongest possible fusion models. Recent selective state-space and flow-based fusion approaches, such as Coupled Mamba~\cite{li2024coupled} and the normalizing-flow-based MANGO~\cite{truong2026mango}, model cross-modal dependencies with greater capacity, and the synchronized multimodal streams of RF-Behavior provide a suitable benchmark for evaluating such methods.

\subsection{Affective behavior recognition}
We evaluate the performance of affective behavior recognition using data from the radar, the IMU, and the statistical feature-based fusion of both modalities. During preliminary analysis, we observed that training a generalized, cross-subject model for this specific task yields unsatisfactory results. Unlike the gesture and activity protocols, which imposed structural constraints on subjects, the affective data collection imposed no limitations on motion. Consequently, the resulting behavioral expressions are highly personalized, varying significantly based on individual personality traits, the specific narrative stimuli, and the intensity of the induced state.

To address this high inter-subject variance, we transition from a cross-subject evaluation to a within-subject evaluation framework. The core advantage of this approach is its ability to demonstrate the representational robustness of the extracted features. 
Specifically, we employ a few-shot learning strategy to develop subject-specific recognition models. For a given subject, a base model is first trained on a limited subset of randomly selected affective states (e.g., focus and relaxation). This pre-trained model is then adapted to classify previously unseen affective states using $k$ reference samples, where $k$ scales from 1 to 10. To ensure computational efficiency during this fine-tuning phase, only the final classification layer is updated.

This within-subject adaptation strategy aligns with realistic deployment scenarios, such as private residences or single-occupancy eldercare monitoring, where the primary subject remains constant. In this within-subject recognition framework, the model can easily be personalized to a specific individual's baseline behaviors and rapidly adapt to their unique expressions of novel affective states without requiring extensive recalibration data.

\paragraph{Preprocessing and Statistical Feature Selection: }
Feature construction uses the same timestamp-based procedure in the lab, industry, and living-room environments whenever the required IMU and radar streams are available. In the configuration analyzed here, the wearable component consists of three IMUs, and the environment contains thirteen radars. The descriptor subset uses two radar channels: the central floor radar and the central ceiling radar. Windows are defined on the chest IMU timeline (12\,s, 75\% overlap). Each IMU sensor is sliced on the same wall-clock interval, and each of the two radar channels is aligned separately to the chest IMU magnitude signal by cross-correlation.

Windows pass quality control only if IMU coverage is at least 85\% with no internal gap larger than 0.25\,s, and if the lag-corrected radar interval retains enough valid frames for stable statistics. The quality-control counts are computed on the lab affective-behavior subset used in the current benchmark analysis. After quality control, the reference table contains 4,976 aligned windows from 23 subjects and 133 sessions. The stricter subset used in later descriptor-based experiments contains 4,130 windows (83.0\%) from 22 subjects. Counts remain adequate across all six affective conditions, although the sitting classes are imbalanced (e.g., distraction: 1,362 windows; relaxation: 738 windows); downstream training should account for this skew (per-sensor pass rates and window counts in Appendix~\ref{sec:apx_affective_qc}).

To avoid inflating the effective sample size through overlapping windows and repeated measurements from the same subjects, feature screening is performed on subject-level medians with repeated-measures tests: the Friedman test across the four sitting emotions and paired Wilcoxon signed-rank tests across the two standing emotions. The screening confirms discriminative structure in both modalities. Permutation entropy (PE) and jerk RMS dominate among the IMU descriptors, while radar contributes complementary centroid-variability and motion-spread features. The final subset used in the later affective-behavior benchmark contains nine z-scored features, normalized within subject and posture, listed in Table~\ref{tab:final-features}; windows of 12\,s with 75\% overlap are used as a pragmatic default. The full quality-control statistics, per-feature screening scores and rankings, exclusion criteria, and the window-length sensitivity analysis are reported in Appendix~\ref{sec:apx_affective_qc}.

\begin{table}[t]
\centering
\scriptsize
\caption{Compact statistical descriptors used in the later affective-behavior benchmark, either as standalone inputs or fused with raw signals.}
\label{tab:final-features}
\begin{tabular}{@{}>{\raggedright\arraybackslash}p{0.24\columnwidth}
                >{\raggedright\arraybackslash}p{0.16\columnwidth}
                >{\raggedright\arraybackslash}p{0.27\columnwidth}
                >{\raggedright\arraybackslash}p{0.21\columnwidth}@{}}
\toprule
Descriptor & Modality & Repeated-measures support & Interpretation \\
\midrule
Chest permutation entropy & Chest IMU & sit Friedman: $\chi^2=12.44$, $p=6.0\times10^{-3}$ & structural complexity \\
Chest jerk RMS & Chest IMU & stand Wilcoxon: $W=13.0$, $p=6.7\times10^{-4}$ & movement intensity \\
Left-arm permutation entropy & Left-arm IMU & sit Friedman: $\chi^2=29.62$, $p=2.0\times10^{-6}$ & local structure \\
Left-arm jerk RMS & Left-arm IMU & stand Wilcoxon: $W=13.0$, $p=6.7\times10^{-4}$ & local intensity \\
Right-arm jerk RMS & Right-arm IMU & stand Wilcoxon: $W=7.0$, $p=1.5\times10^{-4}$ & complementary arm intensity \\
Radar 2 centroid $x$ variability & Radar 2 & sit Friedman: $\chi^2=12.05$, $p=7.2\times10^{-3}$ & spatial variability \\
Radar 2 centroid $y$ variability & Radar 2 & stand Wilcoxon: $W=22.0$, $p=4.0\times10^{-3}$ & spatial variability \\
Radar 5 mean $y$-spread & Radar 5 & sit Friedman: $\chi^2=14.24$, $p=2.6\times10^{-3}$ & motion spread \\
Radar 2 mean $y$-spread & Radar 2 & stand Wilcoxon: $W=20.0$, $p=2.8\times10^{-3}$ & motion spread \\
\bottomrule
\end{tabular}
\end{table}

\paragraph{Benchmark models: }\label{sec:emotion_fusion} 
To evaluate the quality of the selected features, we utilize several basic models: Convolutional Neural Network (CNN), Long Short-Term Memory (LSTM), and CNN-LSTM models. The CNN evaluates the robustness of local, spatial, and cross-channel representations within the sensor data. The LSTM evaluates the temporal dynamics, testing how well the features capture sequential dependencies over time. Finally, the hybrid CNN-LSTM model is deployed to assess feature quality when both spatial feature extraction and long-term temporal modeling are integrated. We evaluate each model using three input options: raw data, selected features, and a combination of both. 

Tab.~\ref{tab:affective_baseline} presents the within-environment baseline accuracy for affective behavior recognition across three environments, three modalities, and nine input configurations. Overall, CNN consistently outperforms LSTM and CNN-LSTM across all settings, suggesting that local temporal pattern extraction is more effective than sequential modelling for the window lengths used. The Combined input mode (raw signals + extracted features) achieves the highest accuracy in most columns, confirming that handcrafted features and raw signals contain complementary information.
Radar + IMU fusion yields the best results in the Digital Studio and Living Room environments, achieving peak accuracies of 73.7\% and 80.6\% respectively, demonstrating the complementary nature of body movement (IMU) and micro-Doppler signatures (radar) for affective behavior recognition. In the Industrial Workplace, IMU-only achieves competitive performance (82.2\%).

\begin{figure}
\centering
\includegraphics[width=\textwidth]{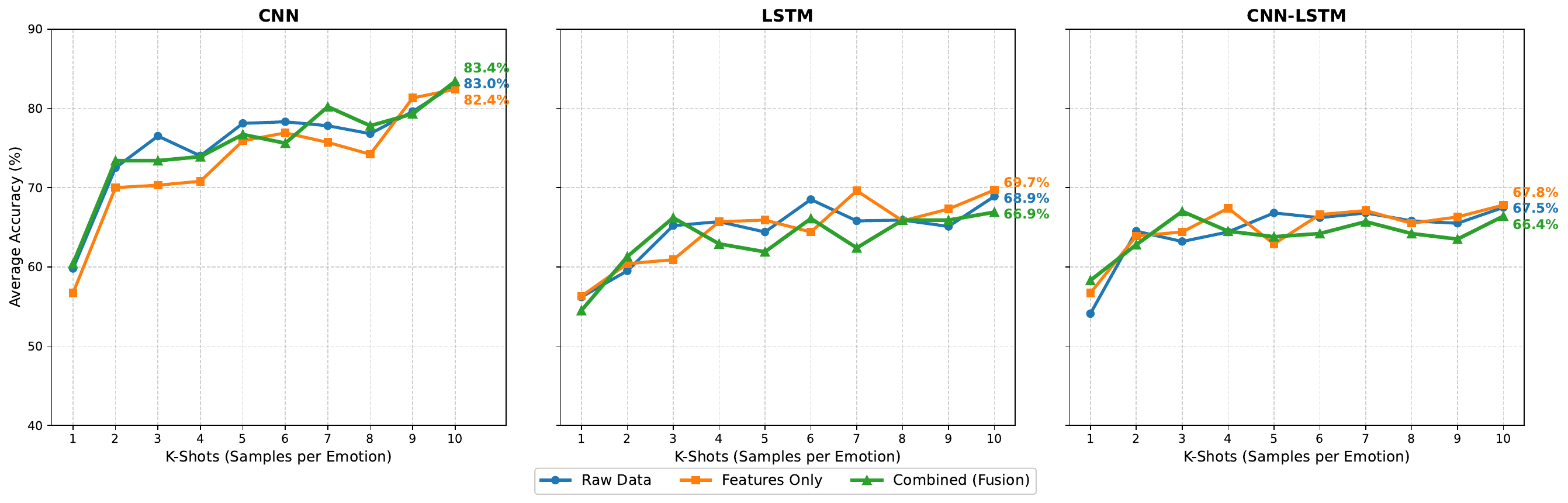}
\caption{Affective behavior recognition: Few-Shot Adaptation across Models and Input options for IMU data.}
\label{fig:imu_emotion_result}
\end{figure}

The reported standard deviations reflect variability across two sources: inter-subject differences and the k-shot range (k\,=\,1 to 10). Since accuracy at low k values (e.g., k,=,1, 2) is naturally lower and less stable than at higher k, averaging across the full k range introduces additional spread into the reported std. Fig.~\ref{fig:imu_emotion_result} shows an example of the average affective behavior recognition accuracy across all subjects using data from IMU sensors. Across all models, learning curves show that accuracy improves most dramatically between one and two reference samples.

Some results are missing, especially in the Industrial Workplace environment. 
This is partly due to the very limited number of subjects (N\,=\,2) in that environment, and partly due to empty point-cloud frames arising from the radar preprocessing pipeline, where no targets were detected during certain recording windows.

\begin{table*}[t]
\centering
\caption{Cross-Environment Baseline Accuracy (\%) for Affective Behavior Recognition.
Evaluation compares Models (CNN, LSTM, CNN-LSTM) and Input Types (Raw, Features, Combined)
across three modalities. Values are mean$\pm$std across subjects.
}
\label{tab:affective_baseline}
\renewcommand{\arraystretch}{1.05}
\resizebox{\linewidth}{!}{%
\begin{tabular}{ll ccc ccc ccc}
\toprule
& & \multicolumn{3}{c}{\textbf{Digital Studio}}
  & \multicolumn{3}{c}{\textbf{Living Room}}
  & \multicolumn{3}{c}{\textbf{Industrial Workplace}} \\
\cmidrule(lr){3-5}\cmidrule(lr){6-8}\cmidrule(lr){9-11}
\textbf{Modality} & \textbf{Model}
  & Raw & Feat. & Comb.
  & Raw & Feat. & Comb.
  & Raw & Feat. & Comb. \\
\midrule
\multirow{3}{*}{Radar Only} & CNN & 66.9$\pm$12.0 & 55.9$\pm$9.8 & 70.6$\pm$9.8 & 63.0$\pm$8.1 & 61.0$\pm$11.7 & 75.8$\pm$12.9 & \textbf{\underline{82.2$\pm$17.9}} & {--} & {--} \\
 & LSTM & 56.6$\pm$11.6 & 48.8$\pm$6.4 & 62.3$\pm$9.7 & 52.1$\pm$13.8 & 56.0$\pm$13.5 & 66.6$\pm$5.1 & 57.1$\pm$14.2 & {--} & {--} \\
 & CNN-LSTM & 58.4$\pm$9.7 & 48.2$\pm$7.4 & 58.7$\pm$8.8 & 53.8$\pm$5.4 & 51.4$\pm$9.8 & 65.9$\pm$11.2 & 56.2$\pm$8.9 & {--} & {--} \\
\midrule
\multirow{3}{*}{IMU Only} & CNN & 65.0$\pm$10.1 & 62.6$\pm$8.4 & 73.0$\pm$8.5 & 63.6$\pm$7.2 & 75.2$\pm$7.0 & 72.8$\pm$5.0 & 72.7$\pm$11.5 & \textbf{\underline{70.7$\pm$14.7}} & \textbf{\underline{82.2$\pm$15.0}} \\
 & LSTM & 45.7$\pm$9.2 & 51.4$\pm$7.0 & 59.7$\pm$7.7 & 51.1$\pm$6.3 & 55.2$\pm$4.2 & 54.8$\pm$10.2 & 49.3$\pm$3.3 & 53.2$\pm$3.4 & 57.7$\pm$7.7 \\
 & CNN-LSTM & 46.0$\pm$8.8 & 52.3$\pm$8.1 & 51.8$\pm$8.0 & 55.8$\pm$4.8 & 58.7$\pm$11.2 & 52.5$\pm$5.8 & 54.7$\pm$15.2 & 53.6$\pm$7.7 & 59.7$\pm$5.6 \\
\midrule
\multirow{3}{*}{Radar + IMU} & CNN & \textbf{\underline{69.9$\pm$10.0}} & \textbf{\underline{64.4$\pm$8.8}} & \textbf{\underline{73.7$\pm$9.7}} & \textbf{\underline{66.2$\pm$10.7}} & \textbf{\underline{77.5$\pm$12.6}} & \textbf{\underline{80.6$\pm$9.6}} & 75.4$\pm$11.0 & {--} & {--} \\
 & LSTM & 53.7$\pm$11.3 & 50.0$\pm$7.3 & {--} & 54.0$\pm$7.1 & 57.4$\pm$11.7 & 53.8$\pm$9.3 & 56.0$\pm$4.9 & {--} & {--} \\
 & CNN-LSTM & 55.0$\pm$8.2 & 51.6$\pm$9.4 & 56.5$\pm$12.5 & 57.9$\pm$5.9 & 62.2$\pm$10.9 & 64.0$\pm$11.4 & 52.6$\pm$4.3 & {--} & {--} \\
\bottomrule
\end{tabular}%
}
\end{table*}

\subsubsection{LoRa-based Affective Behavior Recognition.}\label{sec:lora_affective}
Finally, we quantify how much affective information the single LoRa link carries on its own. Consistent with the observations above, cross-subject generalization of affective expressions is difficult, we therefore report both a within-subject and an unseen-subject evaluation. Each campaign~3 recording is sliced into 32\,s windows with 50\% overlap.
We report window-level metrics and recording-level accuracy via majority voting over window predictions (Tab.~\ref{tab:lora_baseline_C3}). The Random Forest on statistical features is the strongest LoRa model (window-level 40.7\% in both settings; recording-level 39.3\% within- and 31.2\% cross-subject), while the deep models overfit the 133 available recordings, echoing the finding of the radar/IMU benchmark that compact statistical descriptors are effective for this task. The distraction phase, characterized by pronounced macroscopic motion, is detected most reliably, whereas low-motion states (focus, relaxation) are frequently confused. A single amplitude-only link thus provides weak cues for affective behavior analysis.

\begin{table*}[t]
\caption{LoRa-only baseline results on campaign~3. Values in parentheses are recording-level (majority-vote) accuracies.}
\label{tab:lora_baseline_C3}
\centering\small
\setlength{\tabcolsep}{3.5pt}
\begin{tabular}{lcccccccc}
\toprule
\multirow{2}{*}{Model} & \multicolumn{2}{c}{Accuracy (\%)} & \multicolumn{2}{c}{Precision (\%)} & \multicolumn{2}{c}{Recall (\%)} & \multicolumn{2}{c}{F1-score (\%)} \\
\cmidrule(lr){2-3}\cmidrule(lr){4-5}\cmidrule(lr){6-7}\cmidrule(lr){8-9}
 & Within & Cross & Within & Cross & Within & Cross & Within & Cross \\
\midrule
Random Forest (stat.\ features) & 40.7 (39.3) & 40.7 (31.2) & 39.0 & 36.8 & 32.2 & 30.1 & 32.0 & 30.1 \\
CNN-2D (spectrogram) & 21.0 (25.0) & 31.2 (38.4) & 18.6 & 28.3 & 21.6 & 29.6 & 16.7 & 26.3 \\
CNN-1D (temporal) & 20.8 (22.6) & 32.4 (35.0) & 15.6 & 28.2 & 19.3 & 29.1 & 15.1 & 27.7 \\
CNN-BiLSTM & 20.0 (15.5) & 34.6 (32.7) & 15.7 & 30.9 & 15.6 & 27.8 & 14.0 & 27.1 \\
CNN-2D+1D fusion & 17.9 (17.9) & 32.7 (32.7) & 18.7 & 31.3 & 18.0 & 30.4 & 13.4 & 28.0 \\
\bottomrule
\end{tabular}
\end{table*}

\section{Discussion \& Conclusion}
We have presented a comprehensive multimodal dataset, \textit{RF-Behavior}, that integrates eight ground-mounted radars, five ceiling-mounted radars, RFID tags, LoRa, inertial measurement units (IMUs), and infrared cameras to capture human behavior across three distinct temporal and complexity scales: 21~hand gestures, 10~activities, and 6~affective behavior expressions from 61~participants and in three environments. 
By combining complementary sensing and a broad behavioral categories, the dataset enables research on viewpoint-robust RF perception, multimodal fusion under occlusion and poor lighting, and cross-task learning between motion and emotion domains. 
The benchmark results obtained using radars (ground- and ceiling-mounted), RFID tags, LoRa, and IMUs show how these modalities offer complementary strengths across tasks, and the radar and IMU fusion study shows that combining them improves cross-environment robustness. The LoRa baselines delineate the sensing envelope of a single communication link. LoRa supports macroscopic activity recognition (66.7\% under unseen-subject conditions) but remains far below the multi-radar baselines for fine-grained gestures and affective behavior. Whether the high-resolution features learned from radar and RFID can be distilled into such low-resolution LoRa models for ultra-low-power ambient monitoring remains an open cross-modal research direction that the synchronized recordings enable.

This dataset enables rich multimodal fusion and representation learning by combining multi-angle radar, RFID, IMU, and infrared time series dataset, which opens several concrete directions for advancing multimodal sensing, modeling, and deployment. A central opportunity is viewpoint robustness and angle-invariant recognition. With synchronized ground and ceiling radar views, models can be trained to generalize across perspectives, reducing the need for site-specific calibration. Moreover, a proper installation of ceiling radar could possibly remove the need for ground radars, providing angle-invariant recognition with a more user-friendly hardware setup. Large-scale self-supervised objectives—such as masked time–frequency modeling or cross-modal alignment—can learn transferable features that adapt rapidly to new rooms, sensor layouts, or behaviors with few labels. By jointly modeling motion and affective behavior, models can separate the behavioral expression of affect from the ongoing task-related movement, and can track these expressions continuously over time. This helps us understand how affective states modulate human motion in real-life settings. Applications include behavior-aware smart spaces that adapt lighting or notifications to the expressed state of a subject, and privacy-preserving wellbeing monitoring that detects expressions of stress or agitation without identifiable vision information, which is important for healthcare, eldercare, and workplace wellbeing.

There are also limits, which we state explicitly to guide appropriate use of the dataset. \emph{Affective behavior subset:} The affective campaign comprises 23~participants, fewer than the gesture and activity campaigns, because each participant could take part only once to preserve the genuineness of the induced responses. Together with the strongly personalized nature of affective expression, this smaller and demographically narrower subset (see appendix~\ref{sec:appendix_particpantinfo}) is the reason the affective benchmark is framed as within-subject few-shot adaptation rather than cross-subject generalization. Population-level claims about affective behavior recognition may not be well supported by this subset, and results should be read as evidence of per-subject representational feasibility.
The physiological reference subset is smaller still (7 participants in a single environment). It provides individual-level physiological references for the behavioral labels. However, its limited size may not support cross-subject physiological modeling.
The self-reported Likert ratings provide experiential ground truth, but categories such as stress, depression, or excitement can be represented only to the degree that a controlled protocol can induce them. 
The role-played depression/excitement conditions in particular capture deliberately expressed rather than spontaneously occurring behavior. 
\emph{Cross-environment coverage: } 
The living-room and industrial recordings involve fewer participants than the laboratory campaigns. 
They are intended as generalization test-beds rather than training corpora, and the cross-environment results should be interpreted in this role.
\emph{Sensor placement:} For the sports-related activities, participants faced radars 6, 7, and 8 for safety reasons, so these classes have a preferred body orientation. Orientation diversity for them is provided primarily by the ceiling-mounted radars, and models trained on the ground radars may inherit this orientation bias.
\emph{Task continuity:} As gestures, activities, and affective behaviors were collected in separate campaigns, the tasks lack natural temporal continuity. 
However, the shared hardware network, global coordinate frame, and overlapping participant pool still allow models to learn generalized environmental and subject features across tasks.

On the other hand, deploying RF behavior sensing in fully naturalistic environments raises opportunities and challenges beyond what the controlled industrial factory covers. 
Real industrial floors add electromagnetic interference from machinery, dense metallic multipath, and moving equipment. Homes add pets, visitors, and furniture rearrangement. Both require continuous, session-free operation that copes with long-term signal drift, online recalibration, and strict data-protection compliance. The natural next step is to extend the dataset with longitudinal, unscripted recordings in such settings. The living-room physiological reference subset (Section~\ref{sec:physio_subset}) provides an initial reference that links behavioral expressions back to internal state. The released multi-environment benchmark provides the controlled reference point against which such naturalistic data can be calibrated.

\section{Ethical Approval}
For the data collection campaigns we have acquired ethical approval by the Aalto University Research Ethics Committee. 
Subjects were provided with three documents: 
A consent form to be signed by the subjects, an information sheet that provides a general description of the research as well as a privacy notice for the data collection. 
The privacy notice explains in detail how personal data will be used, how we store the data and which rights the participants have according to the GDPR.
Thus, for the entire data collection, we ensured that all participants joined voluntarily and were well informed about the collection procedures and the rights they have. 
To recruit participants, we sent the advertising poster via email lists and put up paper posters at Aalto University.

\section{Dataset Availability}
The use of \textit{RF-Behavior} is limited to academic research purposes.
The dataset is available at \url{https://huggingface.co/datasets/Si-Z/RF-Behavior}.
The project page at \url{https://sizuo.github.io/RF-Behavior/} provides an overview of the dataset, interactive sample visualizations, and usage instructions.
The release contains one archive per trial and modality, named \texttt{<node>\_<modality>\_<environment>\_<user>\_<class>\_<repetition>\_<time>.zip} and organized in the folders \texttt{radar}, \texttt{lora}, \texttt{rfid}, \texttt{mocap}, and \texttt{imu}, together with metadata tables.
A selective download script, a data loader, and a demonstration notebook are released as part of the dataset. Subsets can be retrieved by environment, campaign, participant, class, and modality.
The physiological reference subset (Section~\ref{sec:physio_subset}) is documented on the dataset page.

\bibliographystyle{ACM-Reference-Format}
\bibliography{ref}

\appendix
\section{Details of the data collected}
\subsection{Participants}\label{sec:appendix_particpantinfo}
A total of~61 subjects participated in the data collection, consisting of 30 male and 31 female participants (see Table~\ref{tab:participant_info} and Table~\ref{tab:extended_participant_info}). 
Not all subjects participated in all campaigns or environments. 
Across all settings, there were~37 unique subjects for campaign~1, 33 subjects for campaign~2, 31 subjects for campaign~3, and 17 subjects for campaign~4. 
The participant group was predominantly in the 21-30 age range (44 subjects). 
The study also included three younger participants under the age of 18 (for whom we obtained signed parental consent), three participants aged 18-20, and nine individuals between the ages of~31 and~50. 
Two participants chose to not disclose their age. 
Additionally, we collected arm length information from all participants.

\begin{table}
\centering 
\caption{Participant Information and campaign Completion Status For Lab Setting.}
\label{tab:participant_info}
\resizebox{0.8\textwidth}{!}{
\begin{tabular}{c c c c c c c c c}
\toprule
\textbf{UserID} & \textbf{Gender} & \textbf{AgeRange} & \textbf{ArmLength (cm)} & \textbf{Campaign~1} & \textbf{Campaign~2} & \textbf{Campaign~3} & \textbf{Campaign~4}\\
\midrule
U01 & F & 41-50   &  57   & \checkmark & \checkmark & -         &   -\\
U02 & F & 21-30       &  57  & -          & -          & - & \checkmark \\
U03 & M & 21-30   &   58 & \checkmark & \checkmark & \checkmark &  -\\
U04 & M & 21-30   &  54  & \checkmark & \checkmark & \checkmark &  -\\
U05 & M & 21-30   &  54  & \checkmark & \checkmark & -          &  -\\
U06 & M & 21-30   &  58  & \checkmark & \checkmark & \checkmark &  -\\
U07 & F & 21-30   &   53  & \checkmark & \checkmark & \checkmark &  -\\
U08 & F & 21-30   &  52  & -          & -          & \checkmark &  -\\
U09 & M & 21-30       &  62  & -          & -          & - & \checkmark \\
U10 & M & $<18$   &  43 & \checkmark & \checkmark & \checkmark  &  -\\
U11 & F & 21-30   &  50  & \checkmark & \checkmark & \checkmark &  -\\
U12 & F & 31-40   &   58 & \checkmark & \checkmark & \checkmark &  -\\
U13 & M & 21-30   &  56   & -          & \checkmark & \checkmark &  -\\
U14 & F & 18-20   &   52 & \checkmark & \checkmark & \checkmark &  -\\
U15 & M & 21-30       &  62  & -          & -          & - & \checkmark \\
U16 & M & 21-30       &  64  & -          & -          & - & \checkmark \\
U17 & M & 21-30   &  52  & \checkmark & \checkmark & \checkmark &  -\\
U18 & M & 21-30       &  70  & -          & -          & - & \checkmark \\
U19 & F & $<18$   &  55  & \checkmark & \checkmark & -          &  -\\
U20 & F & 21-30       &  67  & -          & -          & - & \checkmark \\
U21 & M & 21-30   &  52   & \checkmark & \checkmark & \checkmark &  -\\
U22 & M & 21-30   &  58  & \checkmark & \checkmark & \checkmark &  -\\
U23 & F & 21-30       &  58  & -          & -          & - & \checkmark \\
U24 & M & 21-30   & 57   & -          & -          & \checkmark &  \checkmark\\
U25 & M & $<18$   &  39  & \checkmark & \checkmark & \checkmark &  -\\
U26 & M & 21-30       &  64  & -          & -          & - & \checkmark \\
U27 & M & 21-30       &  70  & -          & -          & - & \checkmark \\
U28 & M & 41-50   &  55 & \checkmark & \checkmark & -          &  \checkmark \\
U29 & F & 31-40   &   49  & \checkmark & \checkmark & \checkmark &  -\\
U30 & M & 21-30   &   53  & -          & \checkmark & \checkmark &  -\\
U31 & F & 21-30       &  57  & -          & -          & - & \checkmark \\
U32 & M & 21-30   & 59   & \checkmark & \checkmark & \checkmark &  -\\
U33 & F & 21-30   &  48  & \checkmark & \checkmark & \checkmark &  -\\
U34 & F & 21-30       &  52  & -          & -          & - & \checkmark \\
U35 & M & 18-20   &  50  & \checkmark & \checkmark & \checkmark &  -\\
U36 & M & 21-30   &  -  & \checkmark & -          & -          &  -\\
U37 & M & 21-30       &  67  & -          & -          & - & \checkmark \\
U38 & M & 21-30   &  57  & \checkmark & \checkmark & \checkmark &  -\\
U39 & M & 21-30       &  74  & -          & -          & - & \checkmark \\
U40 & M & 21-30       &  75  & -          & -          & - & \checkmark \\
U41 & F & 21-30   &  54  & \checkmark & \checkmark & \checkmark &  -\\
U42 & F & 21-30   &  52   & \checkmark & -          & -          &  -\\
U43 & F & 21-30       &  61  & -          & -          & - & \checkmark \\
U44 & M & 21-30   &  56  & \checkmark & \checkmark & \checkmark &  -\\

\bottomrule
\end{tabular}
}
\end{table}

\begin{table}
\centering
\caption{Participant Demographics and campaign Completion Status Across Extended Environments (Living Room and Industrial Workplace).}
\label{tab:extended_participant_info}
\resizebox{0.8\textwidth}{!}{
\begin{tabular}{l c c c | ccc | ccc}
\toprule
 & & & & \multicolumn{3}{c|}{\textbf{Living Room (LR)}} & \multicolumn{3}{c}{\textbf{Industrial (IND)}} \\
\textbf{UserID} & \textbf{Gender} & \textbf{Age} & \textbf{Arm (cm)} & \textbf{C1} & \textbf{C2} & \textbf{C3} & \textbf{C1} & \textbf{C2} & \textbf{C3} \\
\midrule
U19 & F & <18   & 55.0 & -          & \checkmark & -          & -          & -          & -          \\
U24 & M & 21-30 & 57.0 & -          & \checkmark & -          & -          & -          & -          \\
U28 & M & 41-50 & 55.0 & -          & \checkmark & -          & -          & -          & -          \\

U45 & F & 21-30 & 53.5 & \checkmark & -          & -          & -          & -          & -          \\
U46 & F & 31-40 & 51.0 & \checkmark & -          & -          & -          & -          & -          \\
U47 & F & 31-40 & -    & \checkmark & \checkmark & -          & \checkmark & -          & -          \\
U48 & M & 41-50 & 56.0 & \checkmark & \checkmark & \checkmark & \checkmark & -          & -          \\
U49 & F & 21-30 & 59.5 & \checkmark & -          & -          & -          & -          & -          \\
U50 & F & 31-40 & 63.5 & \checkmark & \checkmark & \checkmark & \checkmark & \checkmark & -          \\
U51 & M & 21-30 & 57.0 & \checkmark & -          & -          & -          & -          & -          \\
U52 & F & 21-30 & 50.2 & \checkmark & -          & -          & -          & -          & -          \\
U53 & F & 41-50 & -    & -          & \checkmark & \checkmark & -          & \checkmark & -          \\
U54 & F & 21-30 & 45.5 & -          & \checkmark & \checkmark & -          & -          & -          \\
U55 & F & 18-20 & 52.0 & -          & -          & -          & -          & -          & \checkmark \\
U56 & F & 21-30 & 55.0 & \checkmark & -          & -          & -          & -          & \checkmark \\
U57 & M & 21-30 & 58.5 & \checkmark & -          & -          & -          & -          & -          \\
U58 & F & 21-30 & 53.0 & -          & -          & \checkmark & -          & \checkmark & -          \\
U59 & F & 21-30 & 57.0 & -          & -          & \checkmark & -          & -          & -          \\
U60 & F & -     & -    & -          & -          & -          & \checkmark & -          & -          \\
U61 & F & -     & 56.0 & -          & -          & -          & \checkmark & -          & -          \\
\bottomrule

\end{tabular}
}
\end{table}

\subsection{Statistics on the data}
In this section, we detail the statistical distribution of the data collected in the lab setting, as it serves as the primary, highly controlled baseline featuring the complete suite of sensing modalities. 
While the data from the real-world environments (Living Room and Industrial) share identical temporal and categorical structures, their distinct statistical breakdowns are omitted here.

\begin{figure}
  \centering
  \begin{subfigure}[t]{0.45\textwidth}
    \centering
    \includegraphics[width=\textwidth]{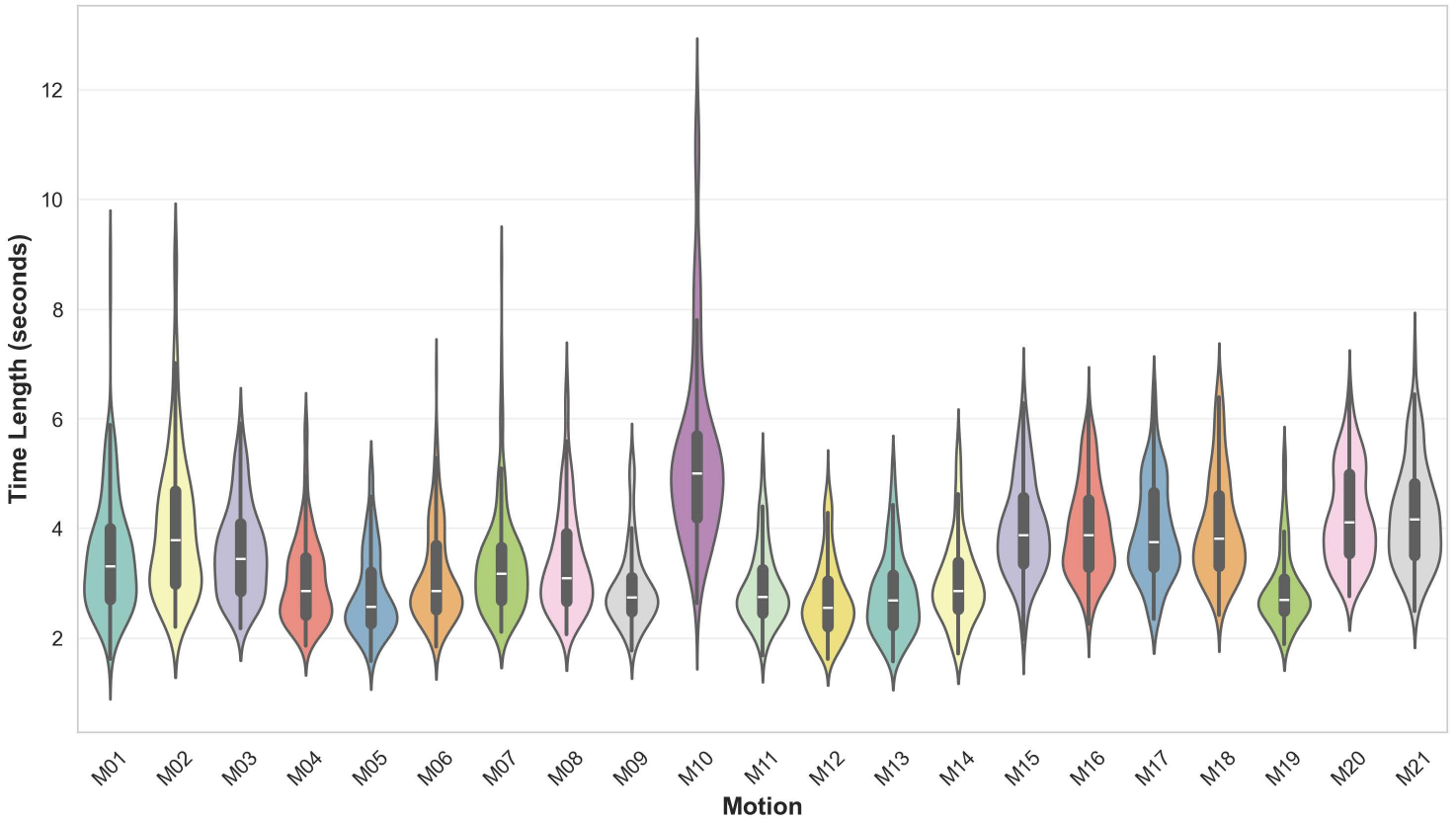} 
  \end{subfigure}
  \begin{subfigure}[t]{0.45\textwidth}
    \centering
    \includegraphics[width=\textwidth]{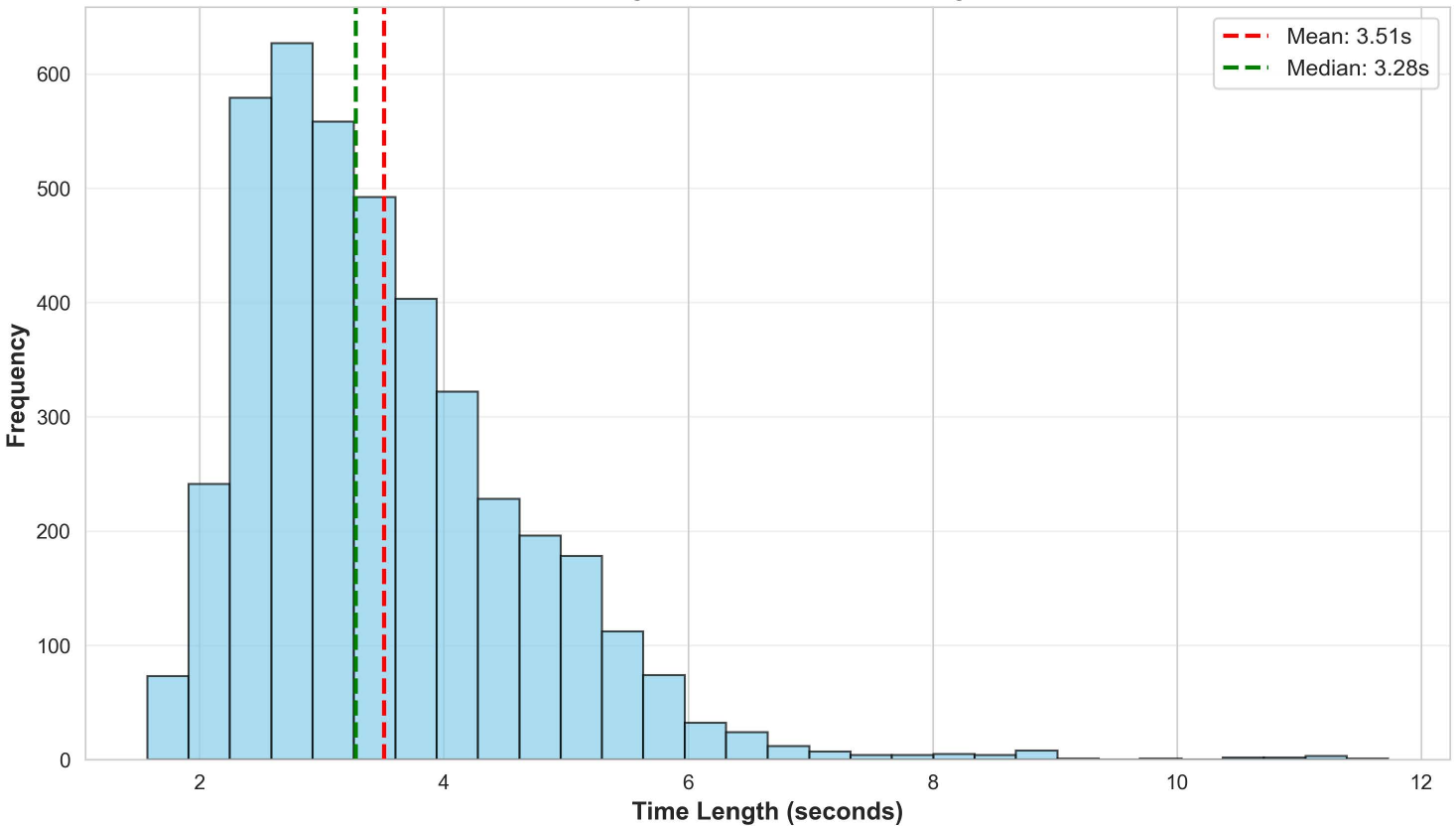} 
  \end{subfigure}
  \begin{subfigure}[t]{0.45\textwidth}
    \centering
    \includegraphics[width=\textwidth]{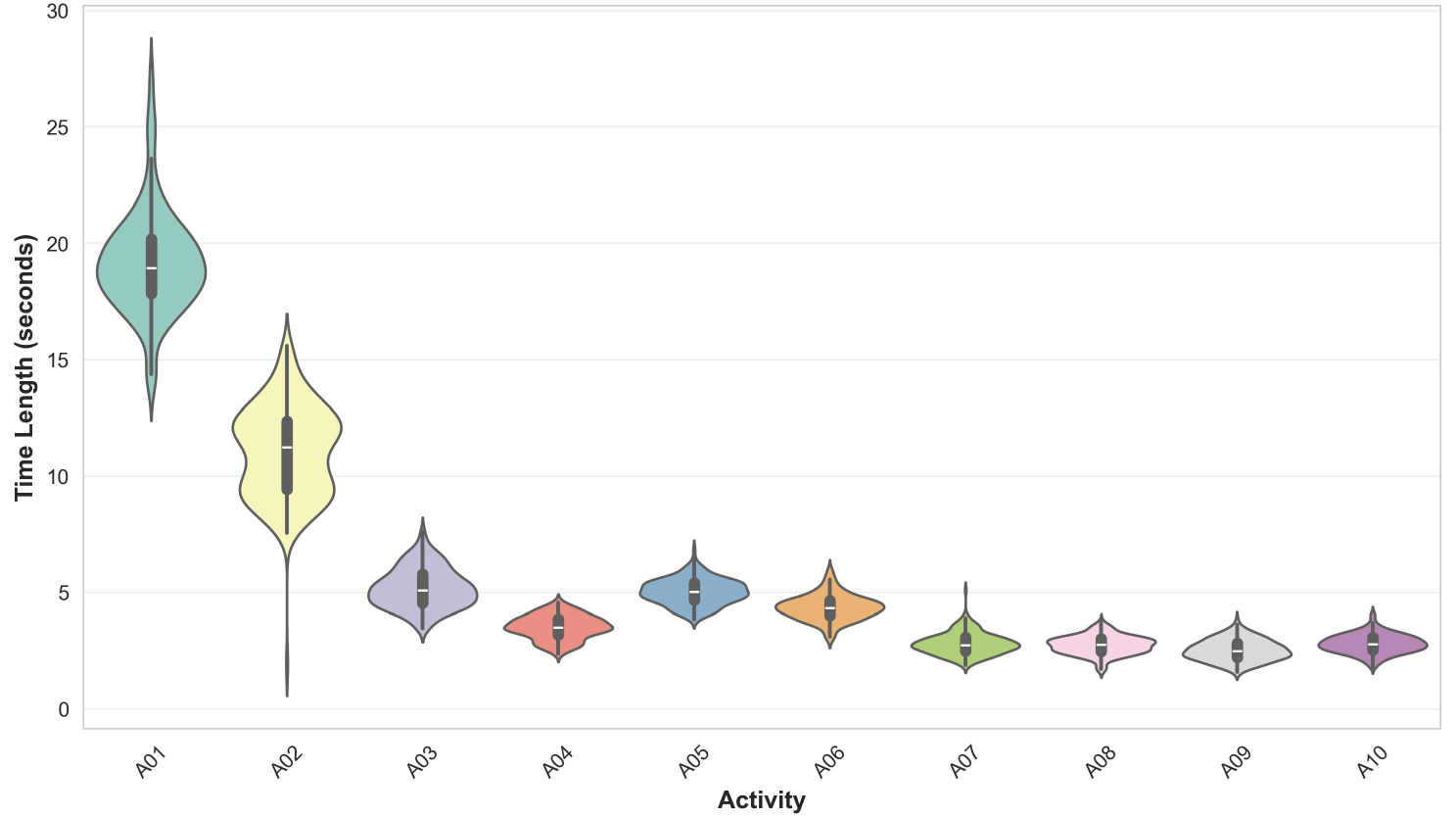} 
  \end{subfigure}
  \begin{subfigure}[t]{0.45\textwidth}
    \centering
    \includegraphics[width=\textwidth]{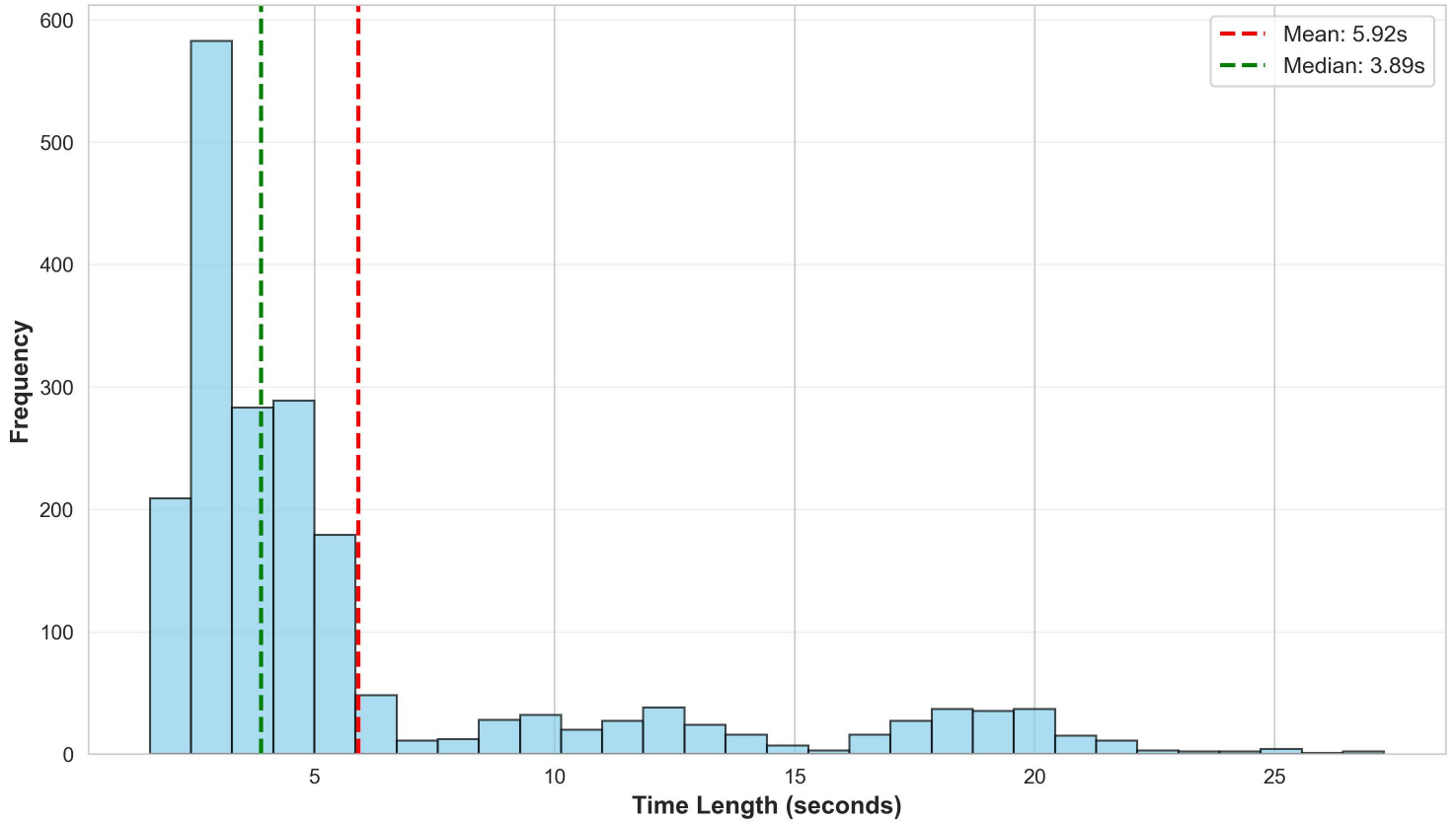} 
  \end{subfigure}
  \begin{subfigure}[t]{0.45\textwidth}
    \centering
    \includegraphics[width=\textwidth]{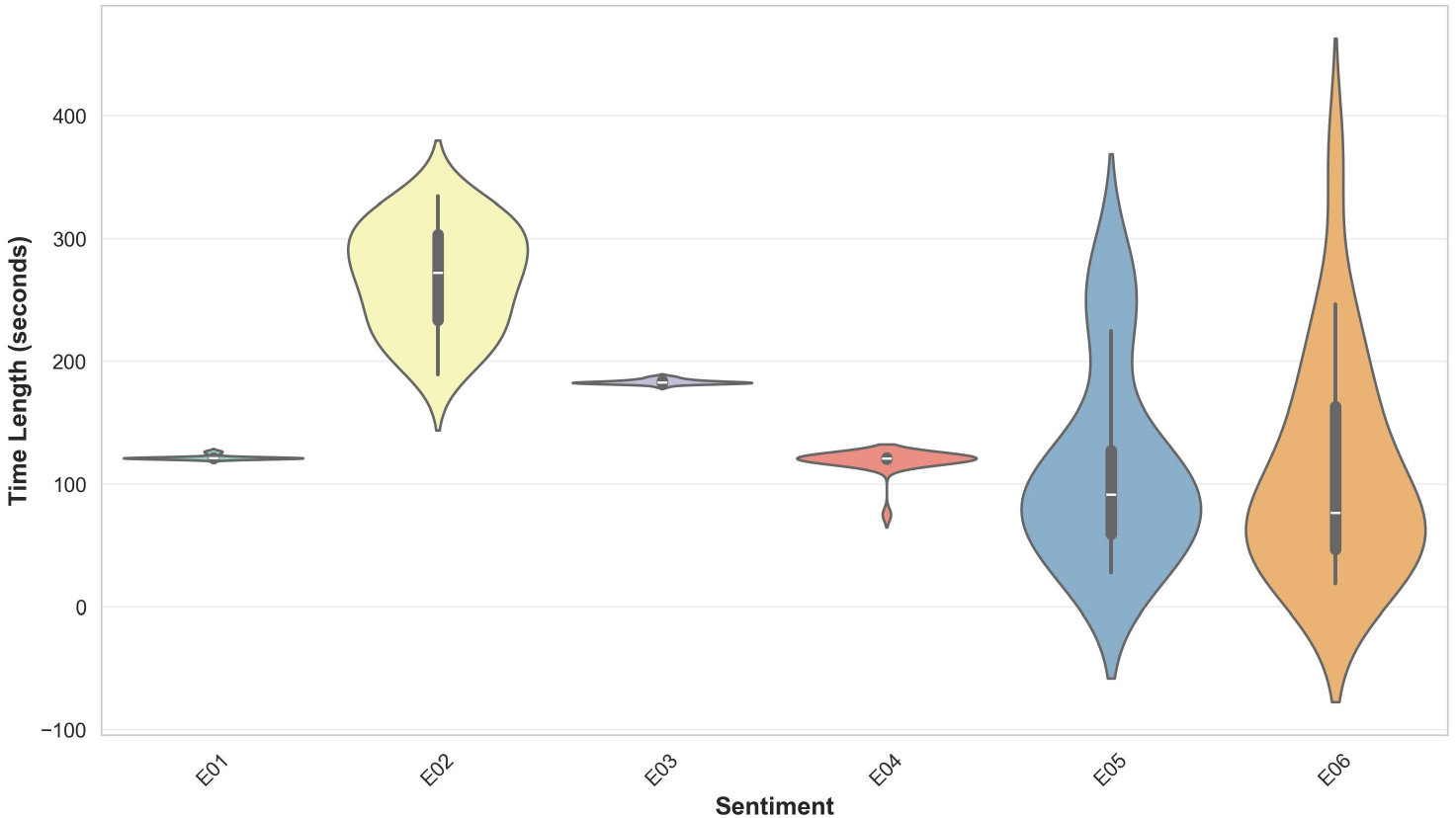} 
  \end{subfigure}
  \begin{subfigure}[t]{0.45\textwidth}
    \centering
    \includegraphics[width=\textwidth]{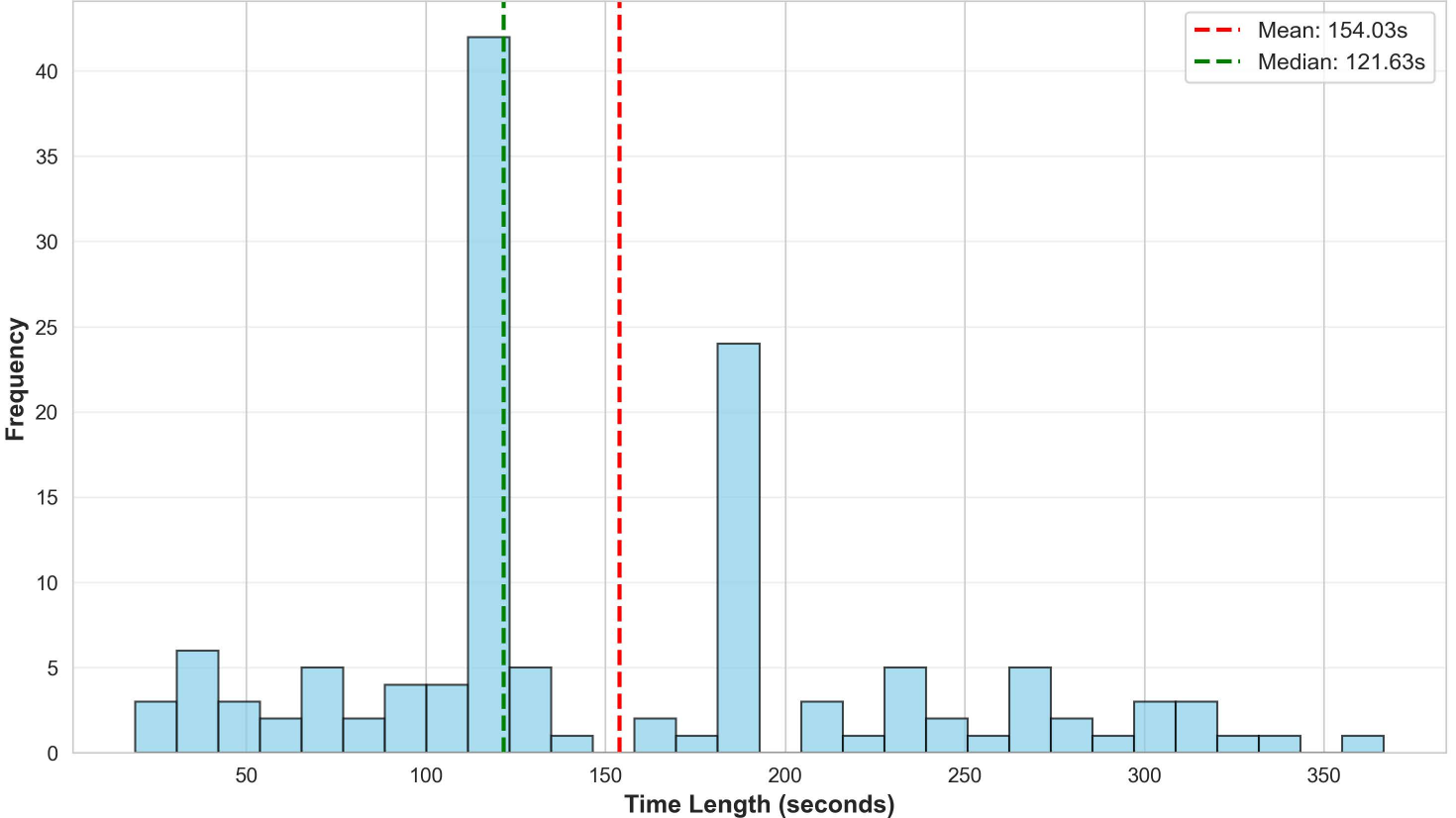} 
  \end{subfigure}
  \begin{subfigure}[t]{0.45\textwidth}
    \centering
    \includegraphics[width=\textwidth]{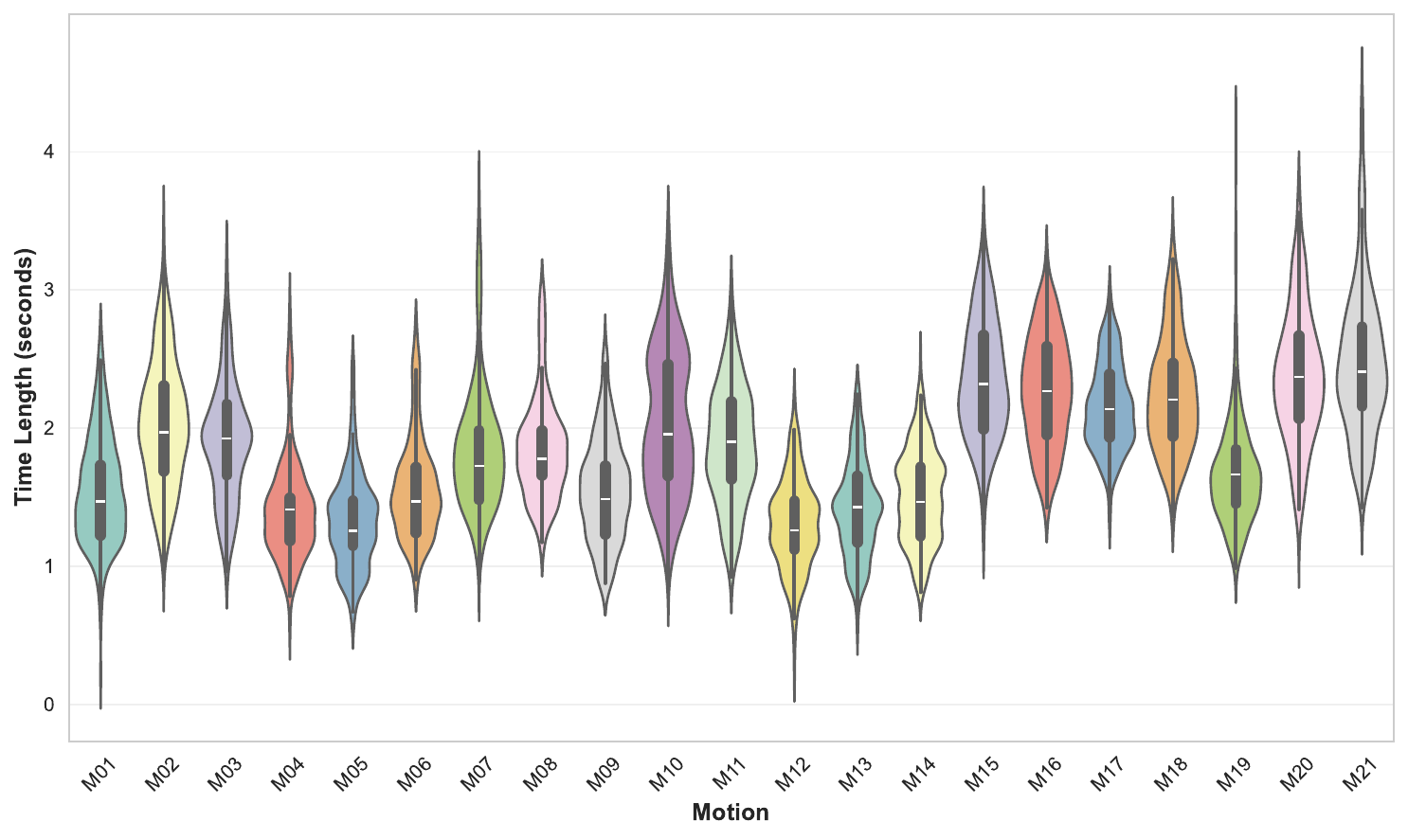} 
  \end{subfigure}
  \begin{subfigure}[t]{0.45\textwidth}
    \centering
    \includegraphics[width=\textwidth]{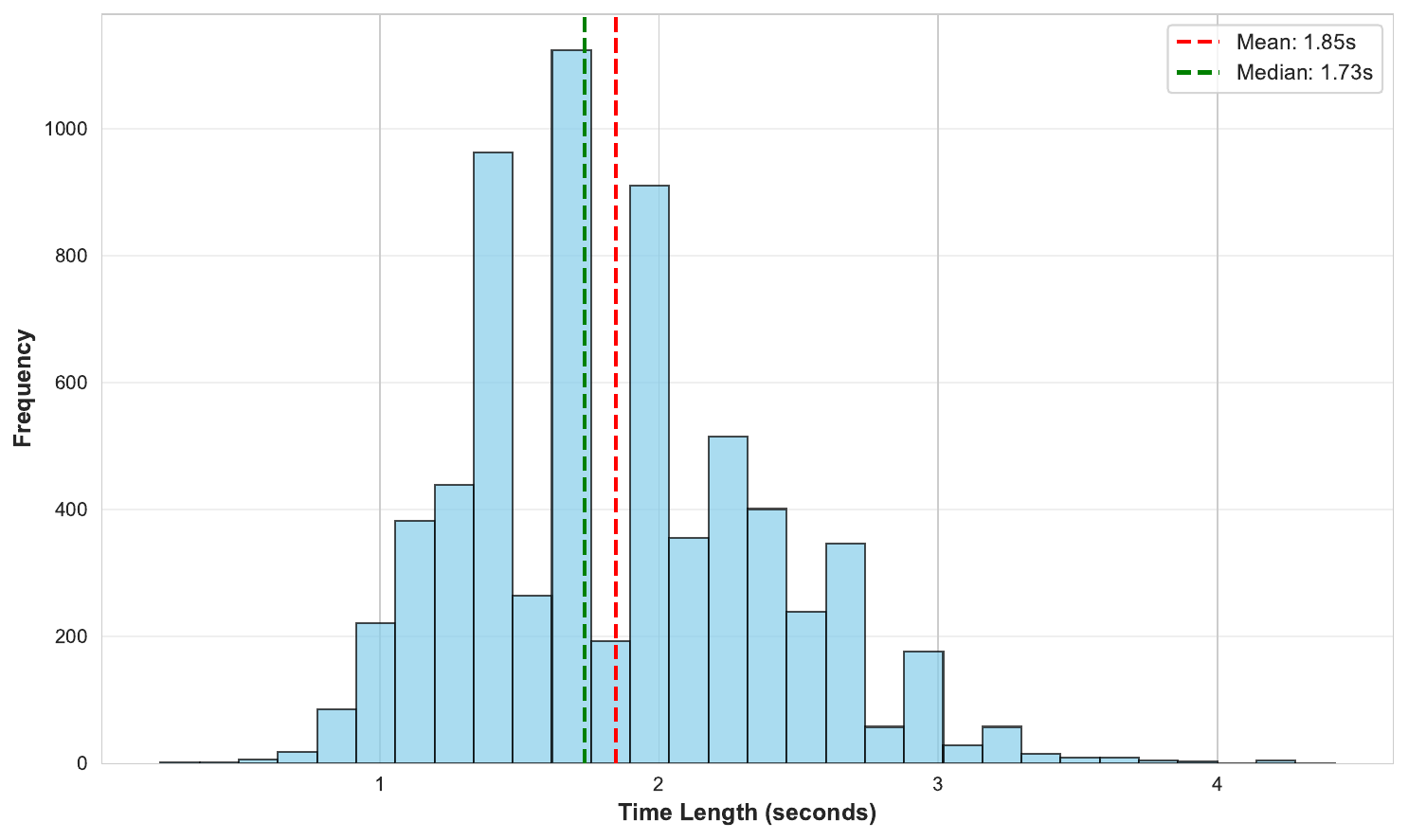} 
  \end{subfigure}
  \caption{From Top $\rightarrow$ Bottom: Time length distribution for data from campaign~1, 2, 3, 4.}
  \label{fig:all_violinhistg}
\end{figure}

Fig.~\ref{fig:all_violinhistg} reveals that the campaigns operate on entirely distinct time scales and exhibit unique distribution patterns. 
Campaign~1 (gestures) shows tight clustering with a mean of~3.51 seconds and median of~3.29 seconds, representing quick, discrete physical movements from the arm. 
Campaign~2 (activity) displays moderate durations with a mean of 5.92 seconds and median of 3.89 seconds, indicating more complex behaviors that still remain relatively brief. 
Campaign~3 (affective behavior) operates on a dramatically different scale with a mean of 154.03 seconds and a median of 121.63 seconds, representing sustained emotional expressions lasting over two minutes on average, which also aligns with the design of the experiment. 

For campaign~1 (Motion Gestures), the violin plots reveal that most individual gestures maintain narrow distributions. The majority of gestures cluster in the 2-4 second range across the violin plots, with only a few motions (M10 - arm swing, M20 - circle-clockwise, M21 - circle-counter-circles) requiring extended durations, suggesting these represent either more complex multi-step movements or gestures requiring greater spatial displacement.

For campaign~2, the histogram shows a dominant peak around 3-4 seconds but also reveals a significant secondary distribution extending from 15-25 seconds. 
This bimodality is strikingly apparent in the violin plots, where A01 (walking) and A02 (running) form a clearly separate cluster with distributions centered around 18-20 seconds and 10-12 seconds, respectively. 
This is also expected as for walking and running, the participants were asked to walk/run in an "S" shape within the recording area.

Campaign~3 shows an extremely wide distribution ranging from approximately~20 seconds to over~400 seconds. 
The histogram reveals a highly uneven distribution with the peak around~120 seconds, but with many instances extending well beyond~200 seconds. 
The large difference between the mean (154.03s) and median (121.63s) indicates that longer durations significantly pull the average upward. 
The violin plots reveal heterogeneity across emotional state types, reflecting different experimental protocols for each emotion. 
E01 (focus) shows a concentrated distribution around 120 seconds, while E03 (stress) displays a concentrated distribution around 180~seconds. These patterns directly correspond to the experimental design, where we imposed time limits of 2~minutes for focus and 3~inutes for stress. 
E04 (relaxation) similarly shows a relatively narrow, consistent pattern around 120 seconds. 
In contrast, E05 and E06 display moderate durations with wide distributions spanning 50-250 seconds. 
This variability is reasonable and expected, as data collection for depression and excitement involved participants sharing personal stories while experiencing these emotions. 
The recorded length naturally varied depending on each participant's storytelling style, resulting in the observed distribution spread.

Compared with campaign~1, data from campaign~4 shows consistently shorter durations (~47\%). M10 (arm swing) in campaign~4 is comparable to other motions (1.5-3 seconds, median ~2s). These might due to the different participant instructions during the data collection. 

\subsection{Affective behavior quality control and feature screening}\label{sec:apx_affective_qc}
This section reports the quality-control statistics and the repeated-measures feature screening underlying the compact descriptor set of Table~\ref{tab:final-features}.

\begin{figure}[t]
\centering
\includegraphics[width=\linewidth]{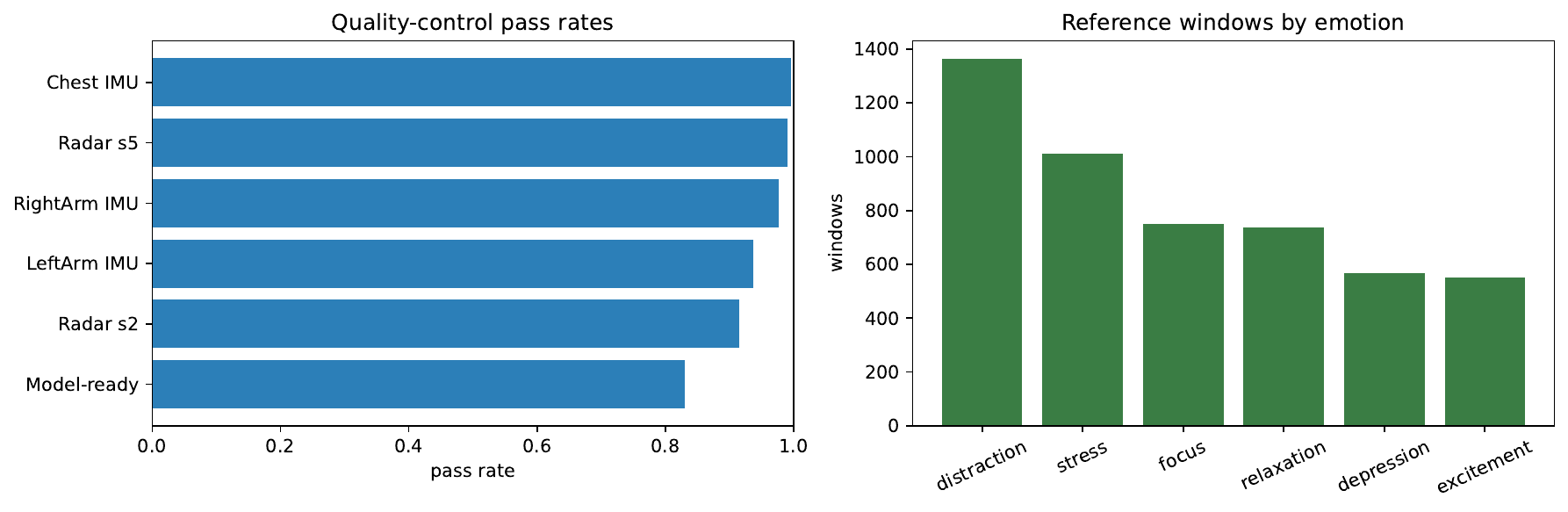}
\caption{Coverage and pass-rate summary for the tightened affective behavior table. The key point is that the dataset remains large after conservative multimodal filtering.}
\label{fig:affective-coverage}
\end{figure}

\begin{table}[t]
\centering
\footnotesize
\caption{Retained affective-behavior windows after timing-aware alignment and quality control.}
\label{tab:affective-coverage}
\begin{tabular}{lr}
\toprule
Metric & Value \\
\midrule
Reference windows & 4,976 \\
Feature-analysis subset & 4,130 \\
Retention rate & 83.0\% \\
Subjects in reference & 23 \\
Subjects in feature-analysis subset & 22 \\
Sessions & 133 \\
Chest IMU pass rate & 99.6\% \\
Left-arm IMU pass rate & 93.7\% \\
Right-arm IMU pass rate & 97.6\% \\
Radar sensor 2 pass rate & 91.5\% \\
Radar sensor 5 pass rate & 99.1\% \\
\bottomrule
\end{tabular}
\end{table}

The screening is performed on subject-level medians within posture and emotion. 
Sitting features are screened with the Friedman test across the four repeated-measures sitting emotions; 
all~22 subjects in the descriptor subset contribute complete sitting quartets. 
Standing features are screened with paired Wilcoxon signed-rank tests across depression and excitement; these tests are computed on the~18 subjects with complete standing pairs after quality control. 
These tests are used for feature screening rather than final model evaluation.

\begin{figure}[t]
\centering
\begin{minipage}[t]{0.49\linewidth}
    \centering
    \includegraphics[width=\linewidth]{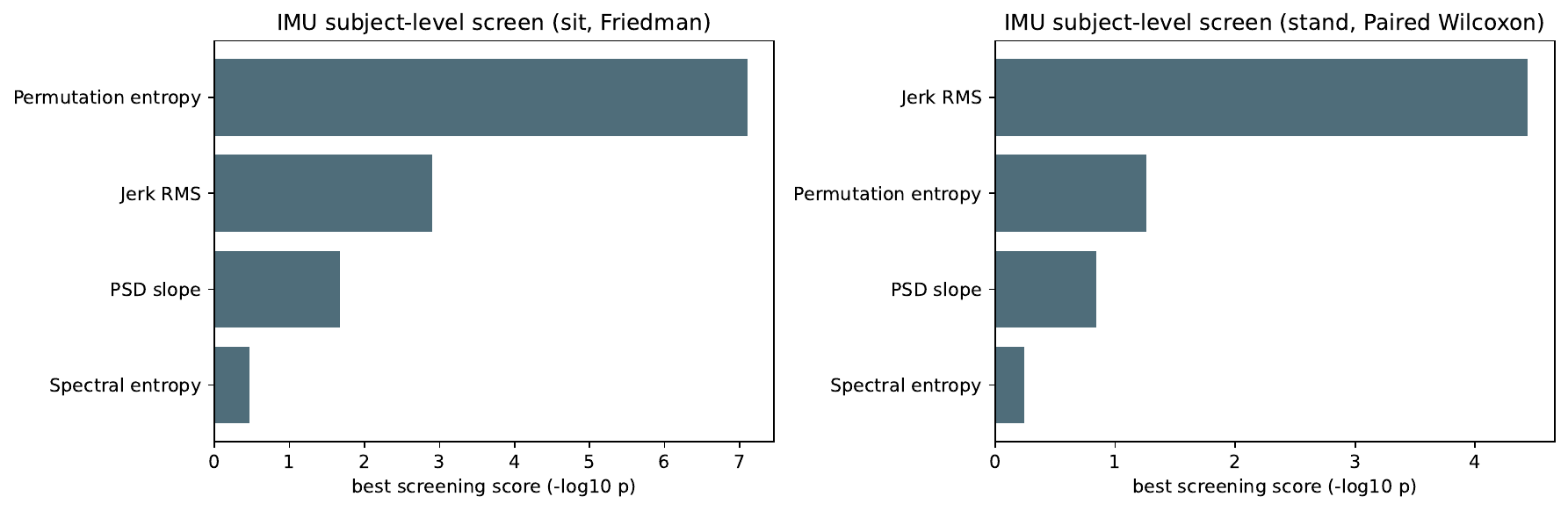}
\end{minipage}\hfill
\begin{minipage}[t]{0.49\linewidth}
    \centering
    \includegraphics[width=\linewidth]{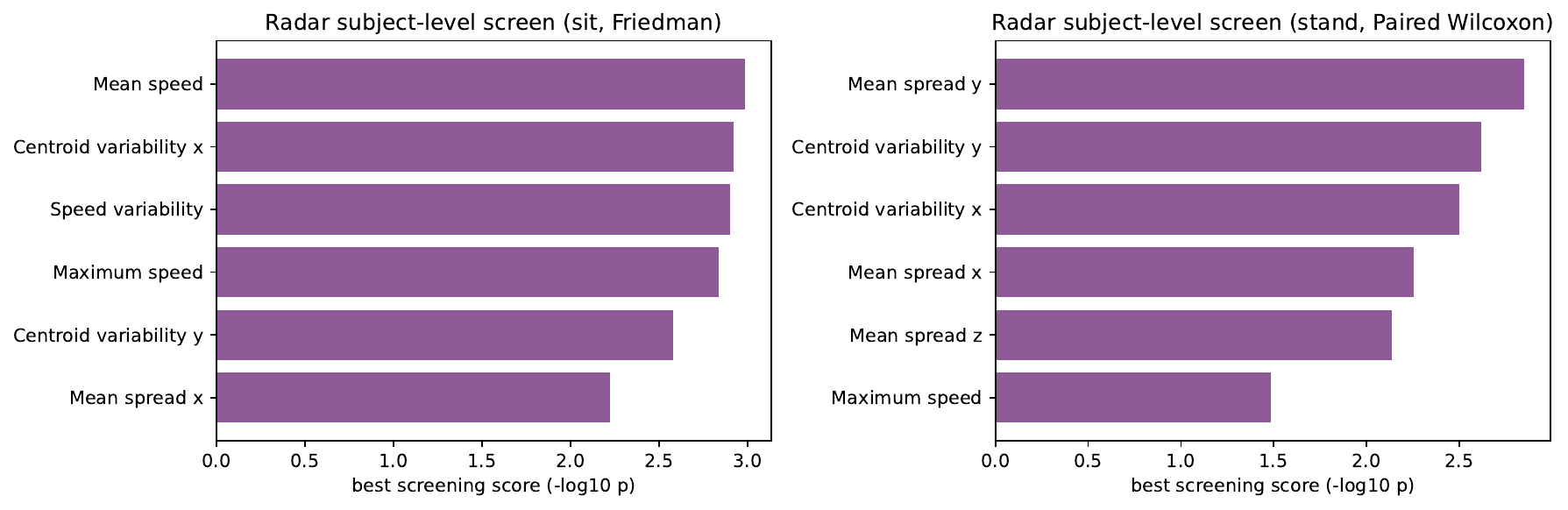}
\end{minipage}
\caption{Subject-level repeated-measures screening of IMU and radar features in the affective behavior recordings. IMU and radar both retain discriminative structure after alignment and quality control.}
\label{fig:screening}
\end{figure}

For IMU, sitting effects are strongest for chest and arm permutation entropy (PE), followed by root-mean-square jerk (jerk RMS), while power spectral density (PSD) contributes more selectively. 
In standing, jerk RMS dominates across the three body locations. Radar contributes complementary centroid-variability and spread features, particularly the variance terms from sensor~2 and the $y$-spread terms from sensors~2 and~5. Figure~\ref{fig:screening} summarizes these rankings.

\begin{figure}[t]
\centering
\includegraphics[width=0.84\linewidth]{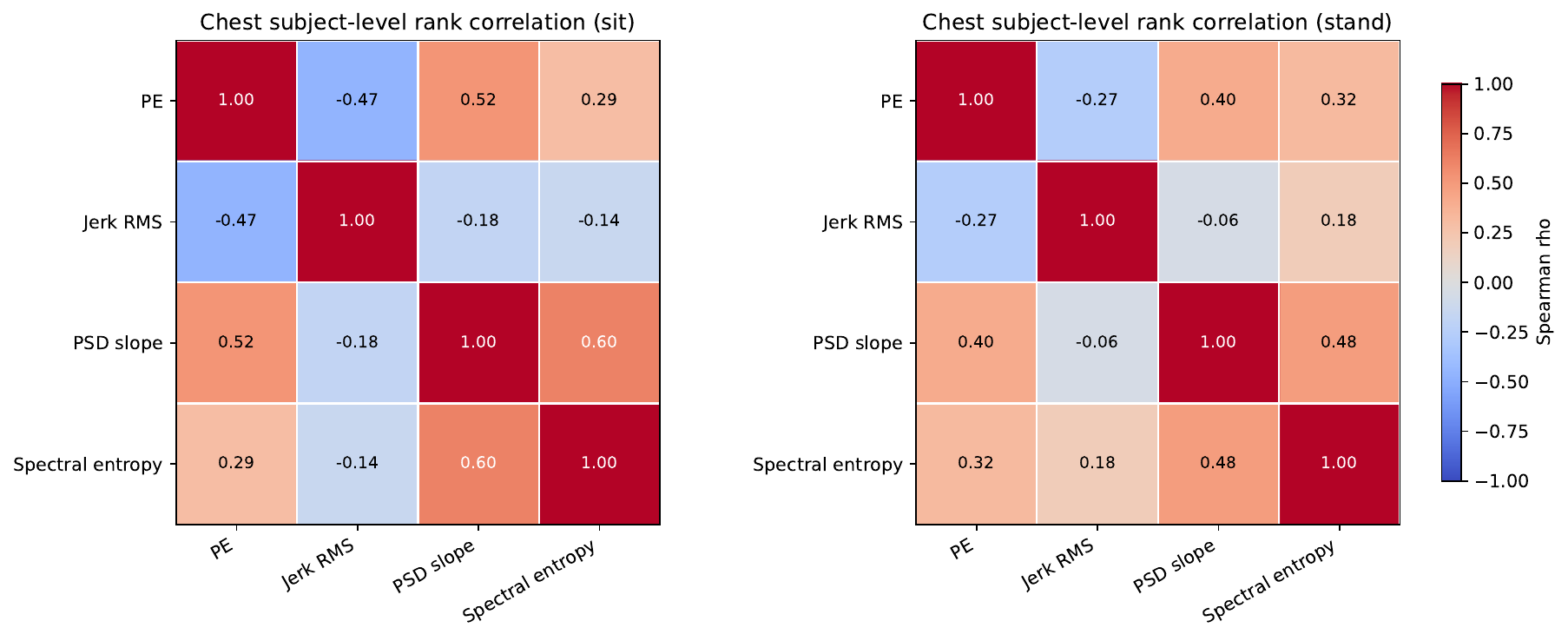}
\caption{Subject-level rank correlations among chest IMU features. In the sitting condition, PE and PSD slope co-vary more strongly with each other than with jerk RMS, consistent with complementary structure-, spectrum-, and intensity-related descriptors.}
\label{fig:chest-corr}
\end{figure}

Chest IMU correlations support the use of complementary descriptors. 
In Fig.~\ref{fig:chest-corr}, PE reflects temporal regularity and complexity, PSD slope spectral organization, and jerk RMS movement intensity. 
In the sitting condition, PE and PSD slope co-vary more strongly with each other than with jerk RMS, indicating that structure- and intensity-related information are not redundant.

From each aligned window, we compute four IMU descriptors per body location (PE, jerk RMS, PSD slope, spectral entropy) and radar point-cloud summaries (centroid means, centroid standard deviations, spread statistics, and speed summaries). 
Features are retained only if they satisfy three criteria: temporal reliability, repeated-measures statistical support, and interpretability. 
Spectral entropy is excluded because it is consistently weaker than PE and jerk. Relative and log-ratio arm features are excluded from the compact subset because they are more sensitive to residual alignment error. 
The PSD slope remains informative, especially in the sitting emotions, but its repeated-measures support is weaker than PE and jerk in the retained subset. 
Right-arm PE is excluded because it does not reach significance in either posture under the repeated-measures screen.

A sensitivity analysis on the lab subset over window lengths of~4, 6, 8, and~12\,s and overlaps of~0, 25, 50, and~75\% indicates that the main selected descriptors remained broadly stable across settings, while 12\,s with 75\% overlap retained the largest quality-controlled multimodal subset. 
We therefore use this setting in the training experiments as a pragmatic default rather than as a narrowly optimized choice.

\subsection{Post-Task Assessment for campaign~3}
Table~\ref{tab:s3_selfassessment} lists the per-subject self-assessment ratings collected immediately after the Campaign~3 recording sessions, as described in Sec.~\ref{sec:c3_protocol}. Each participant rated how strongly they experienced each of the six target states during the corresponding phase, on a 6-point Likert scale from 0 (not in the mood) to 5 (quite in the mood).
\begin{table}
\centering
\caption{Campaign~3 - Emotional \& Cognitive State Self-Assessment.
    \textit{Scale Key: 0=Not in the mood; 1=Slightly; 2=Somewhat; 3=Moderately; 4=Mostly; 5=Quite in the mood.}}
\label{tab:s3_selfassessment}
\resizebox{0.7\columnwidth}{!}{%
\begin{tabular}{c cccc cc}
\toprule
\multirow{2}{*}{\textbf{subject ID}} & \multicolumn{4}{c}{\textbf{Part I}} & \multicolumn{2}{c}{\textbf{Part II}} \\
\cmidrule(lr){2-5} \cmidrule(lr){6-7}
& \textbf{Focus} & \textbf{Distraction} & \textbf{Stress} & \textbf{Relaxation} & \textbf{Depression} & \textbf{Excitement} \\
\midrule
U01 & -- & -- & -- & -- & -- & -- \\
U03 & 4 & 3 & 3 & 4 & 3 & 3 \\
U04 & 4 & 3 & 4 & 2 & -- & 3 \\
U05 & -- & -- & -- & -- & -- & -- \\
U06 & 4 & 4 & 3 & 3 & 3 & 3 \\
U07 & 5 & 3 & 3 & 3 & 5 & 3 \\
U08 & 3 & 2 & 1 & 3 & 4 & 5 \\
U10 & 4 & 3 & 4 & 5 & -- & -- \\
U11 & 5 & 2 & 2 & 3 & 2 & 5 \\
U12 & 4 & 4 & 3 & 3 & 4 & 4 \\
U13 & 4 & 3 & 4 & 5 & 3 & 2 \\
U14 & 4 & 2 & 3 & 5 & 5 & 4 \\
U17 & 4 & 4 & 3 & 3 & 2 & 4 \\
U19 & -- & -- & -- & -- & -- & -- \\
U21 & 4 & 3 & 5 & 3 & 2 & 3 \\
U22 & 4 & 4 & 4 & 3 & 4 & 5 \\
U24 & 4 & 3 & 2 & 5 & 3 & 4 \\
U25 & 4 & 3 & 4 & 4 & -- & -- \\
U28 & -- & -- & -- & -- & -- & -- \\
U29 & 4 & 1 & 1 & 3 & 3 & 1 \\
U30 & 4 & 5 & 4 & 5 & 3 & 4 \\
U32 & 4 & 2 & 1 & 4 & 3 & 2 \\
U33 & 5 & 4 & 4 & 3 & 4 & 3 \\
U35 & 3 & 2 & 2 & 3 & 3 & 2 \\
U36 & -- & -- & -- & -- & -- & -- \\
U38 & 4 & 2 & 2 & 3 & 1 & 2 \\
U41 & 4 & 3 & 3 & 4 & 4 & 3 \\
U42 & -- & -- & -- & -- & -- & -- \\
U44 & 4 & 1 & 1 & 5 & 3 & 5 \\
U45 & -- & -- & -- & -- & -- & -- \\
U46 & -- & -- & -- & -- & -- & -- \\
U47 & -- & -- & -- & -- & -- & -- \\
U48 & 3 & 3 & 2 & 3 & 5 & 4 \\
U49 & -- & -- & -- & -- & -- & -- \\
U50 & 4 & 5 & 5 & 3 & 3 & 1 \\
U51 & -- & -- & -- & -- & -- & -- \\
U52 & -- & -- & -- & -- & -- & -- \\
U53 & 4 & 3 & 2 & 5 & 4 & 4 \\
U54 & 2 & 1 & 2 & 4 & 3 & 4 \\
U55 & 4 & 2 & 1 & 5 & 5 & 1 \\
U56 & 4 & 2 & 3 & 5 & 4 & 0 \\
U57 & -- & -- & -- & -- & -- & -- \\
U58 & 4 & 3 & 4 & 5 & 5 & 4 \\
U59 & 4 & 3 & 4 & 3 & 3 & 4 \\
U60 & -- & -- & -- & -- & -- & -- \\
U61 & -- & -- & -- & -- & -- & -- \\
\bottomrule
\end{tabular}%
}
\end{table}

\subsection{Extended setup and benchmark details}\label{sec:apx_setup_tables_figures}
This appendix collects additional setup illustrations and per-model analysis figures referenced from Sections~3 and~4.

\begin{figure}
\centering
\begin{minipage}{0.45\textwidth}
    \centering
    \includegraphics[width=\textwidth]{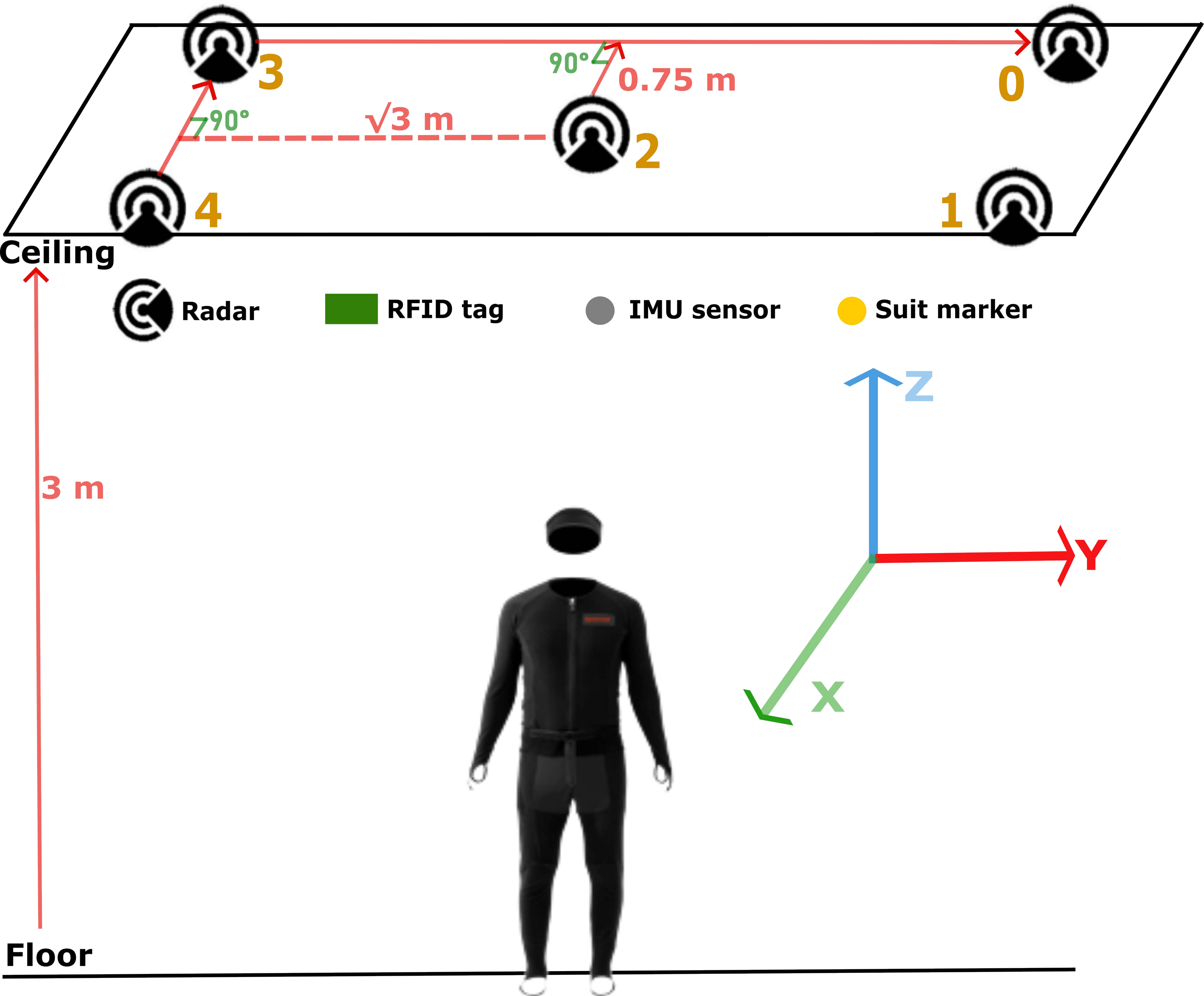}
    \caption{Ceiling setup adjustment.}
    \label{fig:setup_s1_v2}
\end{minipage}\hfill
\begin{minipage}{0.48\textwidth}
    \centering
    \includegraphics[width=\textwidth]{fig/camera_setting.pdf}
    \caption{Infrared camera setup for data collection: The circle represents the recording area for gesture data collection, while the rectangle represents the area for human activities and affective behavior data collection.}
    \label{fig:camera_setup}
\end{minipage}
\end{figure}

\begin{figure}
  \centering
  \begin{subfigure}[t]{0.45\textwidth}
    \centering
    \includegraphics[width=\textwidth]{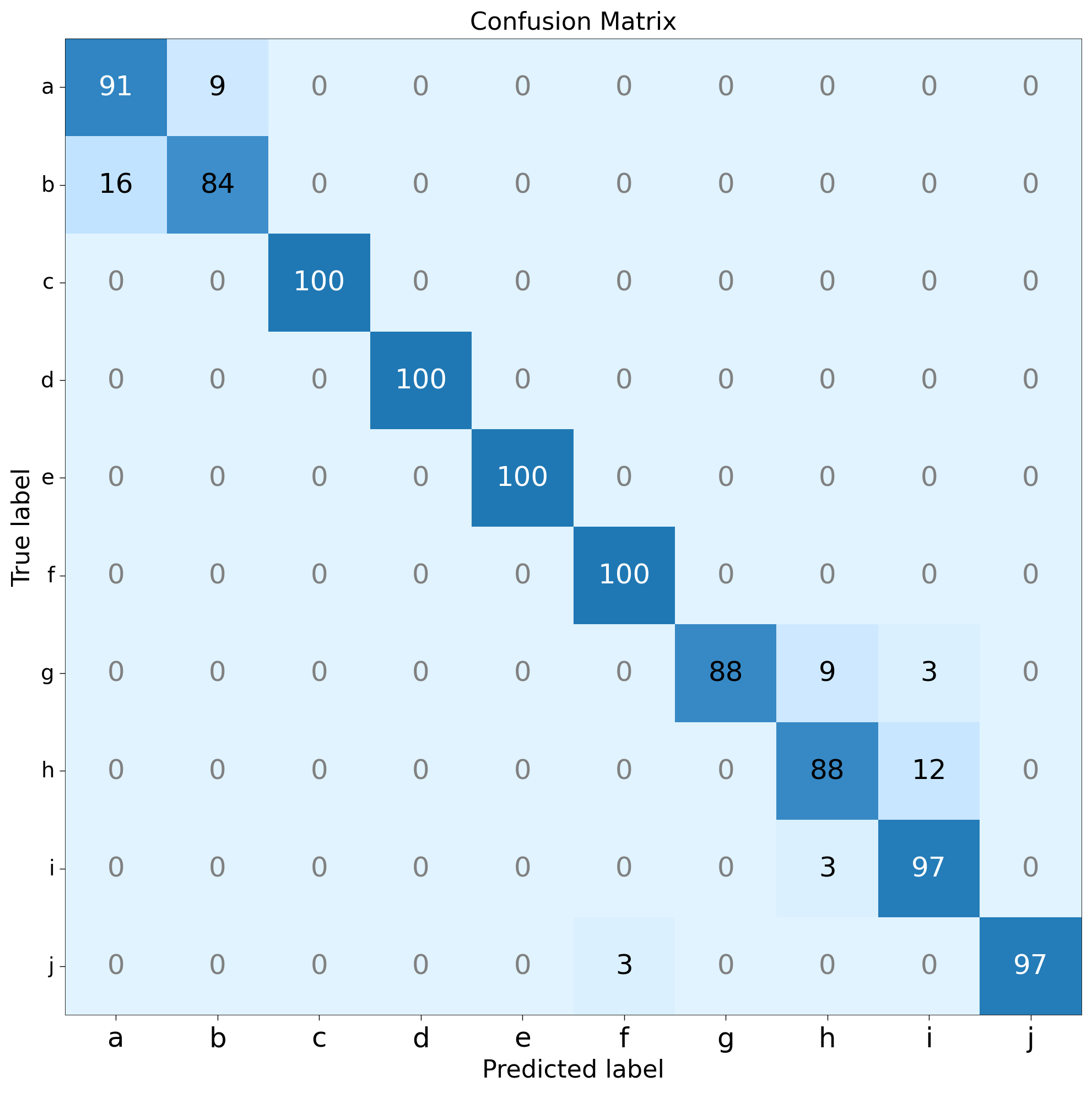}
    \caption{Confusion matrix on Radar 3.}
  \end{subfigure}
  \hfill
  \begin{subfigure}[t]{0.45\textwidth}
    \centering
    \includegraphics[width=\textwidth]{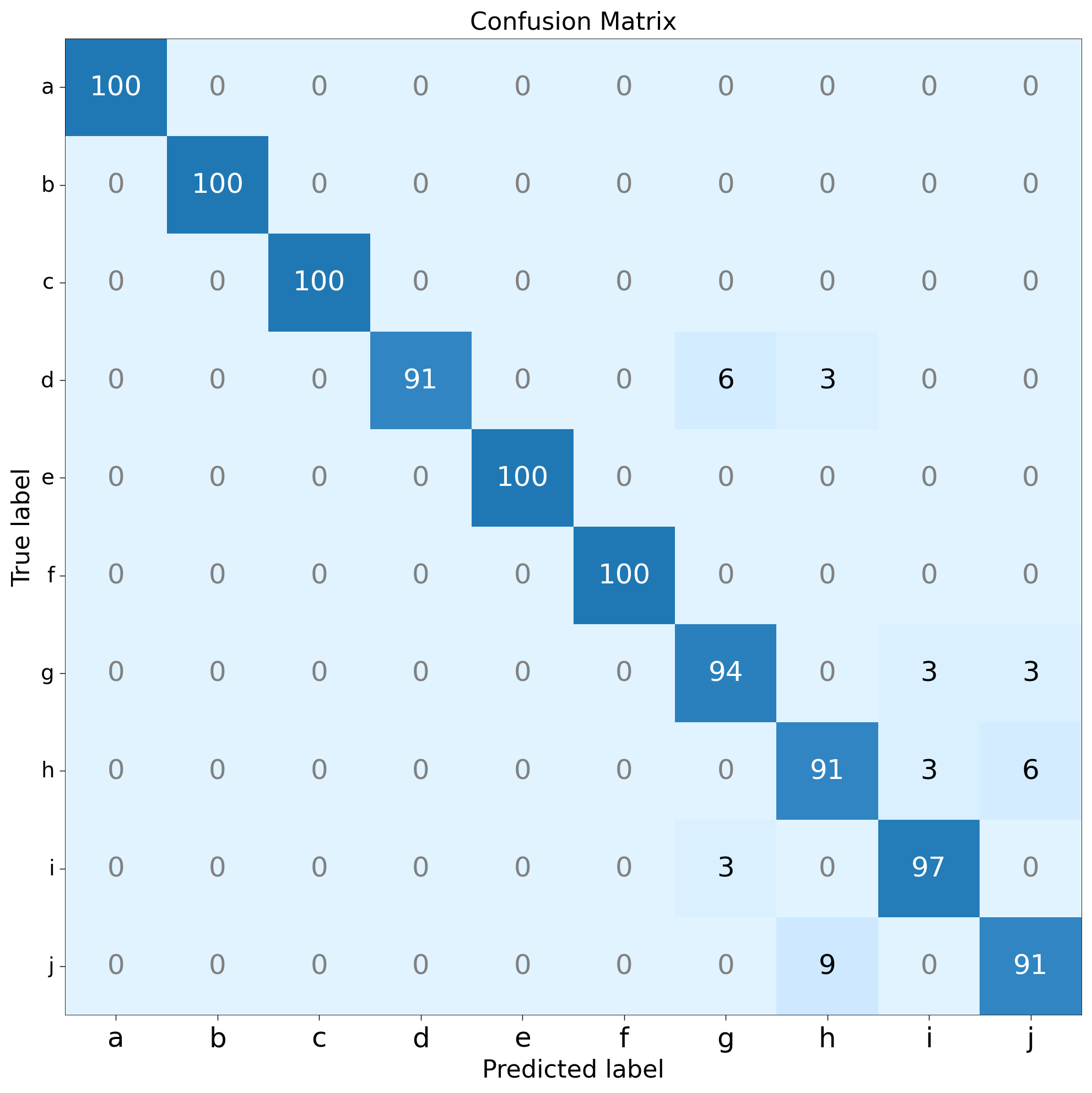}
    \caption{Confusion matrix on Radar 6.}
  \end{subfigure}
  \caption{Confusion matrices for activity recognition results from the Tesla model using 2 different radars.}
  \label{fig:cm_activity}
\end{figure}

\begin{table}
\centering
\caption{Measured 3D coordinates of the radar devices based on the ground-truth data from the infrared cameras in the recorded area.}
\label{tab:radar_coordinates}
\begin{tabular}{lccc}
\toprule
\textbf{Radar Device} & \textbf{X (m)} & \textbf{Y (m)} & \textbf{Z (m)} \\
\midrule
Device 5 & 2.6055 & -0.0940 & 0.9407 \\
Device 6 & 1.8044 & -3.4535 & 0.8953 \\
Device 7 & -0.2786 & -3.1746 & 0.9019 \\
Device 8 & -2.6325 & -3.0253 & 0.8815 \\
Device 9 & -2.9319 & -0.1305 & 0.8704 \\
Device 10 & -1.1499 & 3.2323 & 0.8784 \\
Device 11 & 0.0234 & 3.1029 & 0.9253 \\
Device 12 & 1.1062 & 2.8954 & 0.9261 \\
\bottomrule
\end{tabular}
\end{table}

\begin{table*}[t]
\centering
\caption{Cross-environment activity recognition accuracy (\%) for radar-only baselines and radar and IMU fusion. $N$ denotes the number of Scenario~2 living-room subjects added for adaptation.}
\label{tab:sensor_fusion_cross_env}
\resizebox{\textwidth}{!}{
\begin{tabular}{llcccccccc}
\toprule
\multirow{2}{*}{Protocol} & \multirow{2}{*}{Model / Fusion}
& \multicolumn{4}{c}{Scenario~2: Living Room}
& \multicolumn{4}{c}{Scenario~3: Factory} \\
\cmidrule(lr){3-6} \cmidrule(lr){7-10}
& & $N=0$ & $N=2$ & $N=4$ & $N=6$
  & $N=0$ & $N=2$ & $N=4$ & $N=6$ \\
\midrule
\multirow{2}{*}{Radar Only}
& PointNet++ & 47.4 & 53.0 & 60.0 & 62.0 & 44.7 & 37.0 & 39.0 & 37.0 \\
& Tesla      & 29.8 & 49.0 & 54.0 & 50.0 & 34.2 & 24.0 & 42.0 & 37.0 \\
\midrule
\multirow{6}{*}{\shortstack[l]{radar and IMU,\\Fine-tune}}
& PointNet++ / Cross   & -- & 42.3 & 71.7 & 61.0 & -- & 90.0 & 85.0 & 75.0 \\
& PointNet++ / Feature & -- & 44.9 & 44.7 & 64.9 & -- & 85.0 & 80.0 & 80.0 \\
& PointNet++ / Late    & -- & 100.0 & 100.0 & 98.7 & -- & 90.0 & 90.0 & 85.0 \\
& Tesla / Cross        & -- & 40.7 & 59.9 & 67.5 & -- & 90.0 & 80.0 & 95.0 \\
& Tesla / Feature      & -- & 29.4 & 59.9 & 63.6 & -- & 90.0 & 90.0 & 80.0 \\
& Tesla / Late         & -- & 100.0 & 100.0 & 100.0 & -- & 90.0 & 90.0 & 90.0 \\
\midrule
\multirow{6}{*}{\shortstack[l]{radar and IMU,\\Retrain}}
& PointNet++ / Cross   & -- & 70.1 & 79.0 & 76.6 & -- & 84.2 & 86.8 & 86.8 \\
& PointNet++ / Feature & -- & 54.1 & 61.2 & 71.4 & -- & 68.4 & 86.8 & 86.8 \\
& PointNet++ / Late    & -- & 100.0 & 100.0 & 100.0 & -- & 89.5 & 89.5 & 89.5 \\
& Tesla / Cross        & -- & 61.9 & 51.3 & 62.3 & -- & 76.3 & 31.6 & 81.6 \\
& Tesla / Feature      & -- & 47.9 & 70.4 & 64.9 & -- & 86.8 & 86.8 & 84.2 \\
& Tesla / Late         & -- & 100.0 & 100.0 & 100.0 & -- & 89.5 & 89.5 & 89.5 \\
\bottomrule
\end{tabular}}
\end{table*}

\subsection{Data Samples}\label{sec:apx_datasample}
\begin{figure}
    \centering
    \begin{subfigure}[b]{0.32\textwidth}
        \centering
        \includegraphics[width=\textwidth]{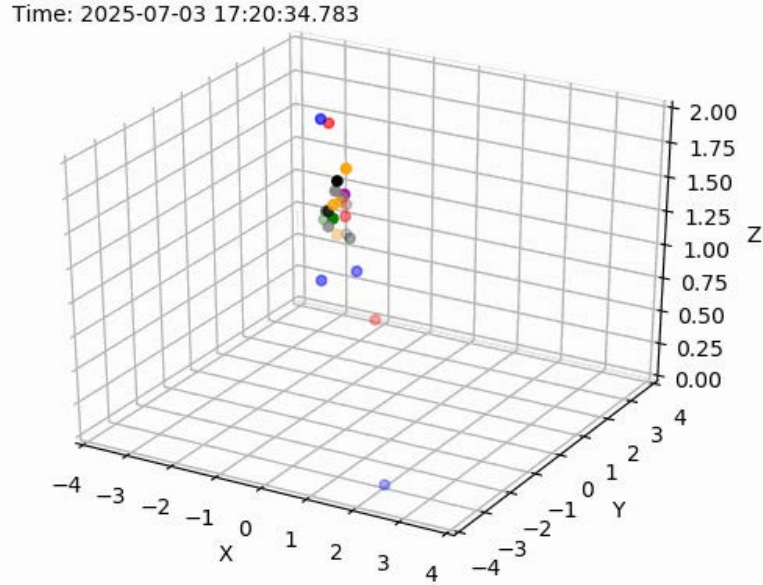}
    \end{subfigure}
    \hfill
    \begin{subfigure}[b]{0.32\textwidth}
        \centering
        \includegraphics[width=\textwidth]{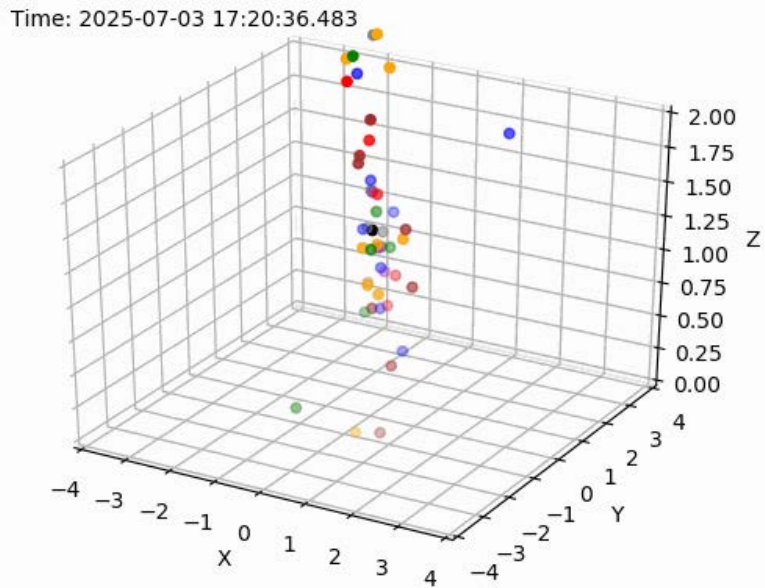}
    \end{subfigure}
    \hfill
    \begin{subfigure}[b]{0.32\textwidth}
        \centering
        \includegraphics[width=\textwidth]{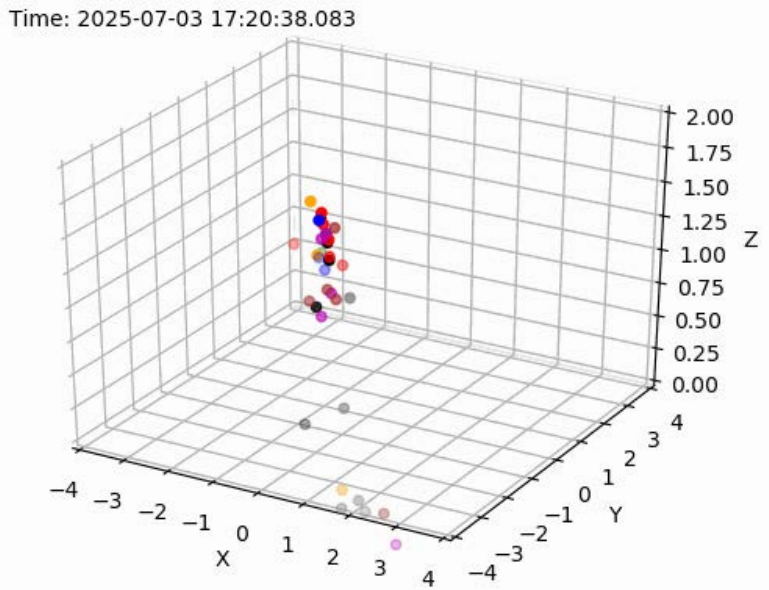}
    \end{subfigure}
    \caption{Examples of aggregated sample data collected by radars mounted on the ground.}
    \label{fig:sample_radar_row1}
    \vspace{10pt} 
    \begin{subfigure}[b]{0.32\textwidth}
        \centering
        \includegraphics[width=\textwidth]{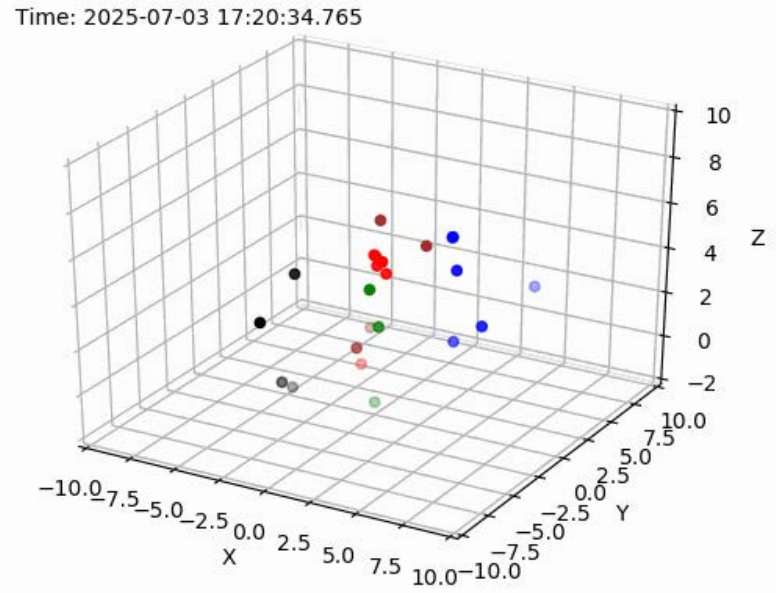}
    \end{subfigure}
    \hfill
    \begin{subfigure}[b]{0.32\textwidth}
        \centering
        \includegraphics[width=\textwidth]{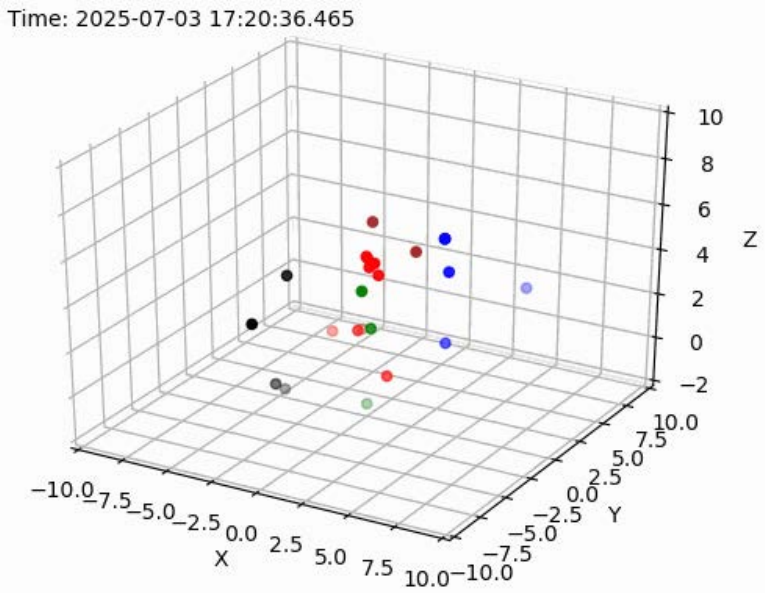}
    \end{subfigure}
    \hfill
    \begin{subfigure}[b]{0.32\textwidth}
        \centering
        \includegraphics[width=\textwidth]{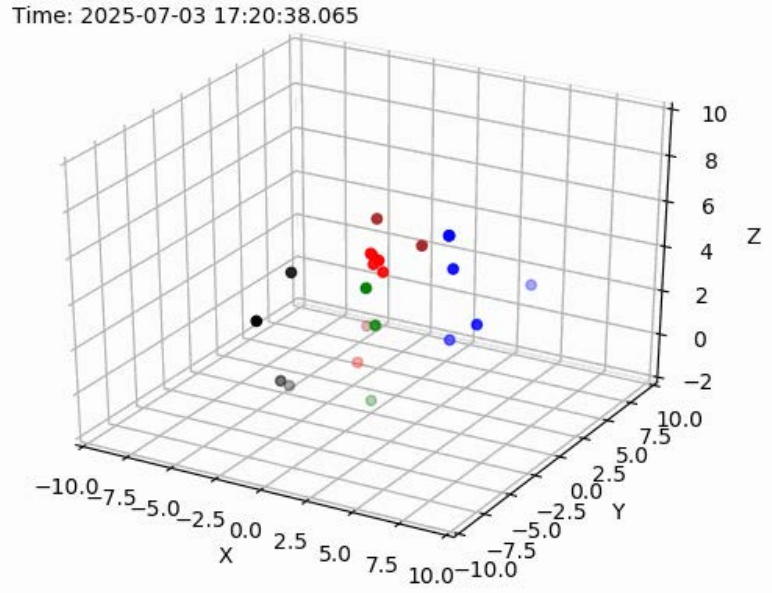}
    \end{subfigure}
    \caption{Examples of aggregated sample data collected by radars mounted on the ceiling.}
    \label{fig:sample_radar}
\end{figure}

\begin{figure}
\centering
\includegraphics[width=\textwidth]{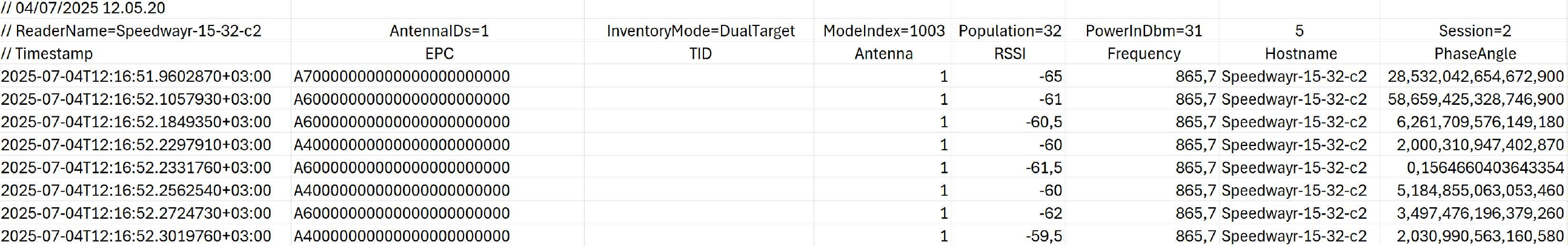}
\caption{A example of collected data from RFID. The key features are timestamp, Electronic Product Code (EPC), the Received Signal Strength (RSS), and the signal phase}
\label{fig:sample_rfid}
\end{figure}

\begin{figure}[!htbp]
	\centering
	\subfloat[A LoRa packet.]{
		\includegraphics[scale=0.3]{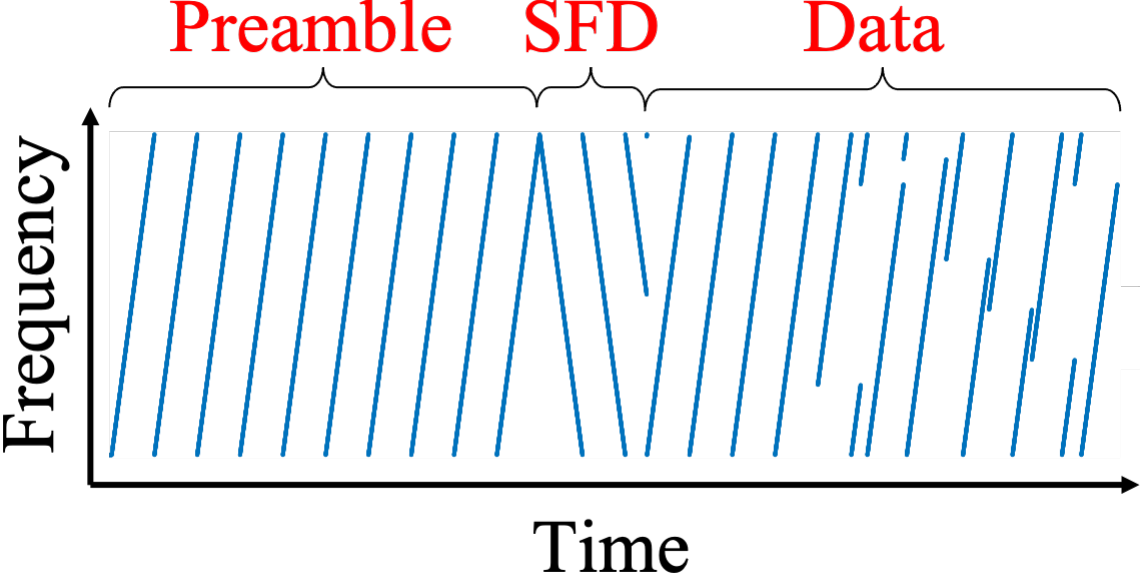}
		\label{fig:lora3a}
	}
	\hfil
	\subfloat[Data encoding.]{
		\includegraphics[scale=0.3]{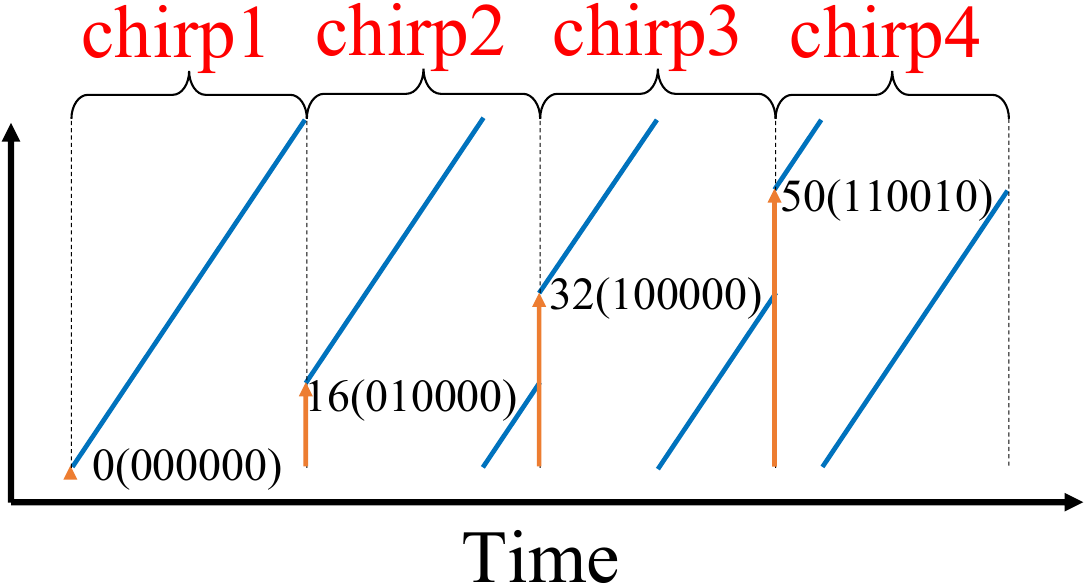}
	}
	\caption{Illustration of LoRa signal spectrum.}
	\label{fig:LoRasignal}
\end{figure}

\begin{figure}
    \centering
    \begin{subfigure}[b]{0.32\textwidth}
        \centering
        \includegraphics[width=\textwidth]{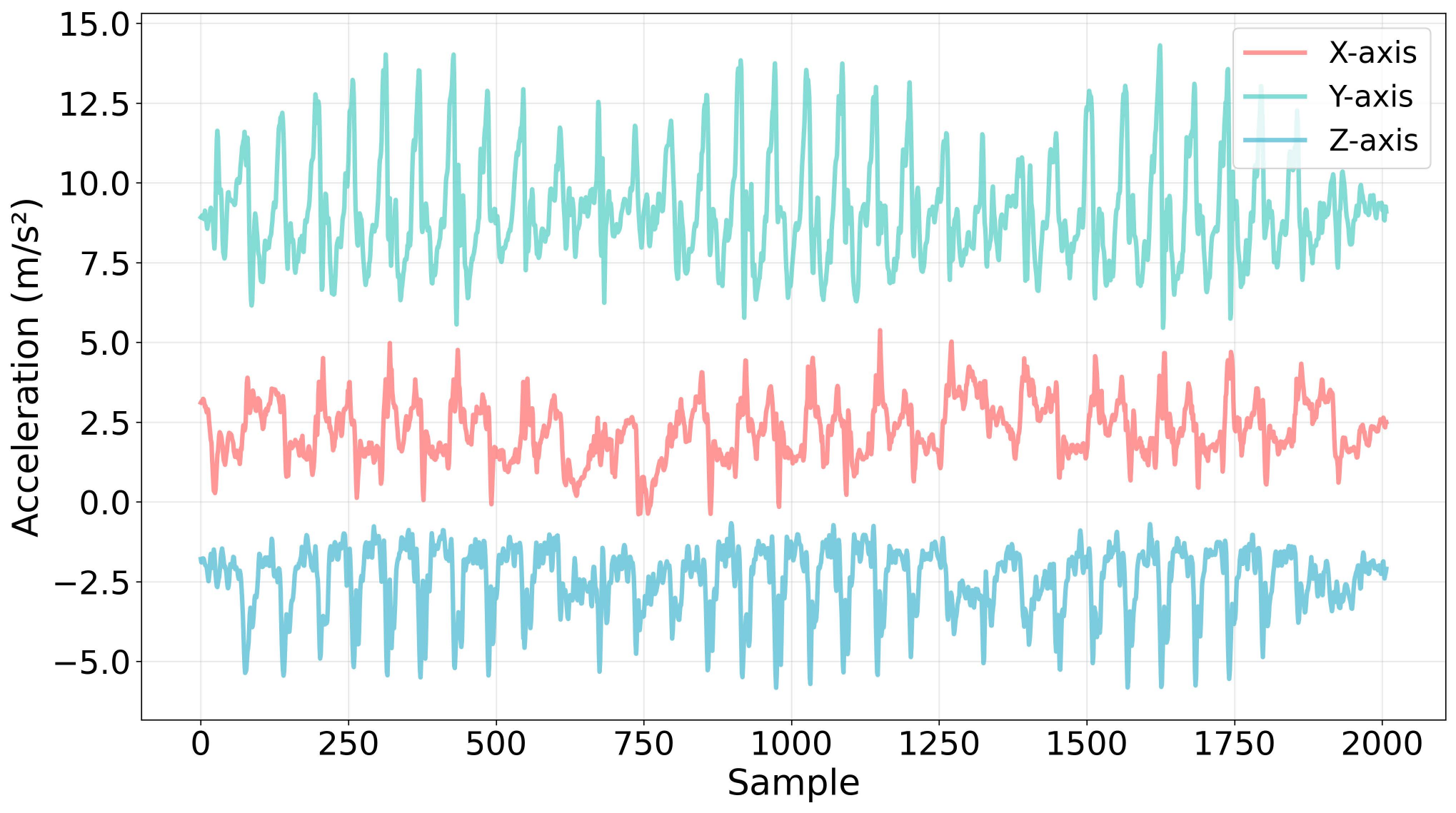}
    \end{subfigure}
    \hfill
    \begin{subfigure}[b]{0.32\textwidth}
        \centering
        \includegraphics[width=\textwidth]{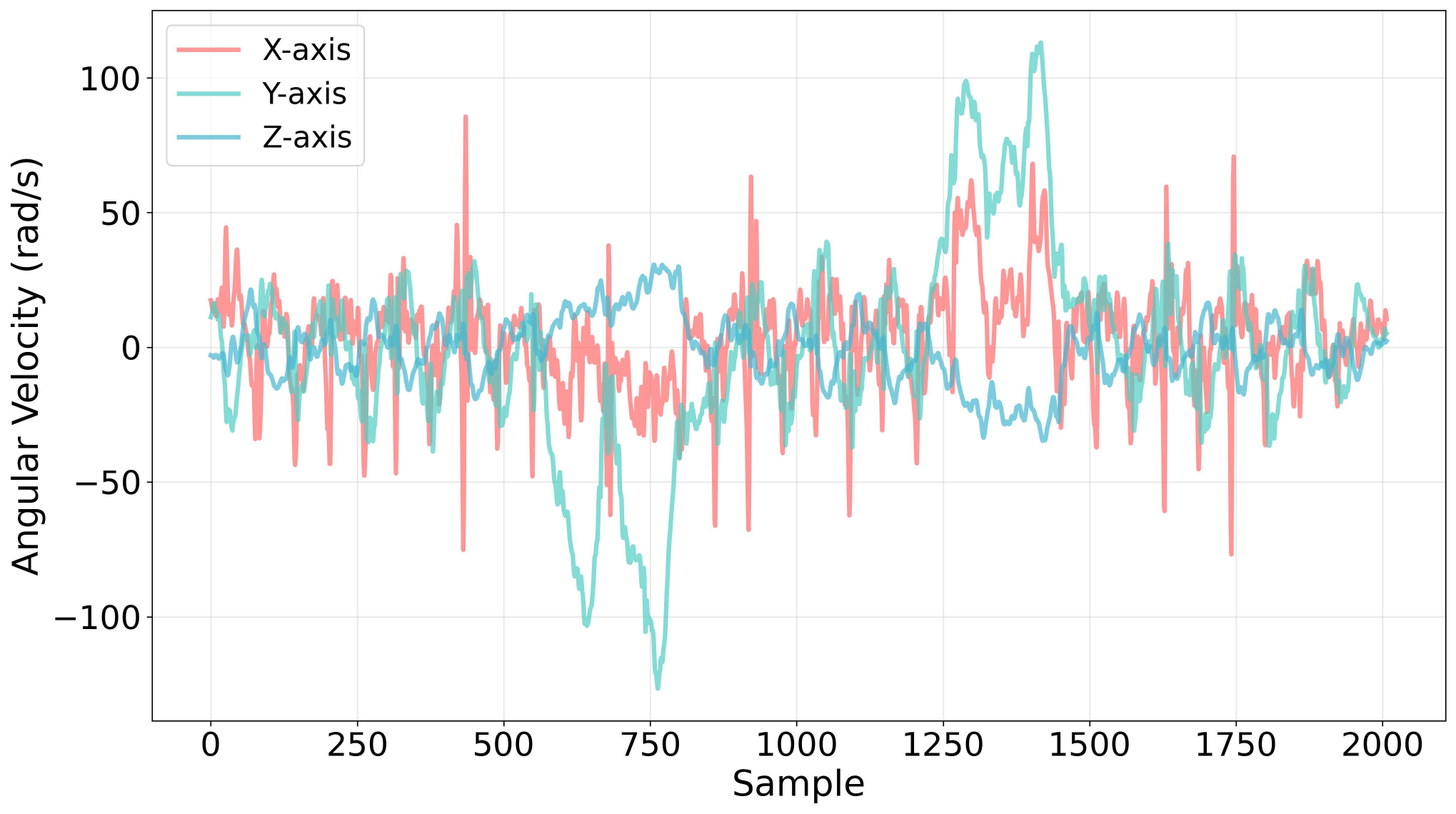}
    \end{subfigure}
    \hfill
    \begin{subfigure}[b]{0.32\textwidth}
        \centering
        \includegraphics[width=\textwidth]{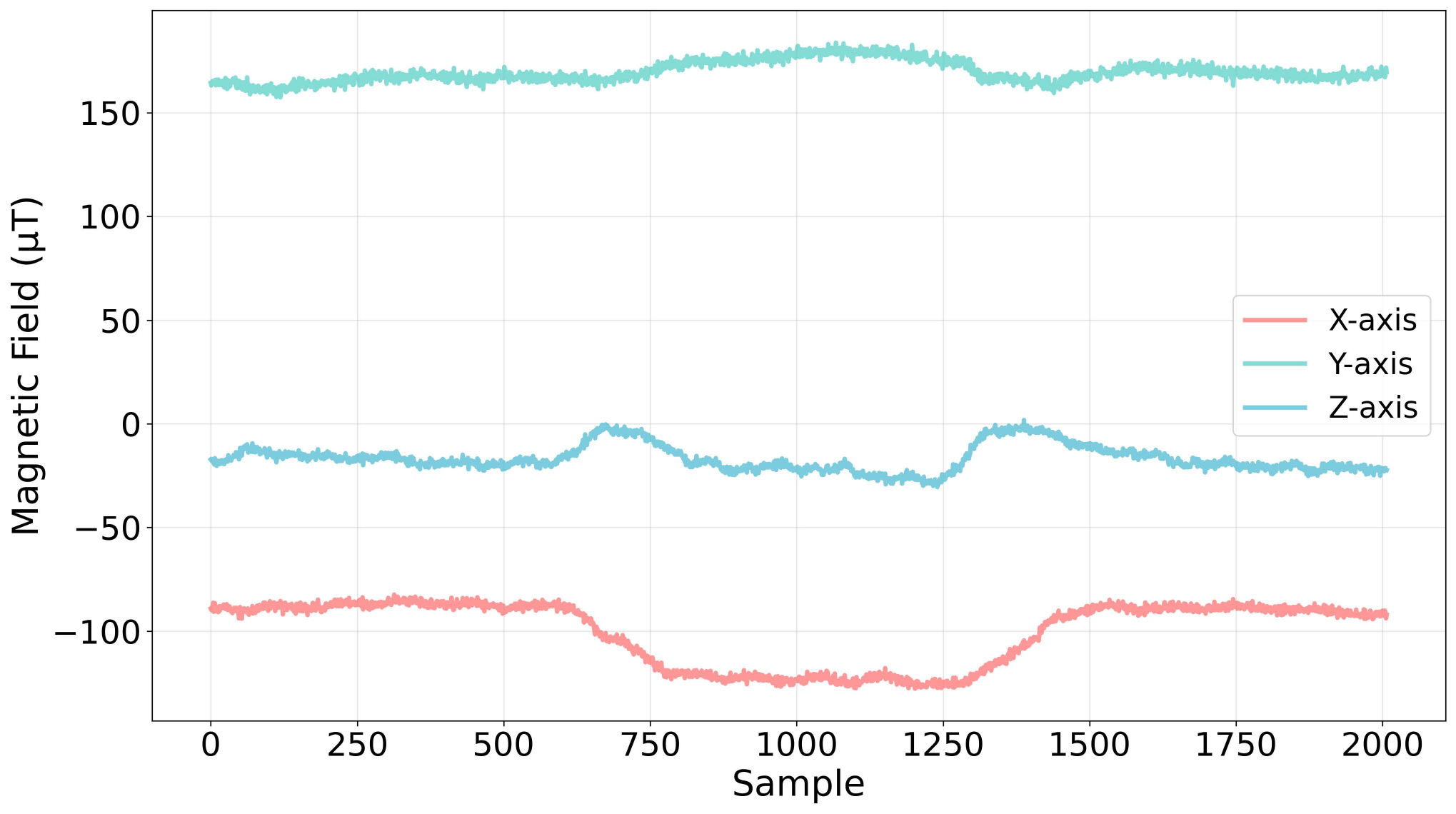}
    \end{subfigure}
    \begin{subfigure}[b]{0.32\textwidth}
        \centering
        \includegraphics[width=\textwidth]{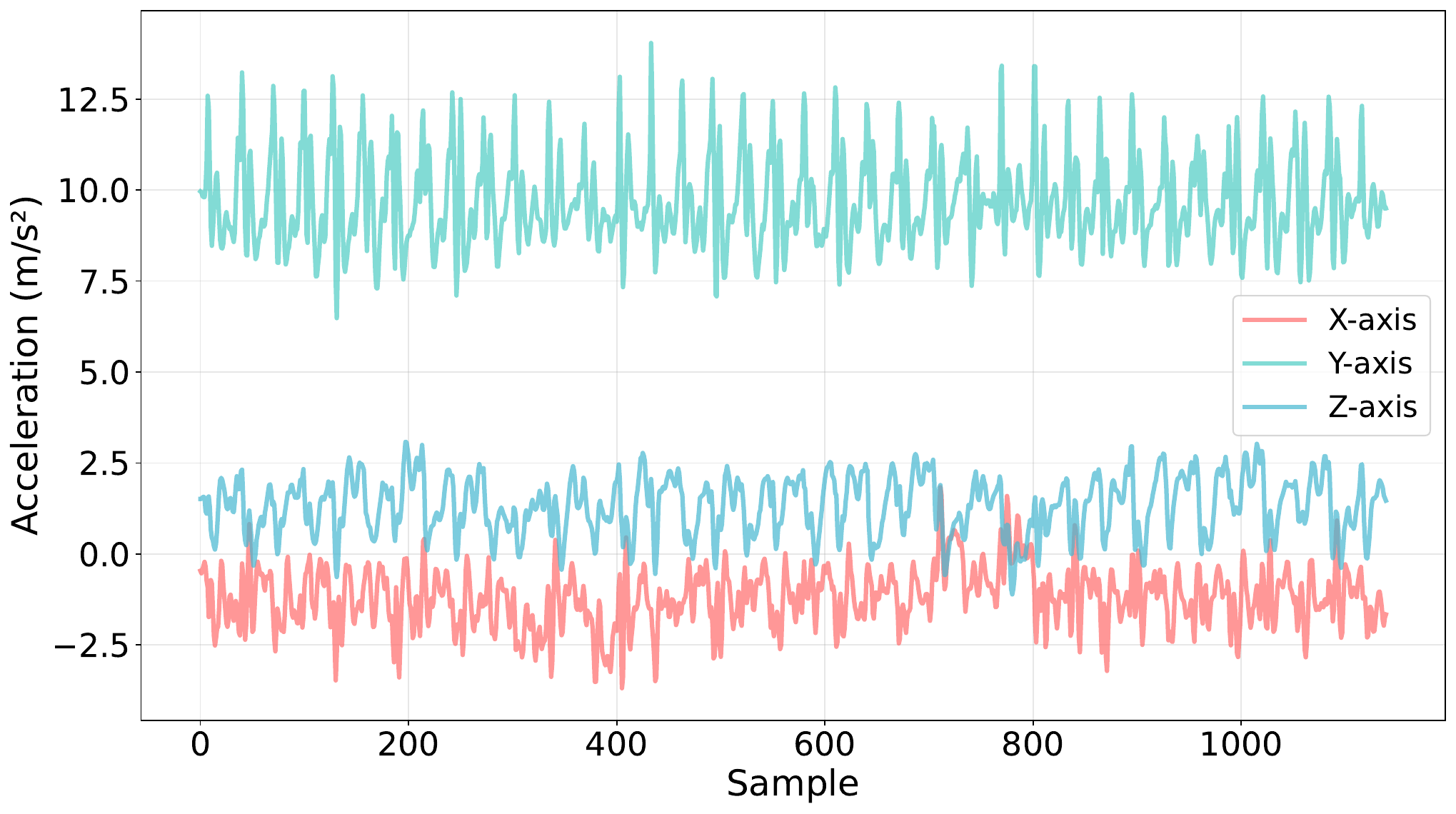}
    \end{subfigure}
    \hfill
    \begin{subfigure}[b]{0.32\textwidth}
        \centering
        \includegraphics[width=\textwidth]{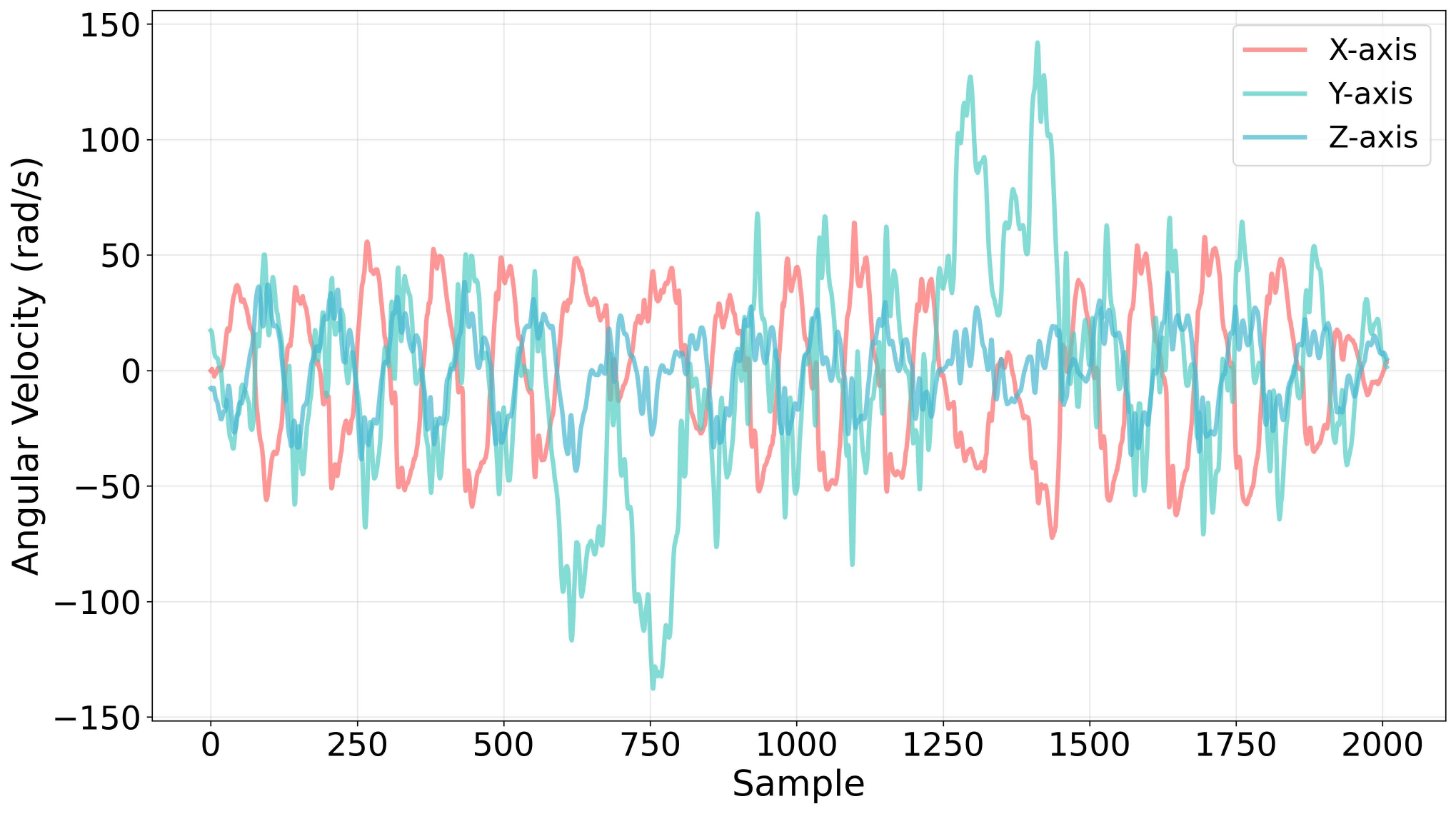}
    \end{subfigure}
    \hfill
    \begin{subfigure}[b]{0.32\textwidth}
        \centering
        \includegraphics[width=\textwidth]{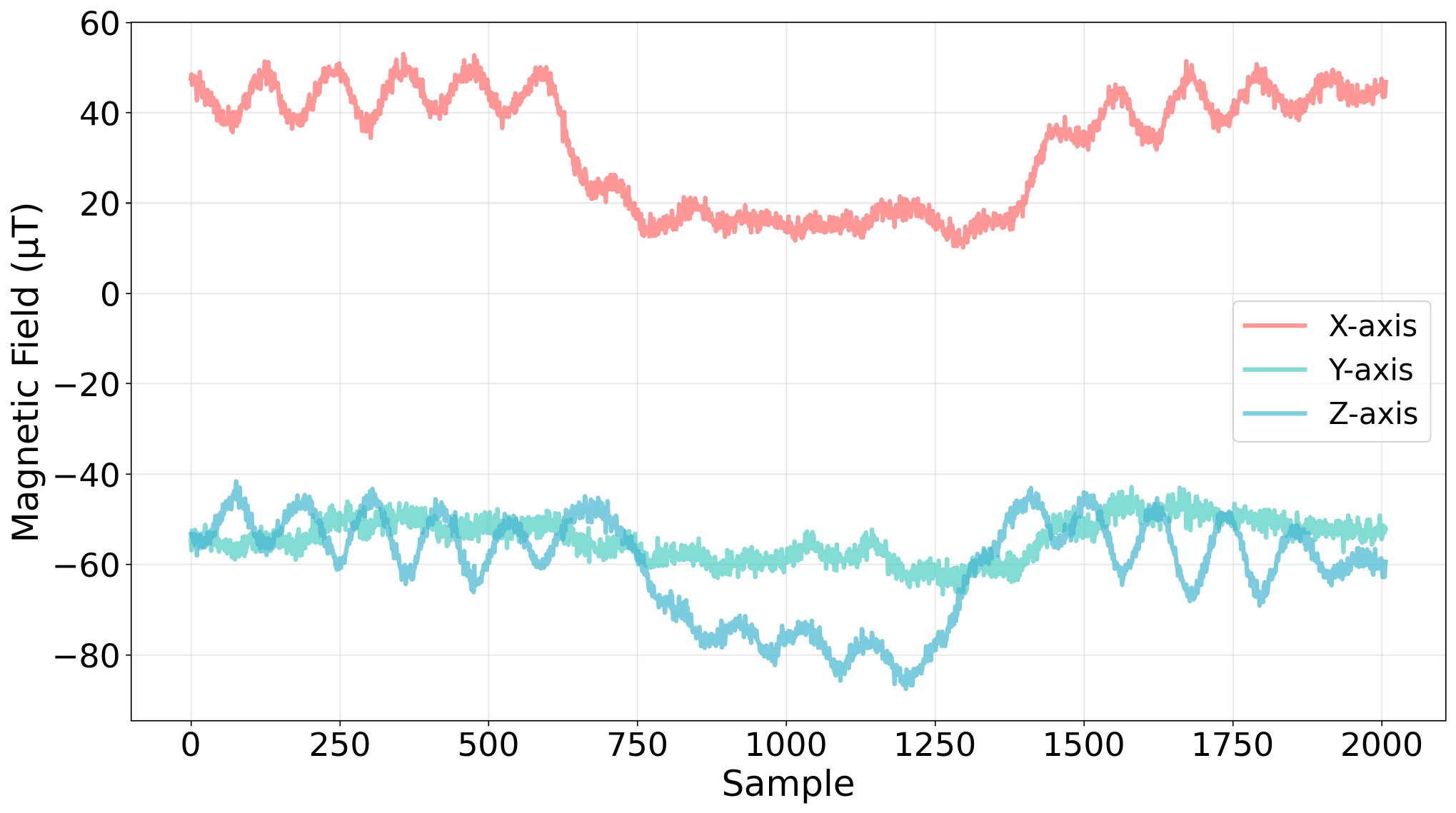}
    \end{subfigure}
    \begin{subfigure}[b]{0.32\textwidth}
        \centering
        \includegraphics[width=\textwidth]{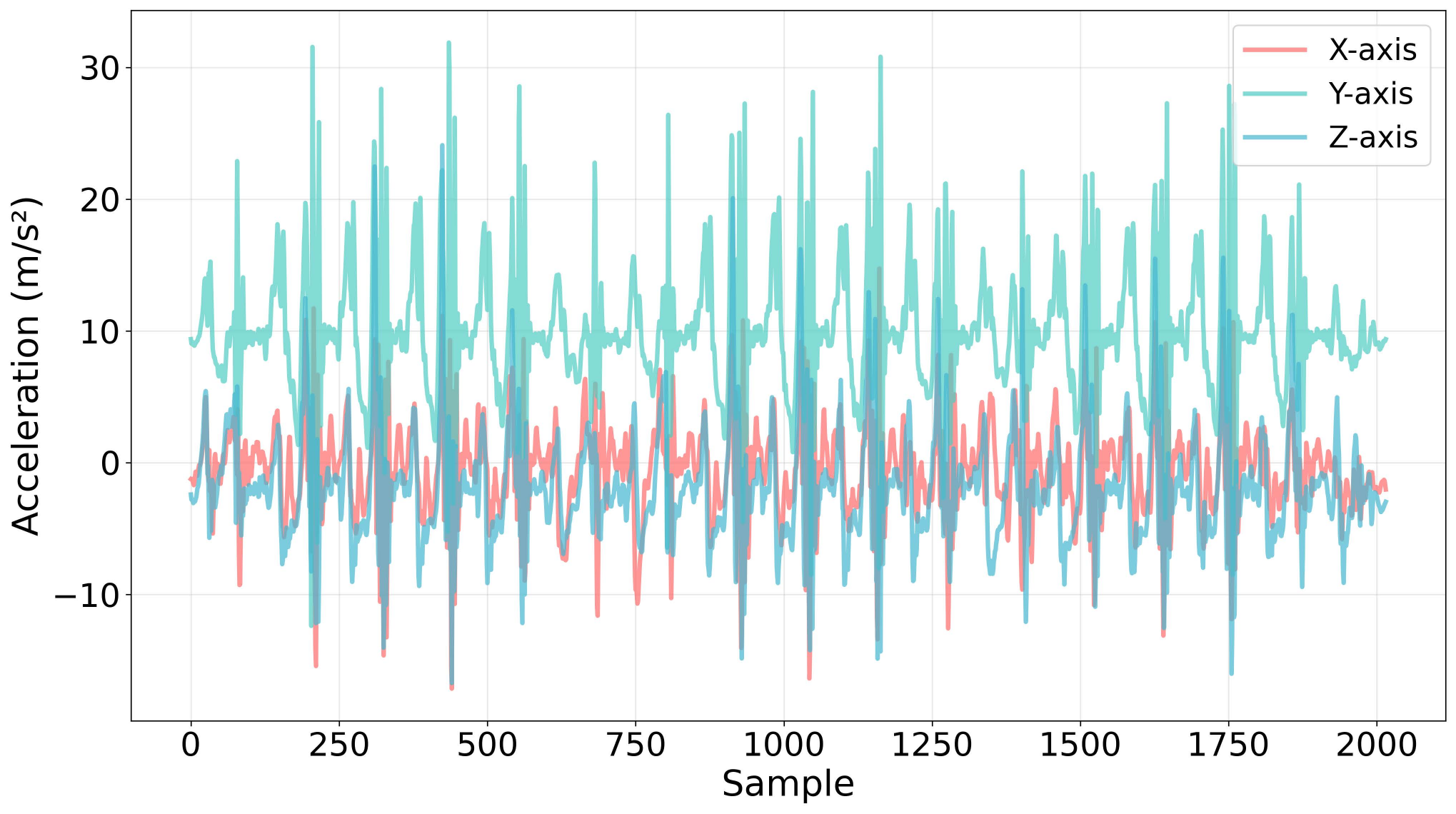}
    \end{subfigure}
    \hfill
    \begin{subfigure}[b]{0.32\textwidth}
        \centering
        \includegraphics[width=\textwidth]{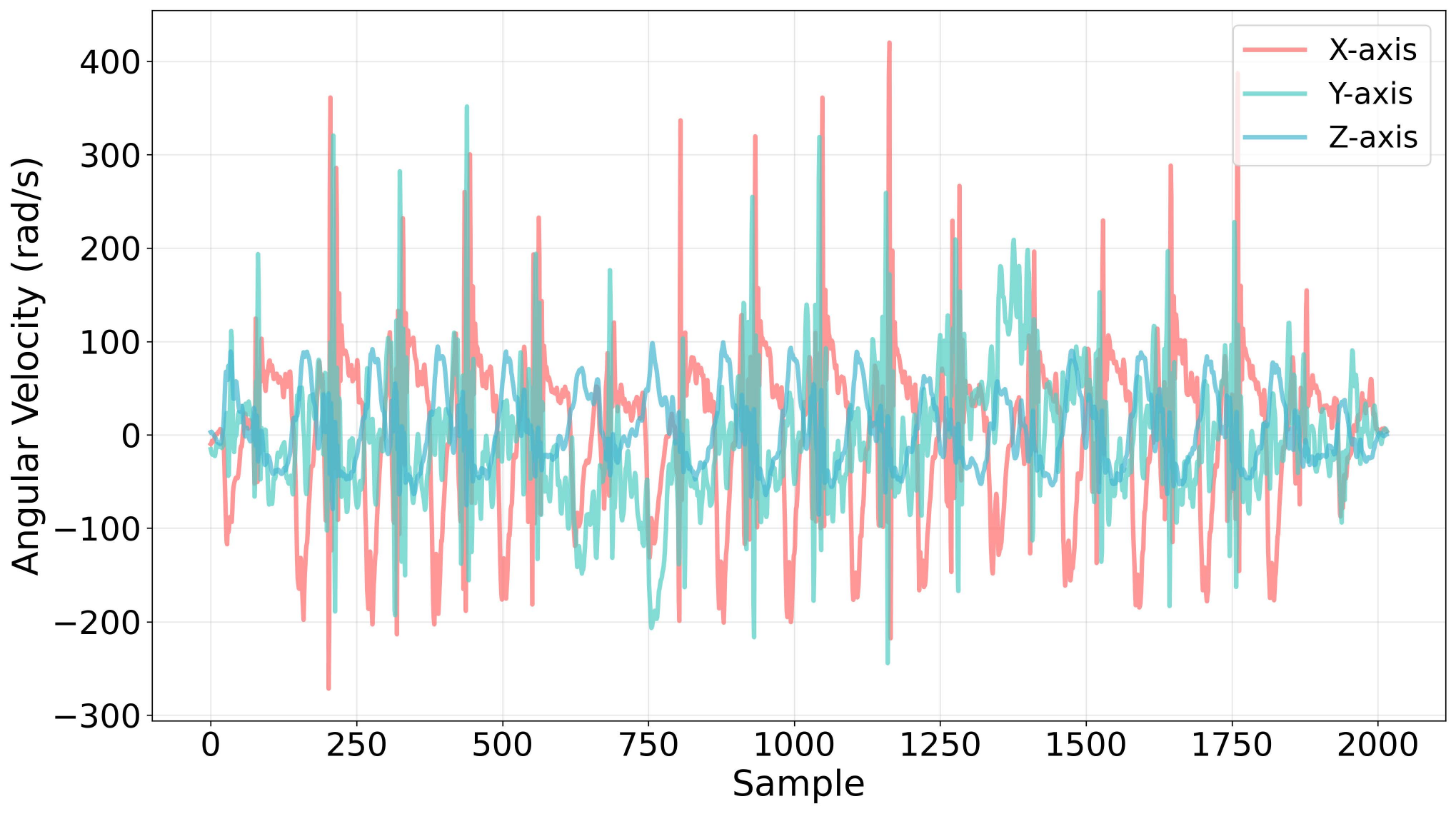}
    \end{subfigure}
    \hfill
    \begin{subfigure}[b]{0.32\textwidth}
        \centering
        \includegraphics[width=\textwidth]{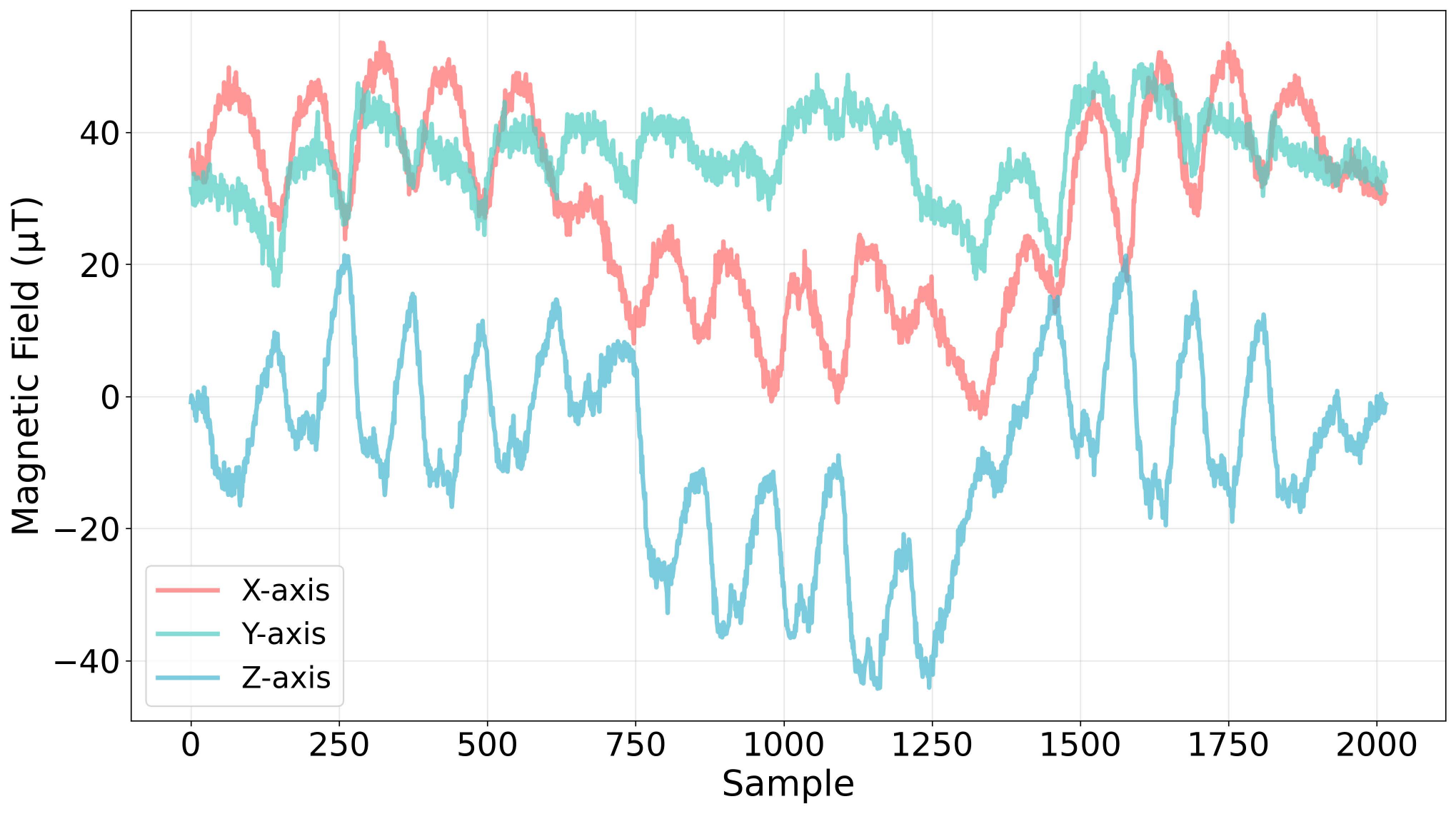}
    \end{subfigure}
    \caption{Sample data collected by the IMU sensors for walking. For each row, from left $\rightarrow$ right: Accelerometer, Gyroscope and Magnetometer. From top $\rightarrow$ bottom: Chest, Right Arm, Right Leg.}
    \label{fig:sample_imu}
\end{figure}

\begin{figure}
    \centering
    \begin{subfigure}[b]{0.32\textwidth}
        \centering
        \includegraphics[width=\textwidth]{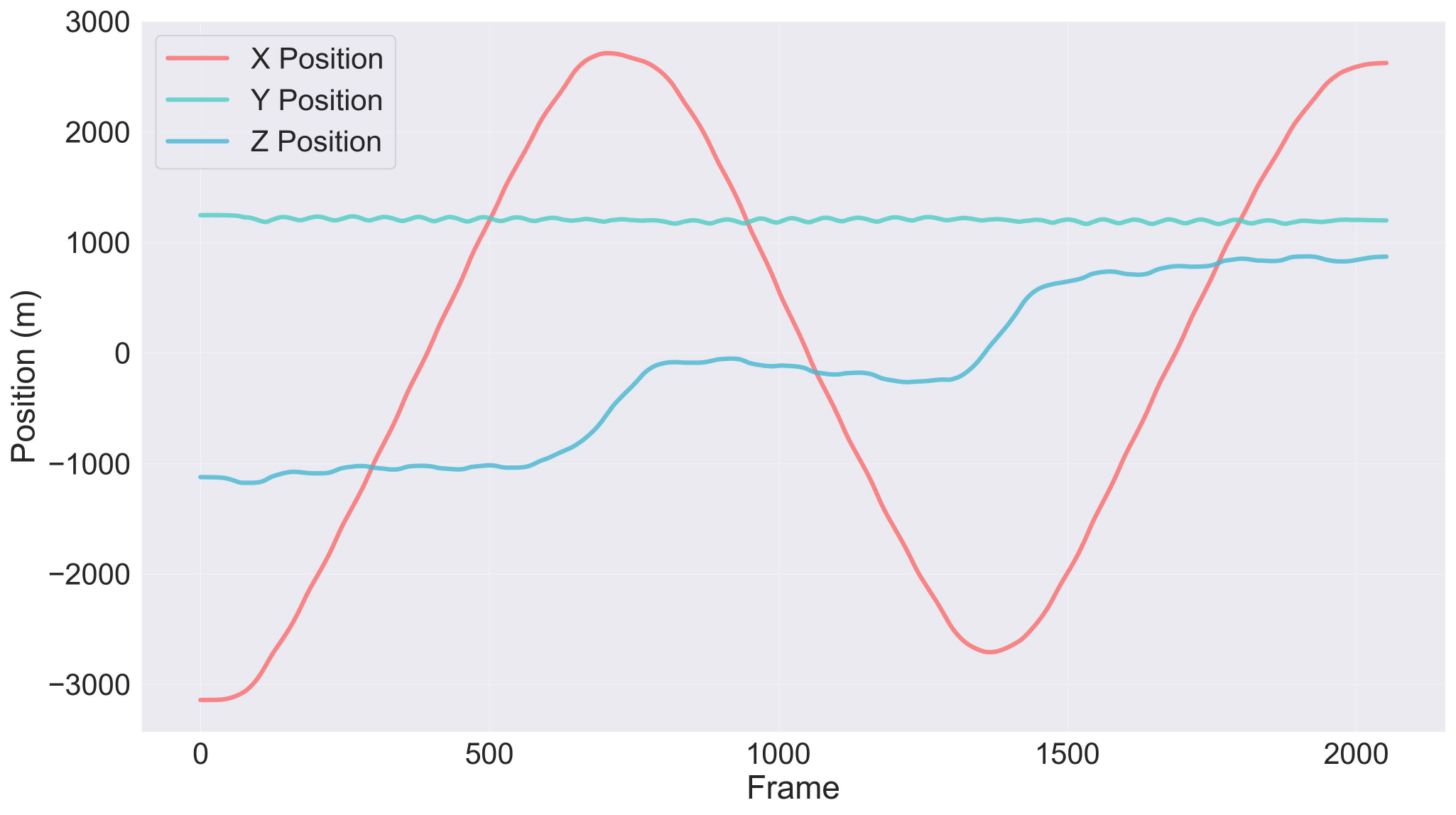}
    \end{subfigure}
    \hfill
    \begin{subfigure}[b]{0.32\textwidth}
        \centering
        \includegraphics[width=\textwidth]{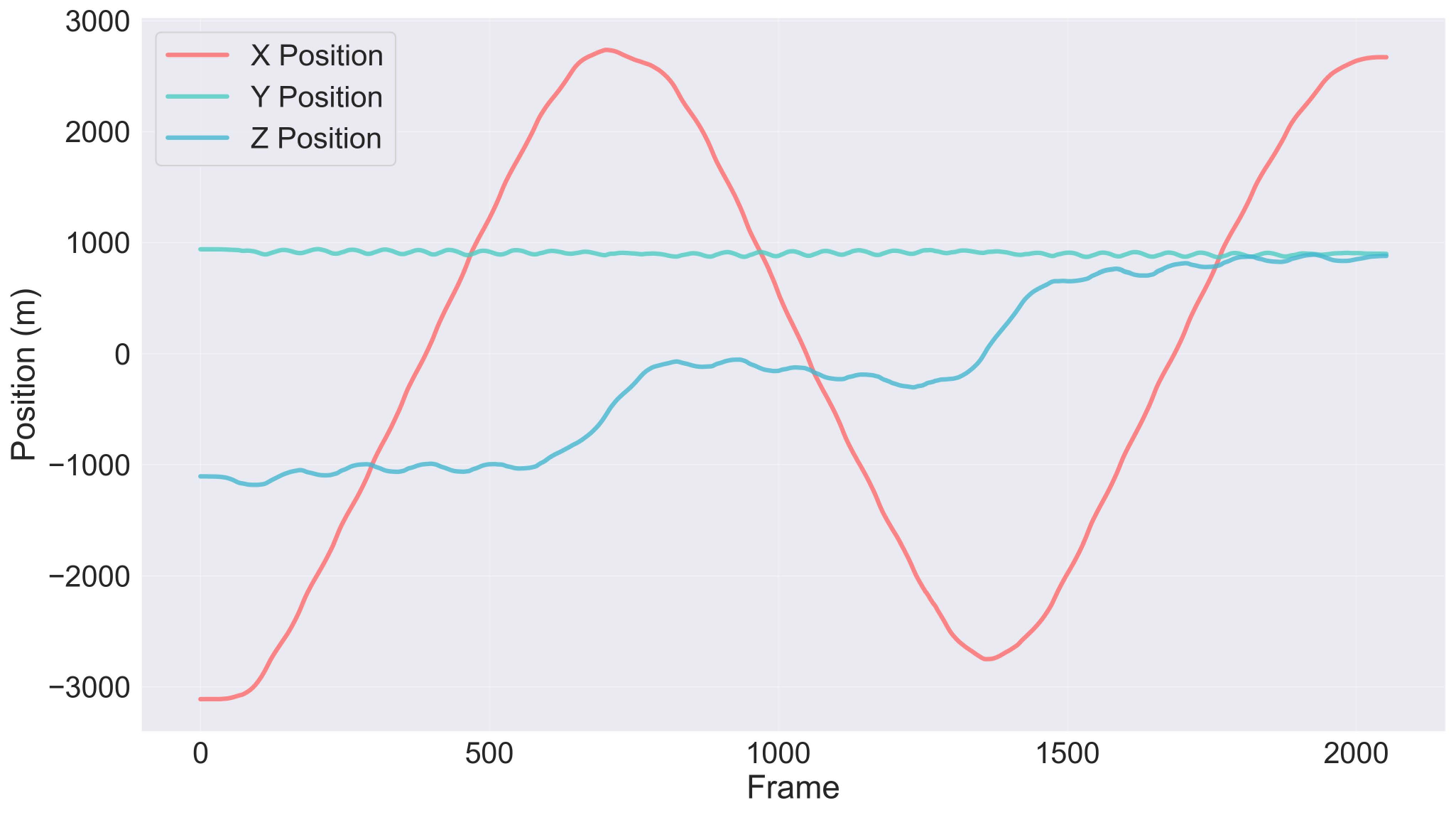}
    \end{subfigure}
    \hfill
    \begin{subfigure}[b]{0.32\textwidth}
        \centering
        \includegraphics[width=\textwidth]{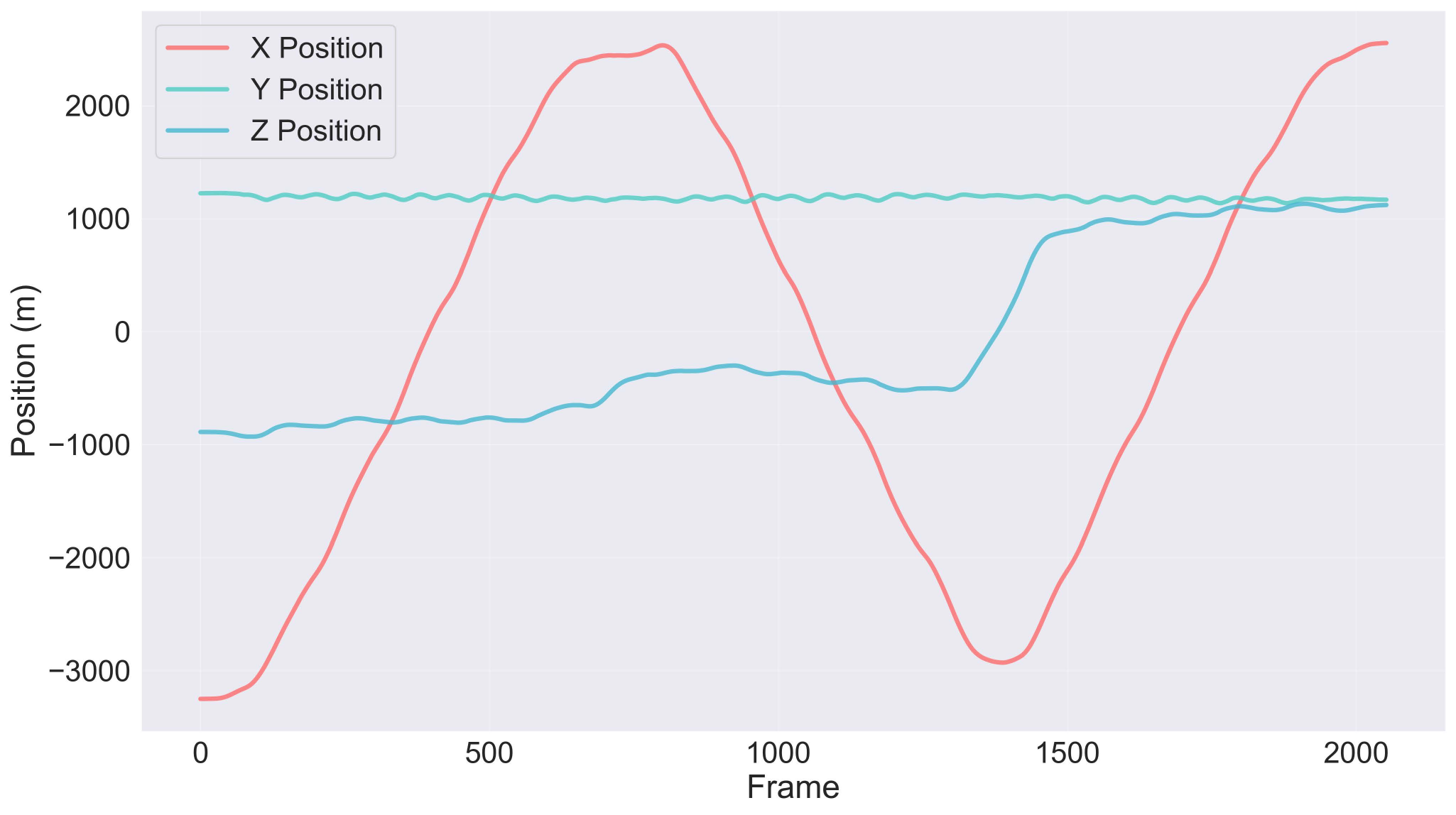}
    \end{subfigure}
    \vspace{10pt} 
    \begin{subfigure}[b]{0.32\textwidth}
        \centering
        \includegraphics[width=\textwidth]{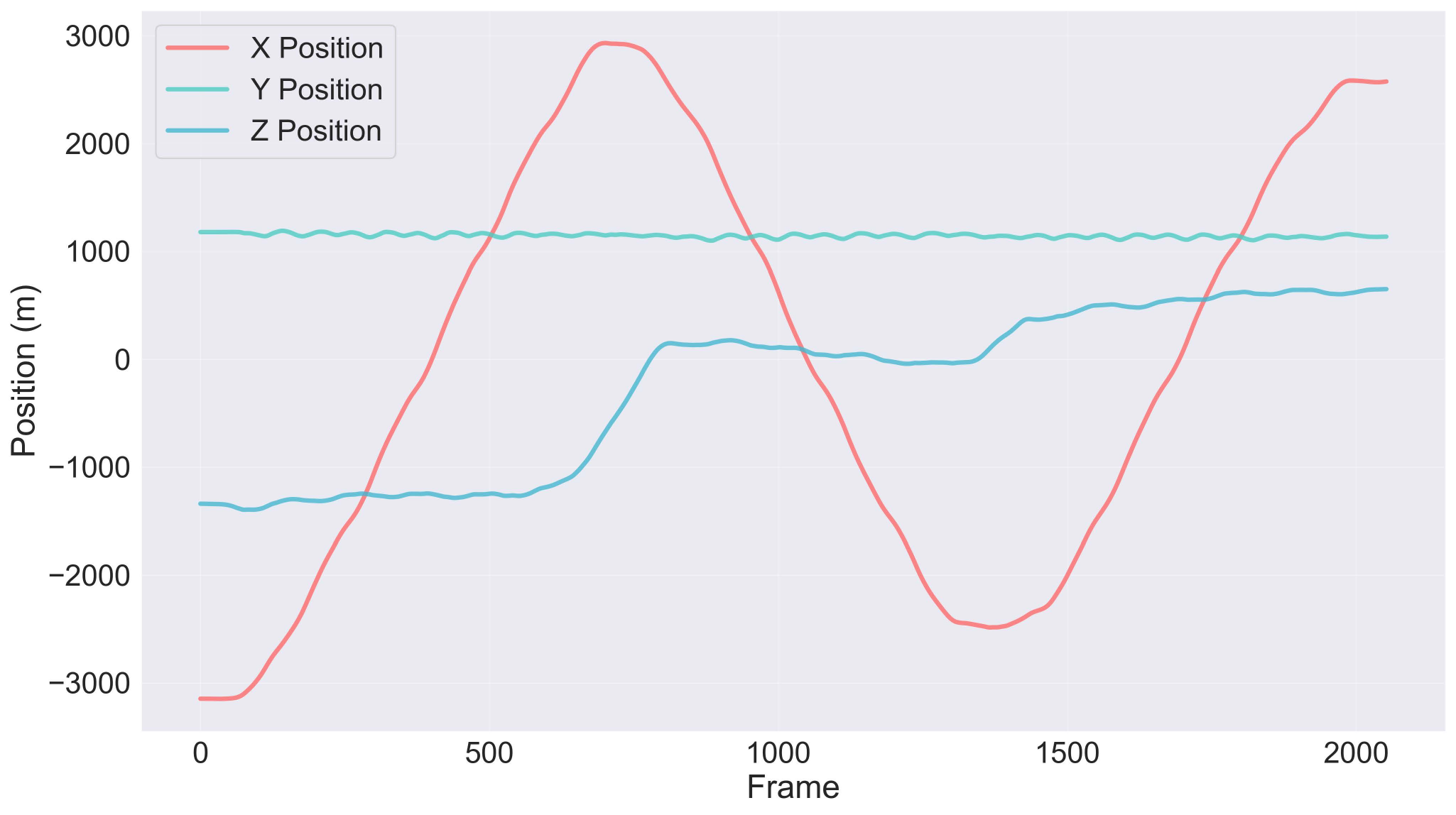}
    \end{subfigure}
    \hfill
    \begin{subfigure}[b]{0.32\textwidth}
        \centering
        \includegraphics[width=\textwidth]{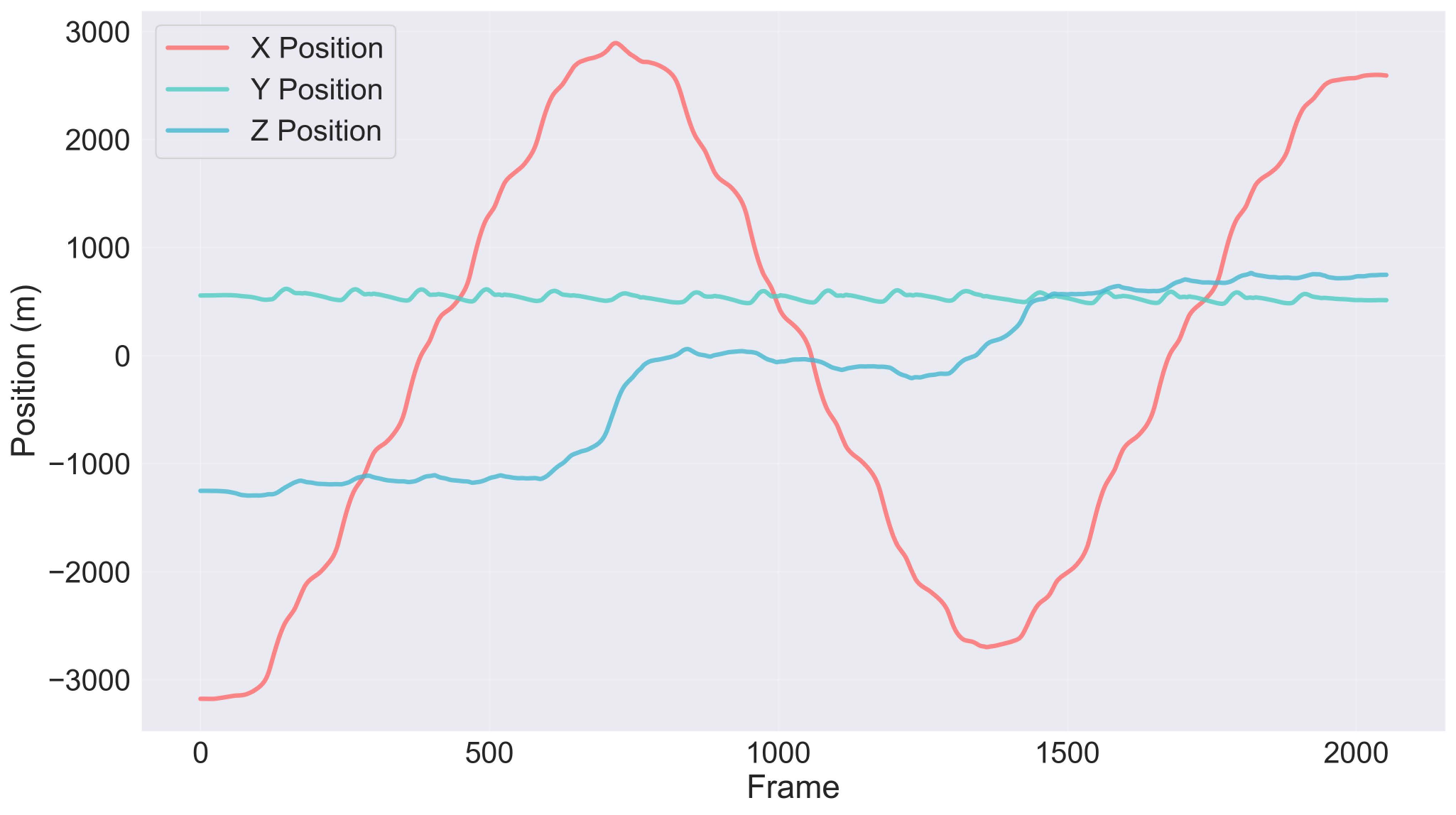}
    \end{subfigure}
    \hfill
    \begin{subfigure}[b]{0.32\textwidth}
        \centering
        \includegraphics[width=\textwidth]{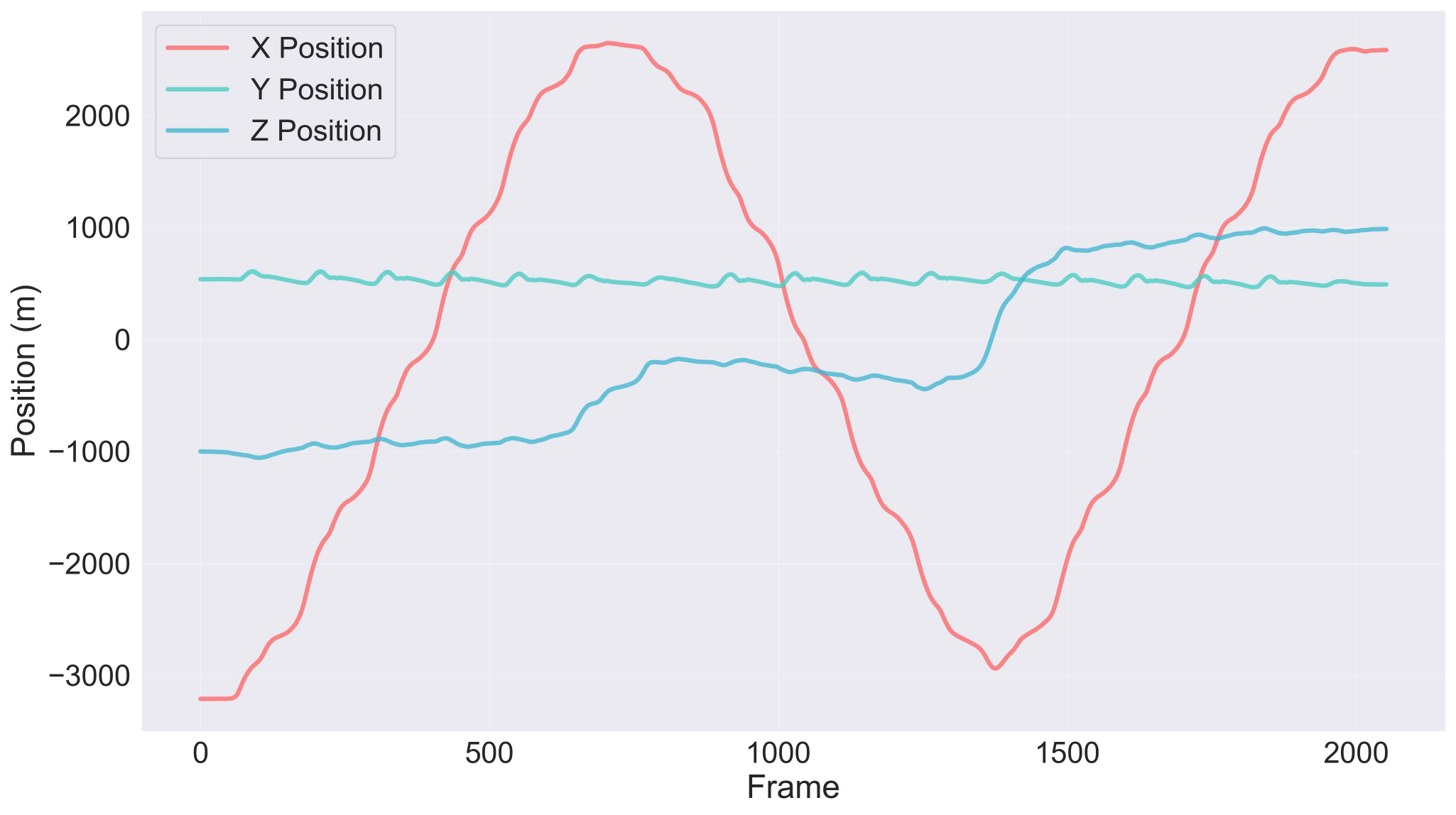}
    \end{subfigure}
    \caption{Examples of sample data collected from different body locations by the infrared camera during walking. Top row: chest, hips, left arm; Bottom row: right arm, left leg, right leg. The camera system tracked markers attached to various body parts within the capture space. The visualizations from different body locations appear similar because all markers remain in close proximity to one another on the body, and their relative distances are small compared to the participant's overall displacement during walking.}
    \label{fig:sample_ir}
\end{figure}

Data samples from Radar, RFID, LoRa, IMU and Infrared Camera are shown in Fig.~\ref{fig:sample_radar}, Fig.~\ref{fig:sample_rfid}
, Fig.~\ref{fig:LoRasignal}, Fig.~\ref{fig:sample_imu} and Fig.~\ref{fig:sample_ir}, respectively. For LoRa, the figure provides an example of data encoding with SF = 6. In this case, a chirp can encode data using~64 different starting frequencies, e.g.,~0 to~63. 

\end{document}